\documentclass[a4paper,fleqn,usenatbib]{mnras}
\usepackage{amsmath}	
\usepackage{txfonts}
\usepackage[T1]{fontenc}
\usepackage{ae,aecompl}
\usepackage{graphicx}	
\usepackage{amssymb}	
\usepackage{caption}
\usepackage{subcaption}
\newcommand{ \Fe }{Fe~{\scriptsize XVII} }
\newcommand{ \sig }{$\sigma$ }
\newcommand{ \kms }{~km~s$^{-1}$}

\title[Measurments of turbulence in elliptical galaxies]{Improved measurements of turbulence in the hot gaseous atmospheres of nearby giant elliptical galaxies}
\author[A. Ogorzalek et al.]{A. Ogorzalek$^{1,2}$\thanks{E-mail: ogoann@stanford.edu}, I. Zhuravleva$^{1,2}$, S. W. Allen$^{1,2, 3}$, C. Pinto$^{4}$,  N. Werner$^{5,6}$,\newauthor A. B. Mantz$^{1,2}$,  R. E. A. Canning$^{1,2}$, A. C. Fabian$^{4}$, J. S. Kaastra$^{7,8}$, and J. de Plaa$^{7}$ \\
$^{1}$Department of Physics, Stanford University, 382 Via Pueblo Mall, Stanford, CA 94305-4060, USA\\
$^{2}$Kavli Institute for Particle Astrophysics and Cosmology, Stanford University, 452 Lomita Mall,  Stanford, CA 94305-4085, USA\\
$^{3}$SLAC National Accelerator Laboratory, 2575 Sand Hill Road, Menlo Park, CA 94025, USA\\
$^{4}$Institute of Astronomy, Madingley Road, CB3 0HA Cambridge, United Kingdom\\
$^{5}$MTA-E\"{o}tv\"{o}s University Lend\"{u}let Hot Universe Research Group, P\'{a}zm\'{a}ny P\'{e}ter s\'{e}t\'{a}ny 1/A, Budapest, 1117, Hungary\\
$^{6}$Department of Theoretical Physics and Astrophysics, Faculty of Science, Masaryk University, Kotl\'{a}\v{r}sk\'{a} 2, Brno, 611 37, Czech Republic\\
$^{7}$SRON Netherlands Institute for Space Research, Sorbonnelaan 2, 3584 CA Utrecht, The Netherlands\\
$^{8}$Leiden Observatory, Leiden University, P.O. Box 9513, 2300 RA Leiden,The Netherlands}
\date{Accepted XXX. Received YYY; in original form ZZZ}

\pubyear{2017}

\begin{document}

\label{firstpage}
\pagerange{\pageref{firstpage}--\pageref{lastpage}}
\maketitle

\begin{abstract} 
We present significantly improved measurements of turbulent velocities in the hot gaseous halos of nearby giant elliptical galaxies. Using deep XMM-\textit{Newton} \textit{Reflection Grating Spectrometer} (\textit{RGS}) observations and a combination of resonance scattering and direct line broadening methods, we obtain well bounded constraints for 13 galaxies. Assuming that the turbulence is isotropic, we obtain a best fit mean 1D turbulent velocity of $\sim$ 110 \kms. This implies a typical 3D Mach number  $\sim$ 0.45 and a typical non-thermal pressure contribution of $\sim$ 6 per cent in the cores of nearby massive galaxies. The intrinsic scatter around these values is modest - consistent with zero, albeit with large statistical uncertainty - hinting at a common and quasi--continuous mechanism sourcing the velocity structure in these objects. Using conservative estimates of the spatial scales associated with the observed turbulent motions, we find that turbulent heating can be sufficient to offset radiative cooling in the inner regions of these galaxies (<~10~kpc, typically 2--3~kpc). The full potential of our analysis methods will be enabled by future X-ray micro--calorimeter observations.
\end{abstract}

\section{Introduction}
\label{sec:intro}

Giant elliptical galaxies are filled with hot ($\sim10^{7}$~K), X-ray emitting gas. Various physical processes, e.g. galaxy mergers, supernova explosions, and AGN activity, can perturb this gas and induce gas motions. These motions transport energy, mass, and metals, influencing the thermodynamic state of galaxies. Therefore, measurements of X-ray gas velocities  are crucial for understanding galaxy formation and evolution \citep[e.g.][]{Sijacki2007, Weinberger2017}, radio--mechanical AGN feedback \citep[e.g.][]{Bruggen2005, Scannapieco2008}, and transport processes \citep[e.g.][]{Roediger2013, Kunz2014}.

A direct way to probe  gas velocity field is by measuring spectral line broadening (differential motions), and  shifts in line centroids (bulk velocities). Recently, the \textit{Hitomi} satellite measured a velocity broadening of 187$\pm$13~km~s$^{-1}$ in the inner 30 kpc of   Perseus cluster, where the galaxy NGC~1275 resides \citep{Hitomi_Perseus}.  If these motions come from turbulence triggered by AGN-blown bubbles (on 20--30~kpc scales), then the associated heat dissipation would be sufficient to offset radiative cooling \citep{Hitomi_Perseus}. This result is  consistent with the velocity structure inferred from the analysis of  X-ray surface brightness fluctuations  in the Perseus Cluster (\citealt{Irina2014}; see also results for the Virgo and Centaurus clusters by \citealt{Irina2014} and \citealt{Walker2015}).

For most giant elliptical galaxies and clusters, however, detailed gas velocity information is lacking.  Additional measurements, extending to more systems, are vital to better understand the role of turbulent heating, as well as other processes, such as e.g. weak shocks and sound waves \citep[e.g.][]{Fabian2006, Randall2015, Fabian2017}, cosmic ray heating \citep[e.g.][]{Guo2008, Sijacki2008, Ruszkowski2017, Jacob2017},  bubble mixing \citep[e.g.][]{Hillel2016}, radiative heating \citep[e.g.][]{Nulsen2000}, and turbulent mixing \citep[e.g.][]{Kim2003}.

The compact, cool, X-ray bright galaxies can be analyzed using high resolution X-ray spectra from \textit{Reflection Grating Spectrometer} (\textit{RGS}). Applying it to a sample of group central galaxies, \cite{Sanders2013} and \cite{Pinto2015} provided velocity upper limits of $\sim$ 300~km~s$^{-1}$. A major systematic uncertainty associated with the  spatial extent of the source, which additionally  broadens the spectral lines, typically limits these constraints to upper bounds only. 

A promising indirect method of probing gas motions is  resonant scattering effect. The X-ray emitting gas in elliptical galaxies is optically thin in the continuum and most line emission (i.e. the optical depth for free-free absorption is $\tau \sim 10^{-29}$,  and is  $\sim 10^{-2}$ for Thomson scattering). However, in the X-ray brightest resonance lines of abundant ions, the optical depth can be greater than one \citep{GILFANOV1987}. This causes photons to be absorbed and then quickly re-emitted, leading to line flux suppression in a galaxy or cluster core \citep[e.g.][]{1998Shigeyama}. Small scale turbulent gas motions broaden the line, lowering its optical depth and reducing this effect. Thus resonant scattering can be used as a sensitive velocity diagnostic tool \citep[for a review, see][]{2010review}. 

\begin{figure}
\includegraphics[width=\columnwidth,trim={4mm 2mm 14mm 7mm},clip]{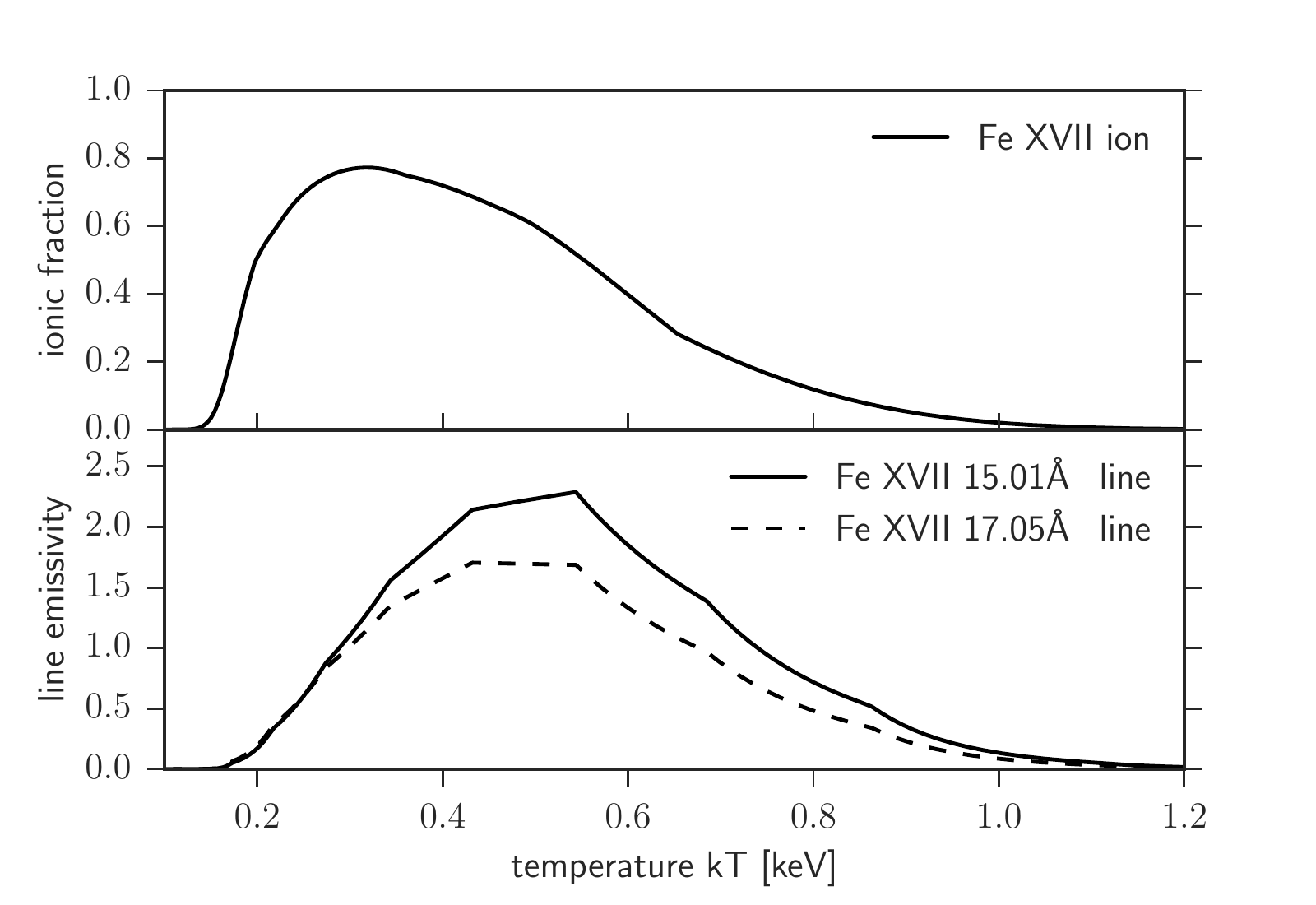}
 \caption{The ionic fraction of \Fe (upper panel) and the emissivity of the considered \Fe  lines (arbitrary units) as a function of plasma temperature, calculated using the APEC model  for optically thin plasmas \citep{APEC1, APEC2} with AtomDB database ver. 2.0.2.}
\label{fig:em_ionbal}
\end{figure}

For galaxies with mean temperatures $\sim$0.3-1.0~keV, the \Fe 15.01~\AA\ (optically thick), 17.05 \AA\ and 17.1 \AA\ (optically thin) lines are the most suitable for resonant scattering analyses, because their emissivity, along with the \Fe ionic fraction, peak in this temperature range (Fig.~\ref{fig:em_ionbal}).  By using the ratio of these lines, velocity amplitudes of the order of $\sim$100~km~s$^{-1}$ have been measured in the cores of 3 elliptical galaxies \citep{Xu2002,  WERNER2009, dePlaa2012}, and Perseus cluster \citep{Churazov2004, Gastaldello2004}.

In this work, we extend this resonant scattering analysis to a sample of 22  massive elliptical galaxies. We combine our resonant scattering measurements with independent line broadening results to obtain the best constraints to date on the turbulent velocity amplitudes, Mach numbers and non-thermal pressure contributions in giant elliptical galaxies, providing bounded constraints  for 13 objects. This enables us, for the first time, to study common kinematic properties of the X-ray gas in elliptical galaxy cores. 

The paper is organized as follows. We describe the \textit{RGS} and \textit{Chandra} data analysis in Section~\ref{sec:data}. The resonant scattering effect and simulations are described in Section~\ref{sec:sim}. The results of the resonant scattering--only and resonant scattering--plus--line broadening analysis are presented in Section~\ref{sec:res}. Section~\ref{sec:unc} discusses the major assumptions and uncertainties in our study. We discuss our findings and their implications in Section~\ref{sec:disc} and provide conclusions in Section~\ref{sec:concl}.

Throughout the paper the abundance of heavy elements is given relative to the proto-solar abundances by \cite{solab}. 


\section{Observations and data analysis}
\label{sec:data}
\subsection{Sample selection}
\label{subsec:sample}
The sources studied in this work constitute a subsample of the CHEmical Enrichment R\textit{GS} cluster Sample (CHEERS) catalog of X-ray bright galaxy groups, clusters and elliptical galaxies. Specifically it includes galaxies with  O~{\scriptsize VIII} 19\text{\AA} line detection \citep{dePlaa2017}, plus two further cool galaxies with \Fe emission detected (NGC~1332 and IC~1459). From this initial group we  selected  22 objects in which the \Fe lines are detected. The properties of the sample are summarized in Table~\ref{tab:sample}. The XMM-\textit{Newton} observations are listed in Table~1 in \cite{Pinto2015}, with addition of $\sim$120 ks of new data for IC~1459, awarded during XMM-\textit{Newton} AO-14. 

\begin{table*}
\begin{minipage}{160mm}
\centering
\caption{Properties of the sample of elliptical galaxies studied in this work.}
\label{tab:sample}
\begin{tabular}{@{ }c@{  \: }c@{ \: }c@{ \: }c@{ \: }c@{\:}c@{\:}c@{\:}c@{\:}c@{\:}} 
\hline
\hline
Source & Alternative &  \textit{Chandra} total &XMM-\textit{Newton} & Distance$^a$  & Average temp. & Sound speed  & $r_{eff}^b$ &  $r_{eff}$  \\
& name &  exposure [ks]& clean time [ks] & [Mpc] & [keV] & [km~s$^{-1}$]& [kpc] & reference$^c$\\
\hline
\hline
IC 1459 & - & 	60	&	156	&	21.1 &0.70 & 430 & 3.5 & B\\
NGC 507 & - & 		44	&	95	&	65.5 &1.19 & 560 & 1.7 & D\\
NGC 533  & A 189 & 	40	&	35	&137 &0.93 & 500 & 14.5  &M \\
NGC 708  & A 262 & 	140	&	173	&69 &1.32 & 590 &  11 & S\\
NGC 1316 & Fornax A &	30	&	166	& 19 & 0.80 & 460& 7.4 & B\\
NGC 1332 & - &	60	&	64	&	 19 &0.80 & 460 & 9.2  & R\\
NGC 1399 & Fornax & 130	&	124	&19 &1.00 & 520 &  32 & Lo \\
NGC 1404 & - &	120	&	29	& 20 & 0.57 & 390&  2.6 & B \\
NGC 3411 & NGC 3402 &	30	&	27	& 66.3 &1.01 & 520 & 31 & F\\
NGC 4261 & - &	100	&	135	& 29.4 & 0.68 & 450 &  5.7 & T\\
NGC 4325 & NGC  4368 &	30	&	22	&108 &0.86 & 480 & 14 & L\\
NGC 4374 & M 84 & 	30	&	92	&18.4 &0.65 & 415 &  4.3 & T\\
NGC 4406 & M 86 & 	30	&	64	&16 & 0.81 & 460& 7.9  & B\\
NGC 4472 & M 49  & 	80	&	82	&17.1 & 0.83 & 480 &  8.0 & T\\
NGC 4552 & M 89 & 	200	&	29	&15.3 & 0.58 & 390 &  2.3 & T\\
NGC 4636 & - &	200	&	103	& 14.3 & 0.61 & 400&  7.0 & B \\
NGC 4649 & M 60 &	310	&	130	&	 17.3 &0.87 & 480&  6.5 & T\\
NGC 4696 & A 3526 & 	760	&	153	&44 &1.14 & 550& 82 & Mi\\
NGC 4761 & HCG 62 &	120	&	165	& 60 &0.83 & 470& 3.8 & Sp\\
NGC 5044 & - &	70	&	127	&	38.9 &0.82 & 480 &  10.4 & T \\
NGC 5813 & - & 	140	&	147	&29.7 & 0.72 & 440 & 8.1 & H\\
NGC 5846 & - & 	170	&	195	&	26.3 &0.63 & 410&  9.8 & H \\
\hline
\hline
\end{tabular}
\\ 
$^a$Adopted from \cite{Pinto2015} for consistency. $^b$The effective radii used here are systematically smaller than the actual ones due to the way that they are measured,  as discussed in depth by \cite{Kormendy}. However, for our purpose of scaling the abundance profiles, such crude measurements are acceptable. $^c$B--\cite{B_reff}; D--\cite{Xreff}; F--\cite{Faber1989};  H--\cite{reff02}; L--\cite{La}; Lo--\cite{L_reff};  M--\cite{reff_a189}; Mi--\cite{Mi};  R--\cite{R}; S--\cite{S_reff}; Sp--\cite{Sp}; T--\cite{reff01}.
\end{minipage}
\end{table*}
\subsection{RGS data analysis}
\label{subsec:rgs}

The XMM-\textit{Newton} satellite is equipped with two main X-ray instruments: the \textit{RGS} and \textit{European Photon Imaging Camera} (\textit{EPIC}). We have used the \textit{RGS} data for our main spectral analysis and the \textit{EPIC} (a MOS detector) data for imaging. The \textit{RGS} spectrometers are slitless and the spectral lines are broadened by the source extent. As in \cite{Pinto2015}, we corrected for spatial broadening through the use of \textit{EPIC}-MOS surface brightness profiles, but used newer calibration files for data reduction and software versions (available from June 2016). We used  the XMM-\textit{Newton} Science Analysis System (SAS) v15.0.0, and 
corrected for contamination from soft-proton flares  using the standard procedure described in \cite{Pinto2015}.

For most objects, we adopted a slit (spectral extraction region) width of  $\sim$ 0.8' ($\sim$ 5\,kpc for most galaxies, cf. Table~\ref{tab:results}), providing good coverage of the inner \Fe bright core, and maximizing the equivalent widths of the \Fe lines with respect to those produced by the hotter gas phase (see e.g. \citealt{Pinto2016}). For a few objects (NGC~4261, NGC~4374, NGC~4472, and NGC~5044)  we extracted \textit{RGS} spectra from regions maximizing the detection of resonant scattering (Sect.~\ref{subsec:turnover}).

We subtracted a model background spectrum, using the standard \textit{RGS} pipeline and a template background file based on the count rate in CCD\,9. We produced MOS\,1 images in the $8-27$\,{\AA} wavelength band and extracted surface brightness profiles to model the \textit{RGS} line spatial broadening with the relation: $\Delta\lambda = 0.138 \, \Delta\theta \, {\mbox{\AA}}$ (see the XMM-\textit{Newton} Users Handbook).

\subsubsection{Baseline model}
\label{sec:baseline_model}

Our analysis focused on the $8-27$\,{\AA} first and second order \textit{RGS} spectra. We performed the spectral analysis with SPEX version 3.00.00, using C-statistics and adopting 68~per~cent errors.

For a full discussion on the spectral modeling we refer the reader to \citet{Pinto2016}. Briefly, we have described the ICM emission with a model for an  isothermal plasma
in collisional ionization equilibrium (\textit{cie}; ionization balance is adopted from \citealt{Bryans2009}). The basis for this model is given by the Mekal code \citep{SPEX}, 
but several updates have been included (see the SPEX manual). The free parameters in the fits were the emission measures $Y=n_{\rm e}\,n_{\rm H}\,dV$,  temperature $T$, and the abundances of N, O, Ne, Mg, and Fe. The nickel abundance was fixed to that of iron. Most objects required two \textit{cie} components to obtain an acceptable fit. In these cases we coupled the abundances of the two \textit{cie} components because the spectra do not allow us to measure them separately. The \textit{cie} emission models were corrected for redshift, Galactic absorption, and line-spatial-broadening through  the multiplicative \textit{lpro} component in SPEX, which uses as input the MOS\,1 surface brightness profile
(Sect.\,\ref{subsec:rgs}).

\begin{figure}
\includegraphics[width=\columnwidth, trim={37mm 35mm 10mm 20mm},clip]{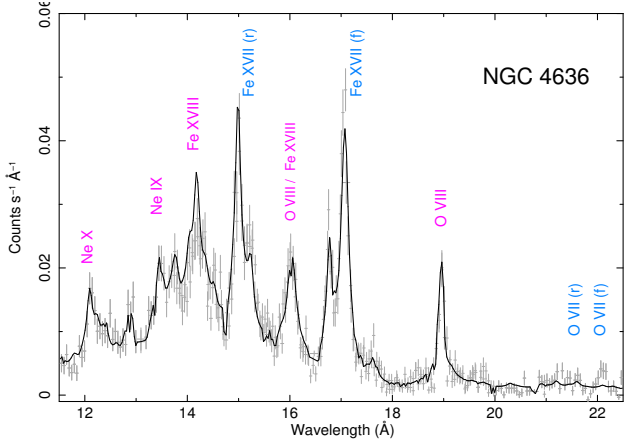}
\caption{The \textit{RGS} spectrum of NGC 4636 (OBS ID = 0111190701). Overlaid is a spectral model with two thermal components. The model underestimates both the O~{\scriptsize VII} (22.1\,{\AA}) and \Fe (17.1\,{\AA}) forbidden lines and overestimates the Fe~{\scriptsize XVIII} (14.2\,{\AA}) resonant line  due to unaccounted resonant scattering effect.}
\label{Fig:NGC4636_spectrum}
\end{figure}

We did not explicitly model the cosmic X-ray background in the \textit{RGS} spectra because any diffuse emission feature would be smeared out into a broad continuum-like component. 
For each source, we have simultaneously fitted the spectra of individual observations by adopting the same model. This gives a good description of the broadband spectrum,
but, as previously shown in \citet{Pinto2015}, underestimates the 17\,{\AA} \Fe line peaks for several sources due to unaccounted for resonant scattering, which we model separately (cf. following Section). In Fig.~\ref{Fig:NGC4636_spectrum} we show an example of a spectrum of NGC~4636.

\subsubsection{\Fe line measurements}
\label{sec:fexvii_line_measurements}

We have measured the fluxes of the \Fe resonant and forbidden lines by removing all lines associated with the \Fe ion from our plasma model (due to unaccounted for resonant scattering, cf. previous Section), and fitting instead four delta lines broadened by the instrumental response fixed at 15.01\,{\AA}, 15.26\,{\AA}, 16.78\,{\AA}, and 17.08\,{\AA},
which are the main \Fe transitions dominating the band considered. The lines were also corrected for redshift, galactic absorption and  spatial broadening following the \textit{cie} components. This method has been successfully used in previous papers  (e.g. \citealt{WERNER2009}, \citealt{dePlaa2012}, \citealt{Pinto2016}). From those fits we calculated the ratio of the \Fe forbidden to  resonant lines for all sources. The results are shown in Table~\ref{tab:results}.

In \citet{Pinto2016} we showed that the low-temperature O~{\scriptsize VII} and \Fe lines have widths narrower than the hotter O~{\scriptsize VIII} and Fe~{\scriptsize XVIII} lines.
Therefore, in our fits, we have used two different \textit{lpro} components: one \textit{lpro} for the (hotter) \textit{cie} component  and the other \textit{lpro} for the cooler  
O~{\scriptsize VII}  and \Fe delta lines. The scale parameters \textit{s} of the two \textit{lpro} components were uncoupled and free parameters in the fits.

\subsubsection{Line broadening analysis}
\label{sec:velocity_broadening}

To provide further constraints on turbulence, we used in our analysis probability distributions on measured line widths using the method described by \cite{Pinto2015}. Starting from our baseline model of two thermal \textit{cie}  components (Sect.\,\ref{sec:fexvii_line_measurements}), and using the   \textit{EPIC} surface brightness profiles, which provide an approximate spatial mapping for the lines in our spectra, we performed spectral fits and evaluated the C-statistics on a fixed velocity grid, with other parameters left free to vary. In particular,  the  \textit{lpro} spatial scale factor \textit{s}=0--1 (i.e no spatial broadening and full \textit{EPIC} broadening) was not fixed. This procedure is asymptotically equivalent to finding the marginalized velocity probability distribution.  

\begin{figure*}
\begin{minipage}{180mm}
\centering
\includegraphics[width=0.47\columnwidth]{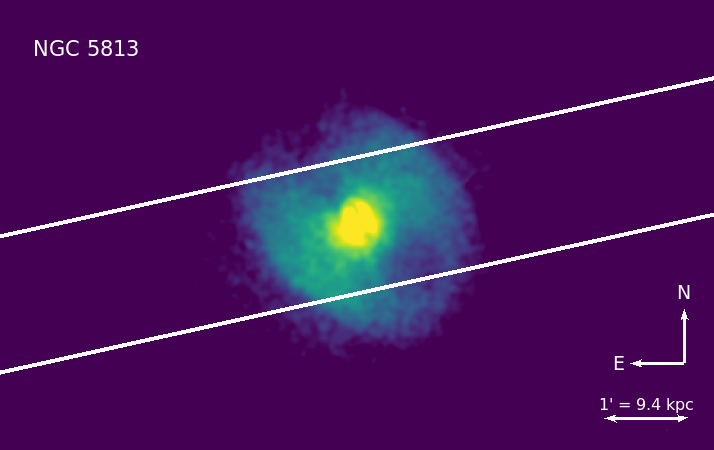} 
\hspace{2ex}
\includegraphics[width=0.47\columnwidth]{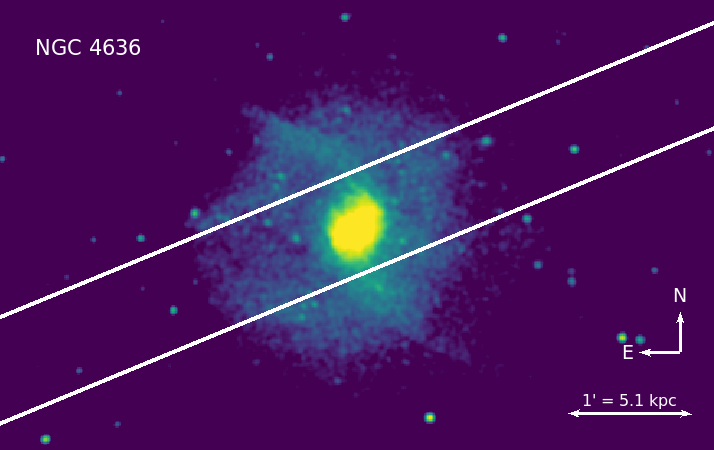} 
 \caption{  \textit{Chandra} images of the elliptical galaxies NGC~5813 (left) and NGC~4636 (right) with the default \textit{RGS} extraction regions over-plotted (0.8$'$ wide; Sect.~\ref{subsec:rgs}). The images are lightly smoothed for visual purposes.}
\label{fig:RGS_slit}
\vspace*{\floatsep}
\includegraphics[width=0.49\columnwidth, trim={0.5mm 9mm 13mm 15mm},clip]{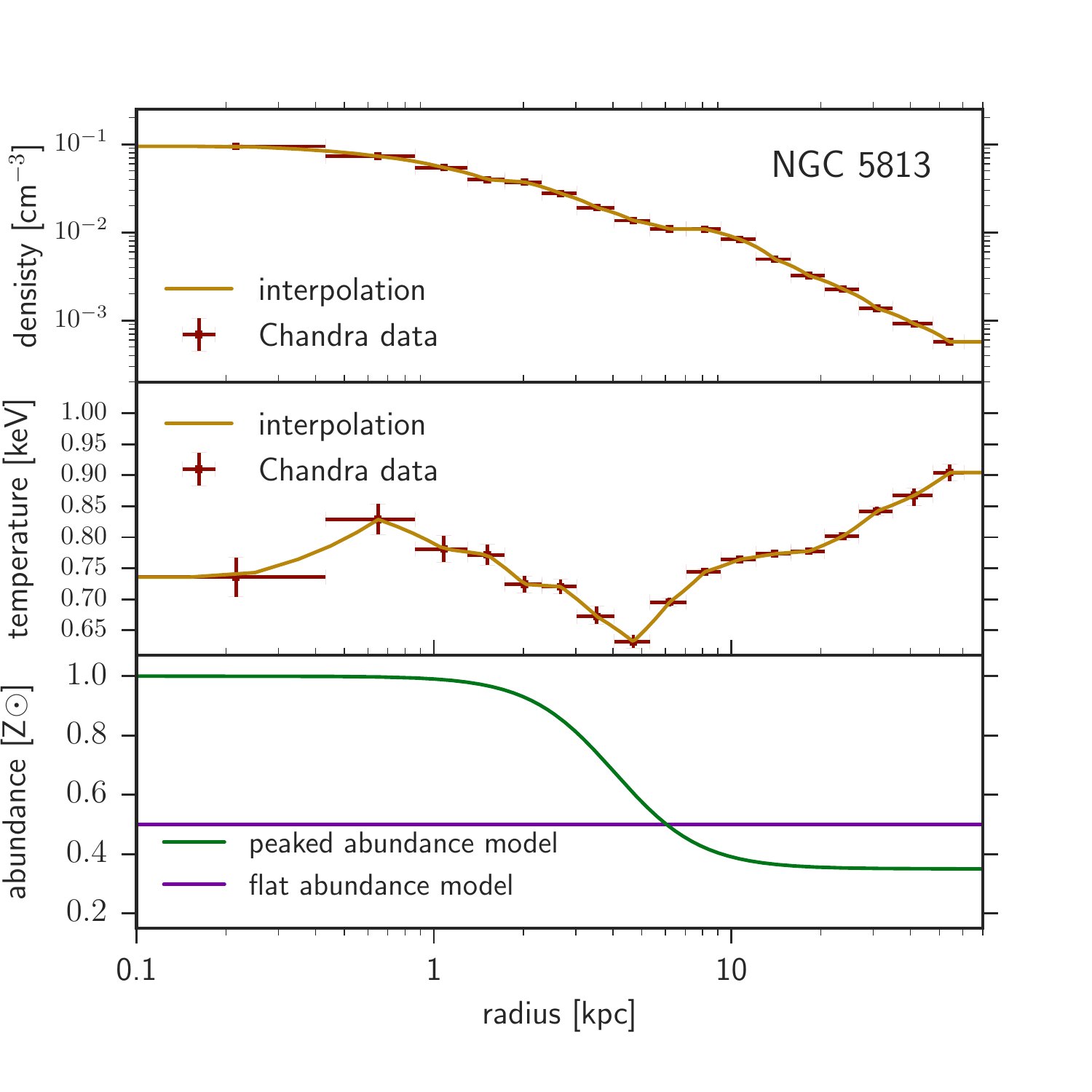}
\includegraphics[width=0.49\columnwidth, trim={0.5mm 9mm 13mm 15mm},clip]{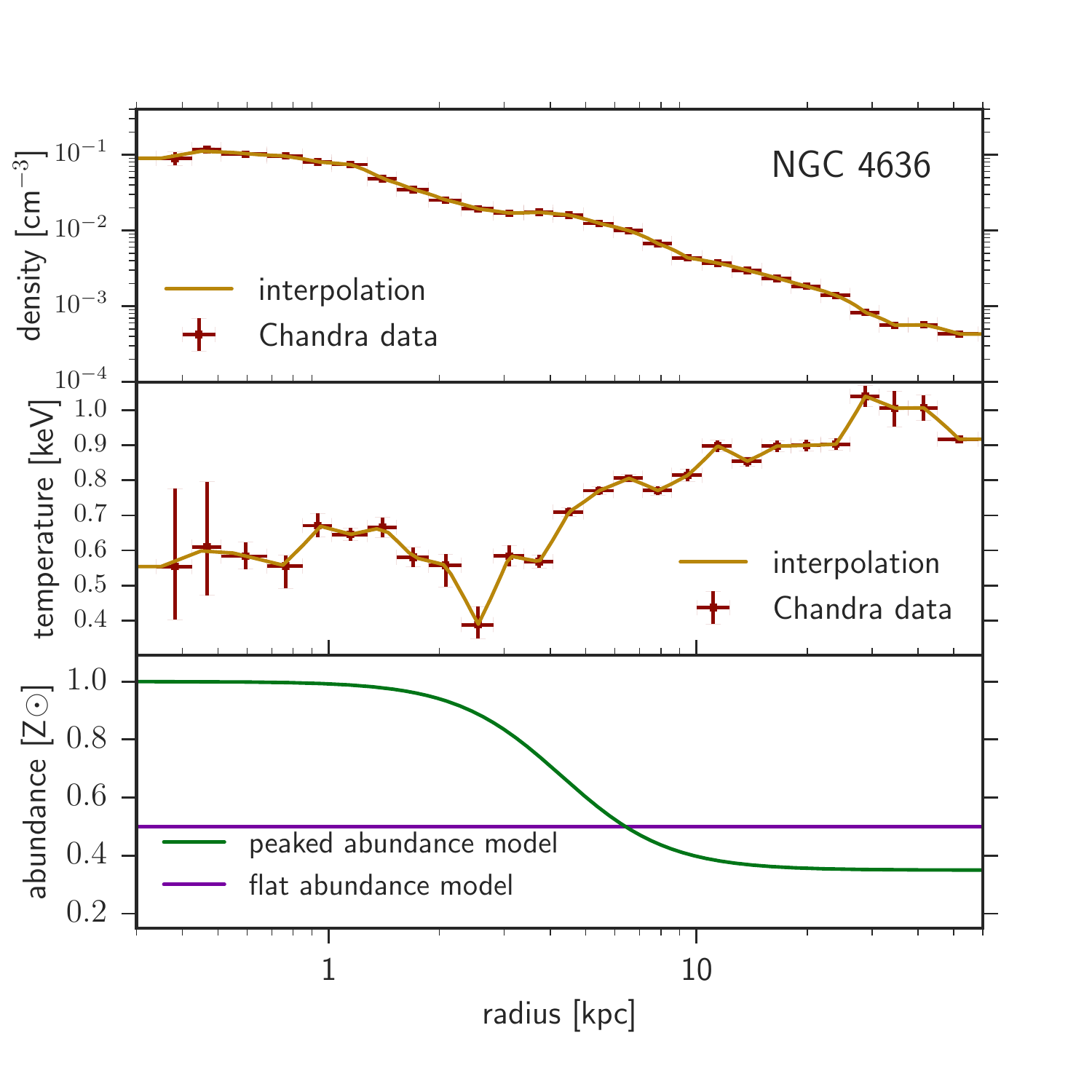}
 \caption{Thermodynamic properties of NGC~5813 (left) and NGC~4636 (right).  Upper panels show the deprojected radial profiles derived from \textit{Chandra} data (red points):  number electron density (top) and gas temperature (center), with 1$\sigma$ error bars (Sect.~\ref{subsec:chandra}); orange curves show the interpolated profiles used for radiative transfer simulations (Sect.~\ref{subsec:input}). Bottom panels present abundance models adapted: fixed at 0.5~Z$_{\sun}$ (purple; default) and centrally peaked (green). }
\label{fig:profiles}
\end{minipage}
\end{figure*}

\subsection{Chandra data analysis}
\label{subsec:chandra}
The initial data processing is done following the standard procedure described in \cite{Vikhlinin2005}. The point sources are detected using the \textit{wvdecomp} tool and excluded from the analysis. Correcting for the exposure, vignetting effects and subtracting the background,  images of the galaxies are produced (Fig.~\ref{fig:RGS_slit}). 

The deprojected number electron density and temperature are obtained from projected spectra using the procedure described in \cite{Churazov2003}. The deprojected spectra are  fitted with an isothermal APEC plasma model \citep{APEC1,APEC2} in the 0.6--8 keV band using XSPEC (ver. 12.8.0). The fitted model takes into account the total galactic HI column density and assumes fixed constant abundance of heavy elements 0.5 Z$_{\sun}$. Examples of the deprojected profiles are shown in Fig.~\ref{fig:profiles}. 

For several galaxies in our sample (NGC~708, NGC~1404, NGC~1399, NGC~4261, NGC~4406, and NGC~4472) we used published deprojected radial profiles from \cite{Werner2012}. The deprojected data for NGC~4696 were provided by \cite{Sanders2016}. 


\section{Simulations}
\label{sec:simulations}
\label{sec:sim}
For all objects in our sample, we conducted radiative transfer Monte Carlo simulations of the resonant scattering (Sect.~\ref{subsec:rsmod}). These simulations predict the fluxes of the 15.01,  17.05 and 17.1 \AA\ \Fe lines, assuming different Mach numbers of turbulence, which are then compared to the measured flux ratios. We illustrate the analysis steps below using NGC~5813 and NGC~4636. Details for the rest of the objects are presented in Appendix~\ref{app:all}.

\subsection{Input models}
\label{subsec:input}
For our resonant scattering simulations we adopted a spherically symmetric model of each galaxy based on the interpolated deprojected profiles of the gas electron number density and temperature (orange lines in Fig.~\ref{fig:profiles}; Sect.~\ref{subsec:chandra}). In order to account for the uncertainty associated with 1$\sigma$  errors in the deprojected data, we vary the models within plotted 68 per cent errors and repeat the calculations.

Since the abundance measurements of the heavy elements  are  uncertain, we performed simulations conservatively assuming  a flat abundance profile fixed at \mbox{0.5 Z$_{\sun}$}. The sensitivity of our results to the choice of the abundance model was checked by  additionally considering a centrally peaked abundance profile, scaled by the galaxy's optical effective (half-light) radius $r_{eff}$, namely:

\begin{equation}
\label{eq:peaked}
Z_{peaked}(r) = 0.65\times\frac{2.0+\left(\frac{r}{0.5r_{eff}}\right)^3}{1.0+\left(\frac{r}{0.5r_{eff}}\right)^3}-0.3\quad [\text{Z}_{\sun} ]. 
\end{equation}
 In this case, the abundance is 1~Z$_{\sun}$ in the galaxy center and drops to 0.35~Z$_{\sun}$ in the outskirts. Both models are shown in the bottom panels of Fig.~\ref{fig:profiles}. Effective radii used are listed in Table~\ref{tab:sample}. As we discuss later, for most galaxies the choice of abundance model does not affect the results (Sect.~\ref{subsec:abun}).

\subsection{Optical depth in the Fe XVII line}
\label{subsec:rs}
\begin{table*}
\begin{minipage}{170mm}
 \caption{Properties of the \Fe line transitions considered in this work, adopted from AtomDB database ver. 2.0.2.}
 \begin{center}
 \label{tab:lines}
 \begin{tabular}{@{}ccccccccc}
  \hline
  \hline
 wavelength & energy & lower & upper & oscillator & \multicolumn{2}{c}{electron configuration} & transition & Rayleigh scattering \\
  $[\text{\AA}]$ &  [keV] & level & level & strength & lower & upper  & type & weight, $w_R$\\
\hline
  \hline
15.01	& 0.826 & 1 & 27 & 2.49  & 2p$^6$, $^1$S, J=0 & 2p$^5$ 3d, $^1$P, J=1 & electric dipole & 1\\
17.05	& 0.727 & 1 & 3 & 0.126 & 2p$^6$, $^1$S, J=0 & 2p$^5$ 3s, $^2$P$_{3/2}$, J=1 & electric dipole & 1\\
17.10	& 0.725 & 1 & 2 & 5.2$\cdot 10^{-8}$ & 2p$^6$, $^1$S, J=0 & 2p$^5$ 3s, $^2$P$_{3/2}$, J=2 &  magnetic quadrupole & -\\
  \hline  
    \hline  
 \end{tabular}
 \\
  \end{center}
\end{minipage}
\end{table*}

The optical depth $\tau$ of a galaxy in a given line is:
 \begin{equation}
\label{eq:tau}
\tau = \int_0^{\infty} n_i  \sigma_0 ds,
\end{equation}
where $s$ is the distance along the photon propagation, $n_i$ is the number density of the ion in the ground state and  $\sigma_0$ is the scattering cross section at the center of the line, given by:
\begin{equation}
\label{eq:sigma}
\sigma_0 = \sqrt{\pi} h cr_e \frac{  f}{\Delta E_D}.
\end{equation}
Here $h$ is Planck's constant, $r_e$ is the classical electron radius, $c$ is the speed of light, $f$ is the oscillator strength of the respective atomic transition, and $\Delta E_D$ is the Doppler line width. $\Delta E_D$ is given by:
 
\begin{equation}
\label{eq:shift}
\Delta E_D =\frac{ E_0 } {c} \left(\frac{2 kT_e}{A m_p} + \frac{2}{3} V^2_{turb} \right)^{0.5},
\end{equation}
where $E_0$ is the rest frame energy of the atomic transition,  $k$ is the Boltzmann constant, $A$ is the atomic mass number, $m_p$ is the proton mass, $T_e$ is the electron temperature,  and $V_{turb}$ is the characteristic 3D turbulent velocity.
We parametrize turbulence by the 3D Mach number\footnote{In this work we assume isotropic turbulence, i.e. $V_{turb}=\sqrt{3}v_{1D}$, where $v_{1D}$ is the 1D differential velocity component. Our simulations assume fixed Mach number everywhere in the galaxy. Running simulations for fixed turbulent velocity instead changes our results by a maximum of 5~\kms\ in only one galaxy.}, $M=V_{turb}/c_s$, where $c_s=\sqrt{\gamma k T / (\mu m_p)}$ is the sound speed ($\gamma=5/3$ is the adiabatic index for an ideal monoatomic gas, and $\mu=0.6$ is the mean particle atomic weight). 

\begin{figure}
\includegraphics[width=\columnwidth,trim={0mm 3mm 15mm 8mm},clip]{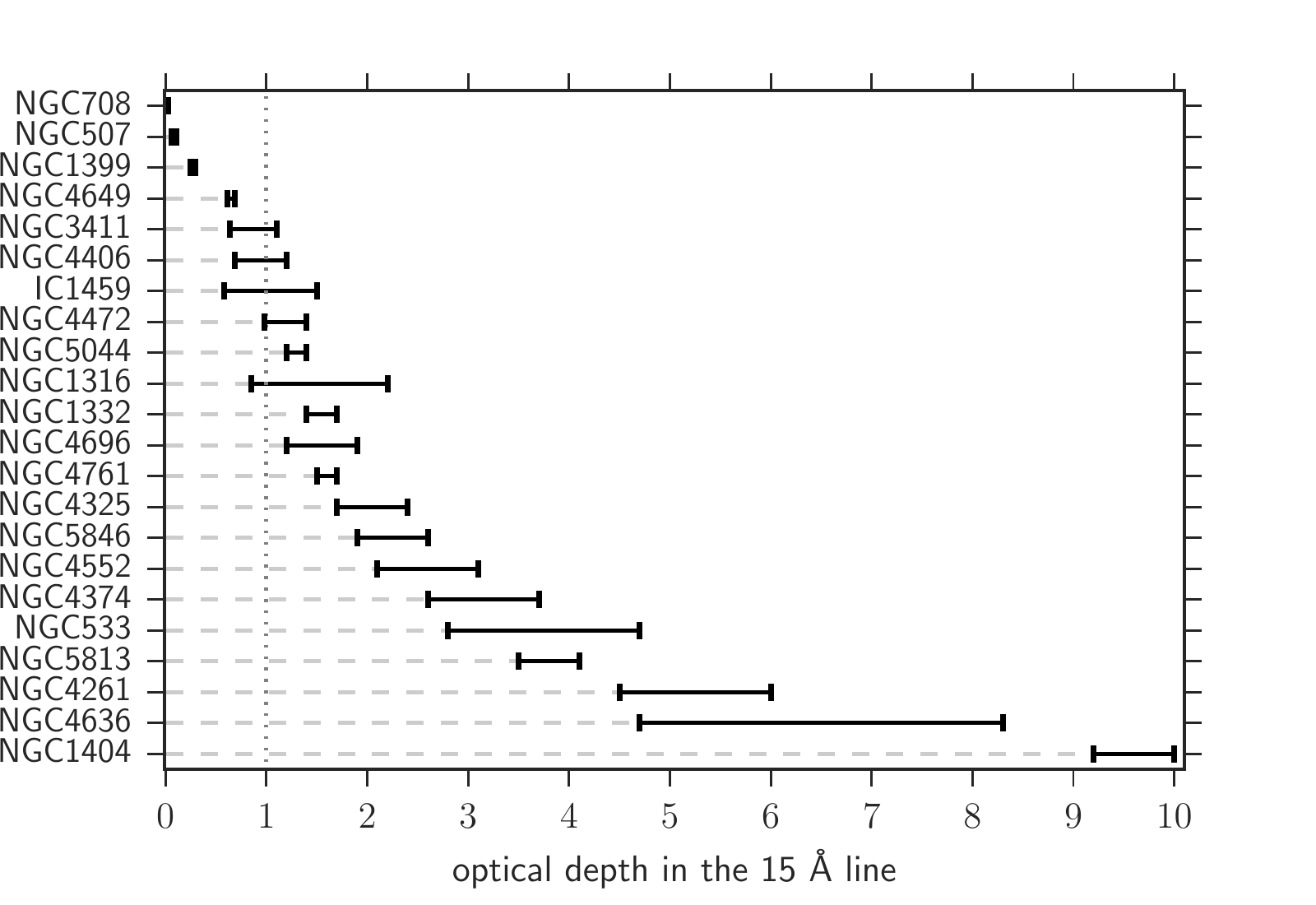}
\caption{Optical depth (Eq.~\ref{eq:tau}) in the optically thick \Fe line at  15.01~\AA\ for all galaxies in our sample. Only thermal broadening is taken into account.   The spread of each value reflects the 1$\sigma$ uncertainty in the deprojected gas temperature and density profiles (see Sect.~\ref{subsec:input} and Fig.~\ref{fig:profiles} for details). The dotted vertical line marks an optical depth of $\tau=1$, beyond which line is considered optically thick. For the majority of galaxies in our sample the 15.01~\AA\ \Fe line is optically thick. NGC~1404 has the largest optical depth due to its high density and low temperature, which results in a high \Fe ionic fraction.}
\label{fig:tau}
\end{figure}

Assuming no differential gas motions, the optical depths  in the 15.01 \AA\ line calculated for all galaxies in our sample are shown in Fig.~\ref{fig:tau}. Most of the objects are optically thick in this line, and for almost a half of the sample $\tau$>2. NGC~1404 has a very large optical depth of $\sim$9--10, due to the combination of its large density and low temperature, resulting in a high ionic \Fe fraction. The other two lines considered in this study, 17.05~\AA\ and 17.10~\AA, are optically thin because of their  low oscillator strengths (cf. Table~\ref{tab:lines}).

\subsection{Resonant scattering modeling}
\label{subsec:rsmod}
Our Monte Carlo simulations of resonant scattering follow the simulations performed by \cite{sazonov2002}, \cite{Churazov2004}, and \cite{irina2010,irina2011}. Namely, we randomly draw the position and propagation direction of each photon (total number of photons >$10^7$), and draw its energy from a Doppler-broadened Gaussian around the line energy. Initially we assign  a unit weight to the photon, which is then reduced after each scattering event by a factor of 1-$e^{-\tau}$, until a minimal threshold ($\sim$10$^{-9}$) is reached. Photons can undergo multiple scatterings before they leave the galaxy. After each event a new position and propagation direction are drawn using a phase function, that is a sum of the isotropic and Rayleigh (dipole) scattering phase functions with appropriate weights \citep{Hamilton1947}. We use ionization balance data and line emissivities from the APEC model for optically thin plasmas \citep[][AtomDB ver. 2.0.2\footnote{\url{http://www.atomdb.org/}}]{APEC1, APEC2}. Line transition properties  and  dipole scattering weights are listed in Table~\ref{tab:lines}.

\subsection{Calculating the \Fe  line flux ratio}
\label{subsec:flux}
\label{subsec:ifr}

\begin{figure*}
\vspace{-2ex}
\begin{minipage}{180mm}
\includegraphics[width=0.5\columnwidth, trim={4mm 2mm 12mm 7mm},clip]{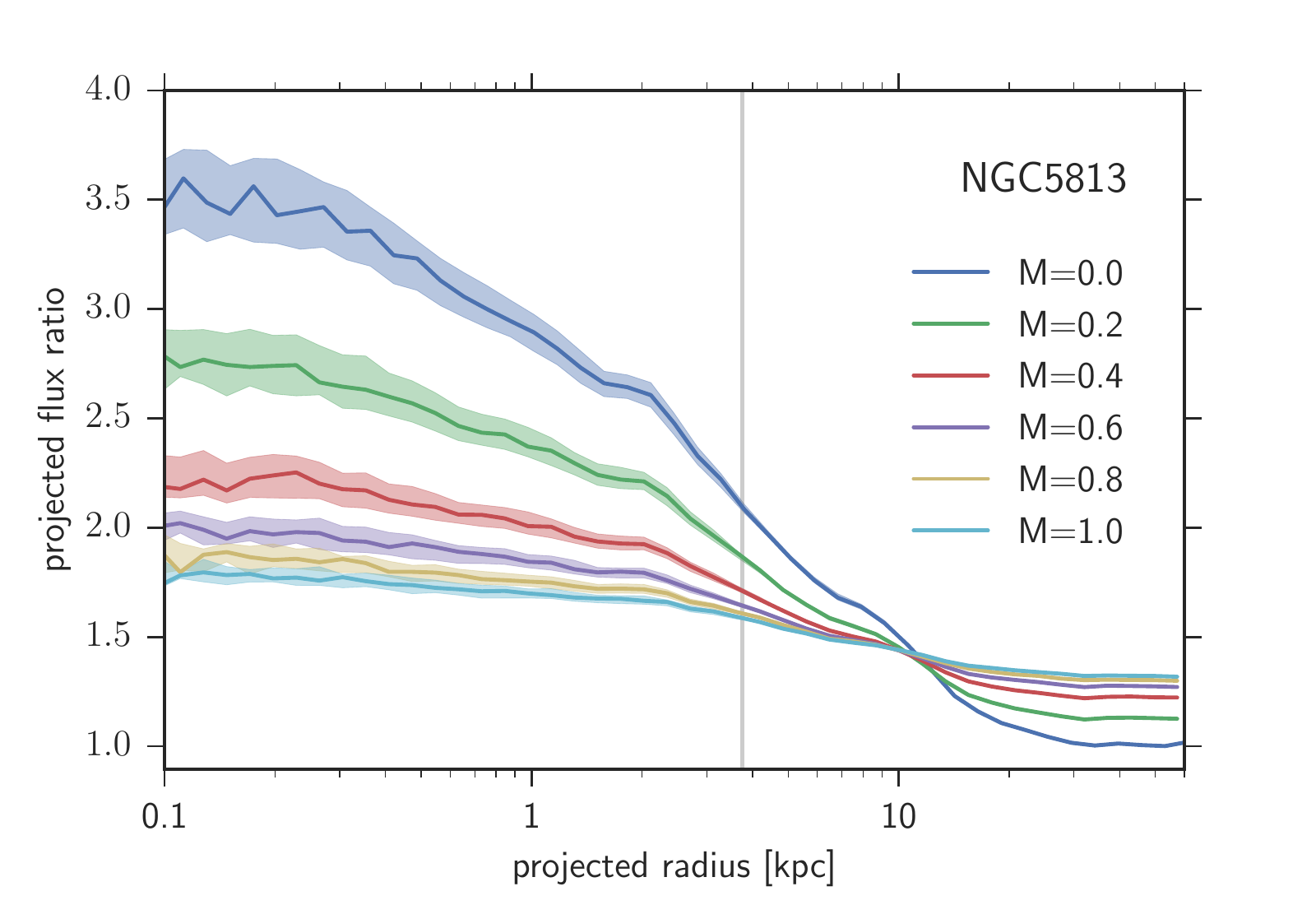}
\includegraphics[width=0.5\columnwidth, trim={4mm 2mm 12mm 7mm},clip]{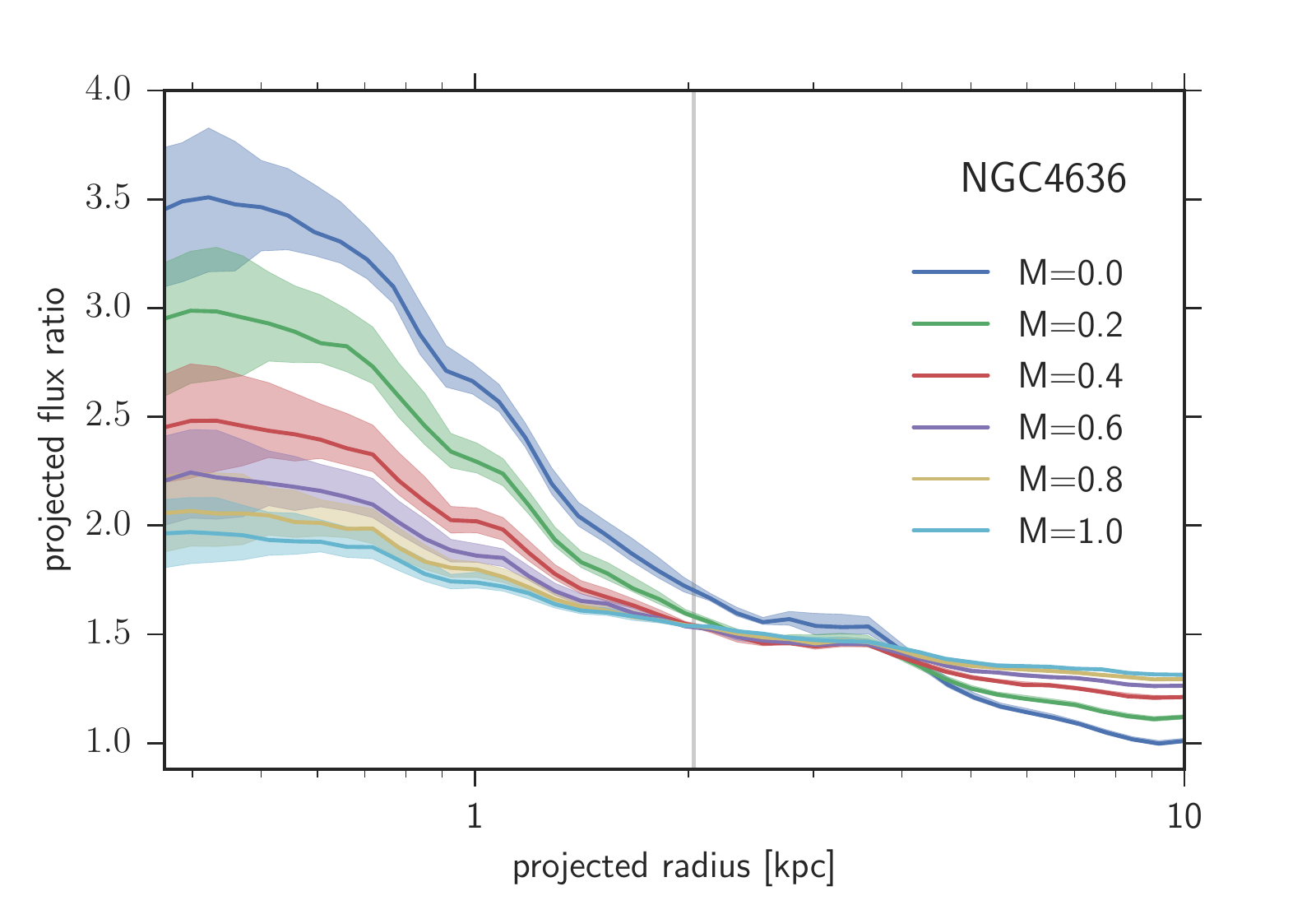}
\vspace{-5ex}
\caption{Predicted projected radial profiles of the \Fe optically thin to thick line flux ratio (Eq.~\ref{eq:fr}) in NGC~5813 (left) and NGC~4636 (right), calculated assuming isotropic turbulence with different 3D Mach numbers $M$, with the uncertainty associated with the 1$\sigma$ errors in the deprojected temperature and density profiles (Sect.~\ref{subsec:input} and Fig.~\ref{fig:profiles}). Grey vertical line marks the size of the spectral extraction region (Fig.~\ref{fig:RGS_slit} and Sect.~\ref{subsec:rgs}). The lower the Mach number, the stronger the suppression of the \Fe optically thick line. This trend reverses at larger radii, where the line becomes optically thin (see Sect.~\ref{subsec:turnover}), which happens at $\sim$10 kpc in NGC~5813 and at $\sim$4 kpc in NGC~4636. } \label{fig:ratio}
\vspace*{\floatsep}%
\vspace{-5.5ex}
\includegraphics[width=0.5\columnwidth, trim={4mm 3mm 12mm 3mm},clip]{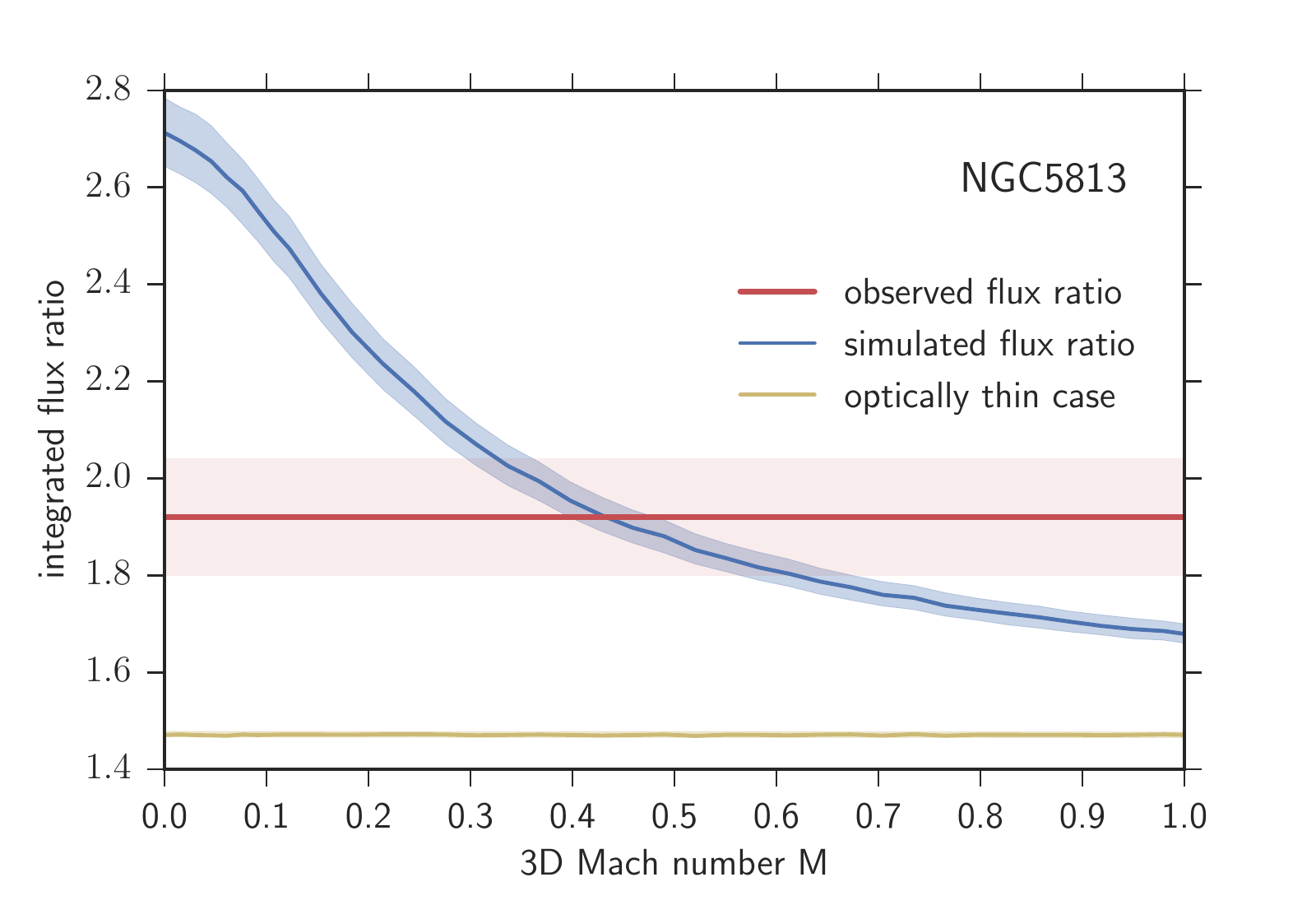}
\includegraphics[width=0.5\columnwidth, trim={4mm 3mm 12mm 3mm},clip]{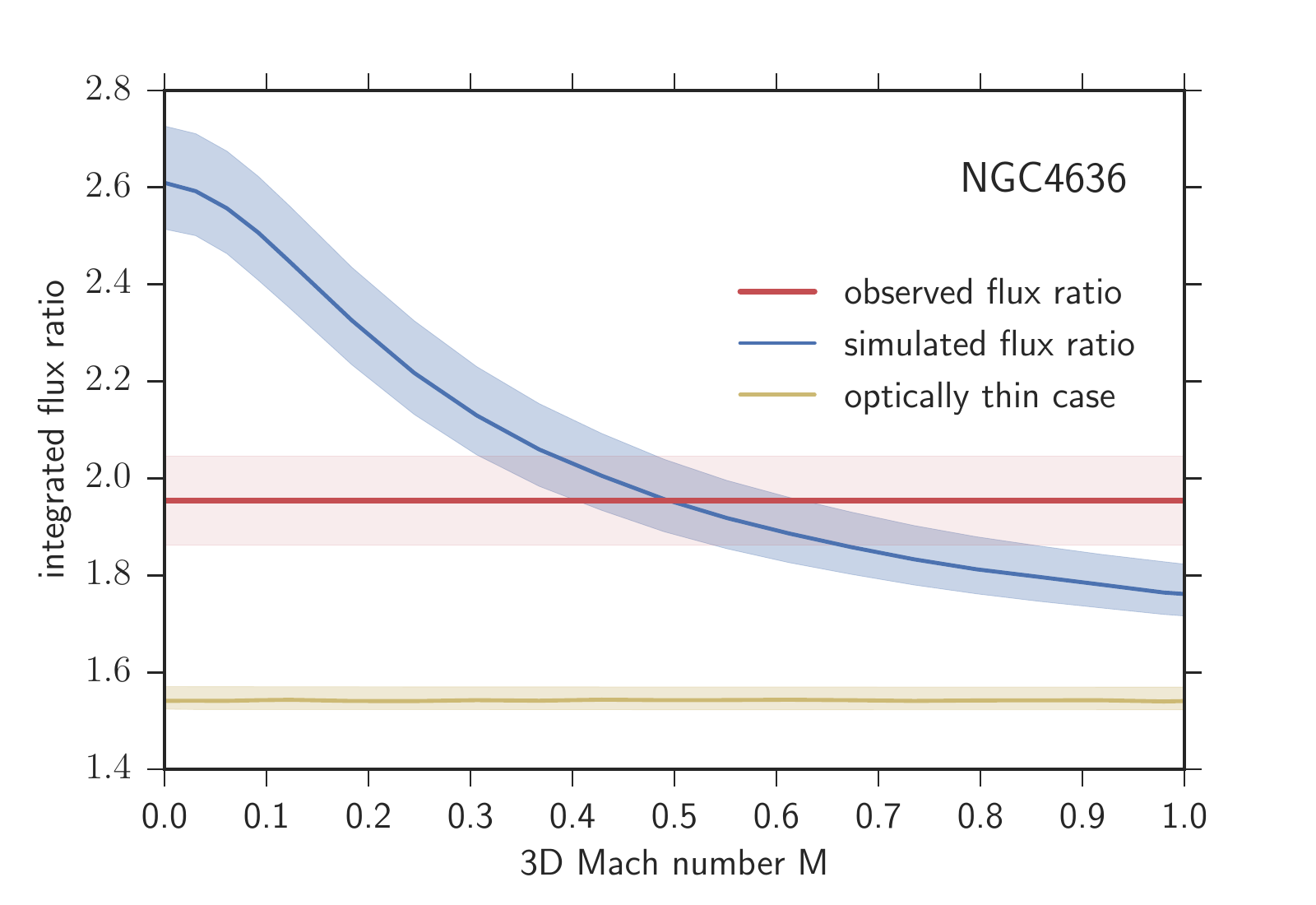}
\includegraphics[width=0.5\columnwidth, trim={4mm 3mm 12mm 7mm},clip]{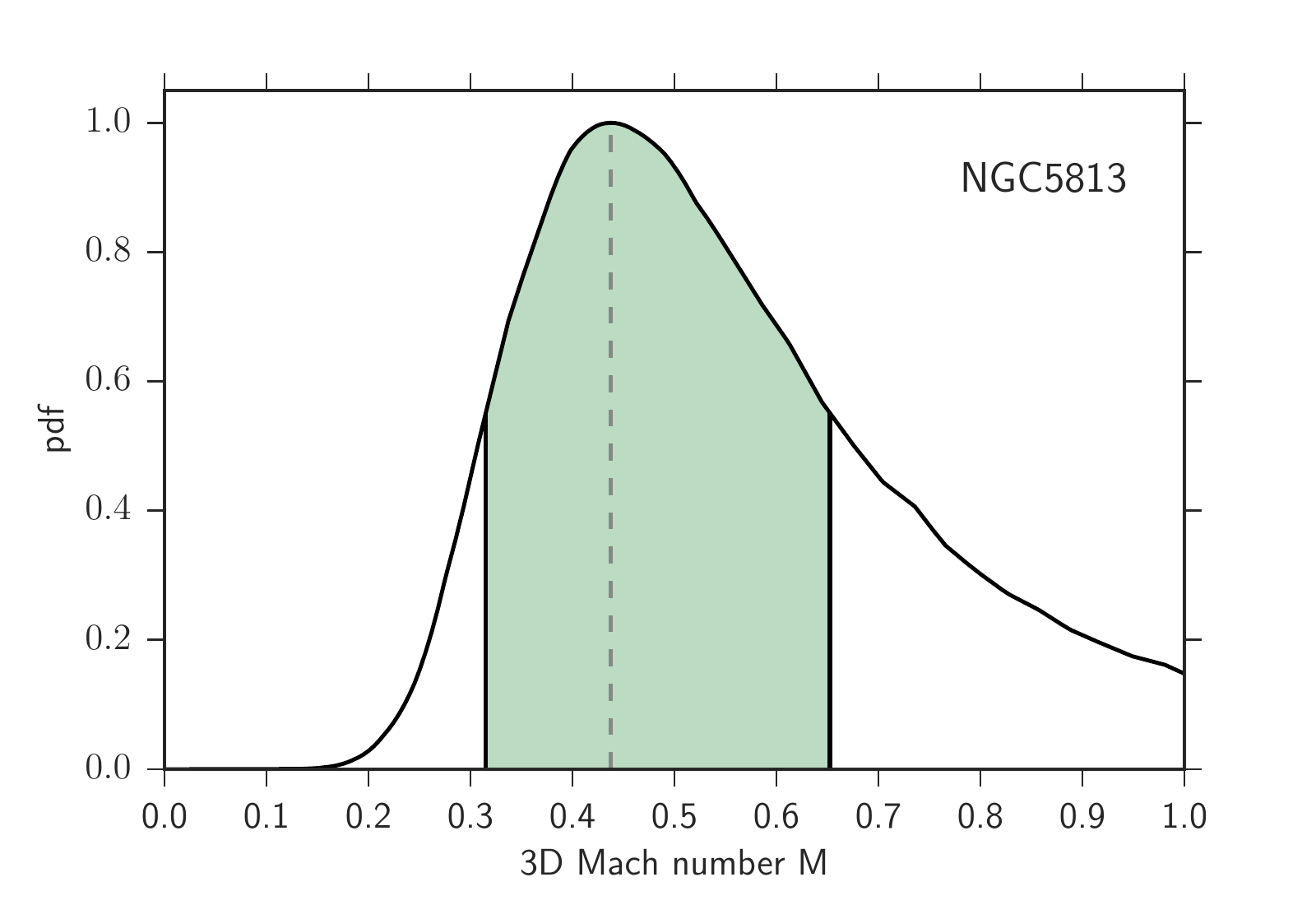}
\includegraphics[width=0.5\columnwidth, trim={4mm 3mm 12mm 7mm},clip]{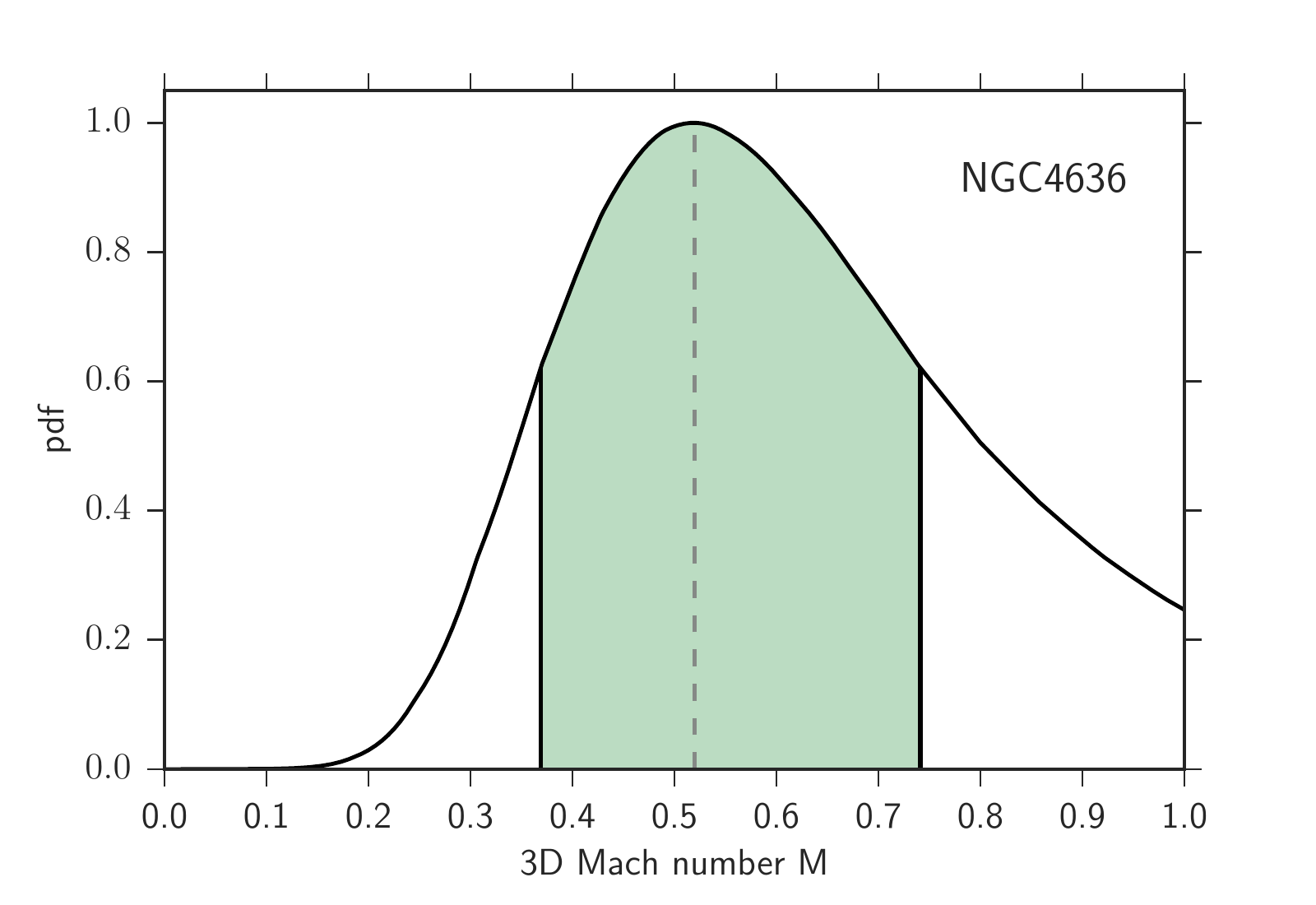}
\vspace{-5ex}
\caption{Resonant scattering results. \textit{Upper panels}: comparison of the observed and simulated flux ratio predicted for  different turbulent motions  for NGC~5813 (left) and NGC~4636 (right; Sect.~\ref{subsec:ifr}). The red line and region show the flux ratio measured from \textit{RGS} observations (plus 68~per~cent uncertainties). The blue line and region show predicted integrated flux ratio as a function of 3D Mach number $M$ (Eq.~\ref{eq:ifr}) with the uncertainty associated with 1$\sigma$ errors in the deprojected thermodynamic profiles (Sect.~\ref{subsec:chandra}). The optically thin limit is shown in yellow. The two regions overlap for Mach numbers  $M\sim$0.30--0.67 in NGC~5813, and for $M\sim$0.30--0.85 in NGC~4636, providing constraints on turbulence. \textit{Lower panels}: Unnormalized probability density function of Mach number in NGC~5813 (left) and NGC~4636 (right; Sect.~\ref{subsec:pdf}). Dashed vertical line marks the most probable Mach number: 0.44 in NGC~5813 and 0.52 in NGC~4636. Green regions represent the 1\sig confidence intervals: $M = 0.32-0.65$ for NGC~5813 and  $M = 0.37-0.74$ for NGC~4636, which translate to $v_{1D}=111^{+54}_{-31}$~km~s$^{-1}$ and  \mbox{$v_{1D}=121^{+51}_{-35}$~km~s$^{-1}$} respectively.}
\label{fig:int_ratio}
\label{fig:pdf}
\end{minipage}
\end{figure*}

Our Monte Carlo simulations provide projected line fluxes $I_M$ as a function of projected radius $r_p$, given a specific 3D Mach number $M$. These profiles take into account resonant scattering, along with thermal and turbulent line broadening. The primary quantity of interest is the projected flux ratio (FR) of optically thin (17.05~\AA, 17.10~\AA) to thick (15.01~\AA) \Fe lines (the 17.05~\AA\ and 17.10~\AA\ lines are observationally indistinguishable, thus their sum is considered). For each object in our sample, we predicted the projected radial profiles of FR varying levels of turbulence (parametrized by the Mach number $M$):
\begin{equation}
\label{eq:fr}
\text{FR}(r_p, M) = \frac{I_{17.05[\text{\AA}]}(r_p, M)+I_{17.10[\text{\AA}]}(r_p, M)}{I_{15.01[\text{\AA}]}(r_p, M)}.
\end{equation}
An example is shown in Fig.~\ref{fig:ratio}.

The observed flux ratio comes from a spatially extended \textit{RGS} spectral extraction region (shown in Fig.~\ref{fig:RGS_slit}). We numerically integrate line fluxes over this region $A$ (a 0.8$'$ wide  slab\footnote{For some objects (NGC~4261, NGC~4374, NGC~4472, and NGC~5044) we actually use narrower slits to maximize the resonant scattering signal (Sect.~\ref{subsec:turnover}).}):
\begin{equation}
\label{eq:ifr}
\text{IFR}(M) = \frac{\int_{slab}  (I_{17.05[\text{\AA}]}(r_p, M)+I_{17.10[\text{\AA}]}(r_p, M)) dA }{ \int_{slab} I_{15.01[\text{\AA}]}(r_p, M) dA }. 
\end{equation}
This integrated flux ratios (IFRs) are then compared to observations, as illustrated in  Fig.~\ref{fig:int_ratio}.


\section{Results}
\label{sec:res}

We present two sets of results: first, based on resonant scattering effect only; and second, using combined resonant scattering and line broadening methods. In the main text we present detailed results for NGC~4636 and NGC~5813. The rest of the sample is shown in Fig.~\ref{fig:pdf1} in Appendix~\ref{app:all}\footnote{We show all galaxies for which the effect of resonant scattering is important.}. Results are summarized in Table~\ref{tab:results}.

\subsection{Resonant scattering--only}
\label{subsec:pdf}
\label{subsec:sub}
\label{subsubsec:rs_ll}

We first compare our observations (red line and region in Fig.~\ref{fig:int_ratio}) to our fiducial model for  the integrated flux ratio, namely simulations run for a flat abundance profile and deprojected thermodynamic profiles without uncertainties (blue line in Fig.~\ref{fig:int_ratio}). This follows the approach of  previous resonant scattering works, e.g. \cite{WERNER2009} and \cite{dePlaa2012}. Lower limits for 8 galaxies can be derived this way\footnote{M89, NGC~1404, NGC~4261, NGC~4325, NGC~4636, NGC~4649, NGC~5813 , and NGC~5846.}, on average  showing $M>0.59$ ($v_{1D}>133$\kms). NGC~4261 has the lowest lower limit on Mach number, $M>0.054$ ($v_{1D}>13$\kms), while NGC~1404 shows the largest lower limit, $M>0.59$ ($v_{1D}>130$\kms). For the remainder of the sample we cannot obtain any limits using this approach.

In order to extract our formal constraints on the Mach number, we approximated the uncertainties on the predicted IFR (blue regions in Fig.~\ref{fig:int_ratio}) as Gaussians. We then convolved this uncertainty with the statistical error on the observed ratio, and calculated the probability density function (pdf) of a Mach number.  The normalized pdfs are shown with black curves in Figs.~\ref{fig:pdf}, \ref{fig:pdf1}. The peaks of the pdfs are indicated with the vertical dashed lines, where appropriate (in 7 galaxies). Note that in most cases the peaks coincide with the intersection of the blue (simulated ratio for fiducial model) and the red (measured ratio) curves.

Assuming that the turbulence in galaxies is subsonic, i.e. $M<1$,  confidence intervals are obtained from the pdfs. Green regions in Fig.~\ref{fig:pdf} show the 68~per~cent confidence intervals, providing constraints of \mbox{$M=0.44^{+0.21}_{-0.12}$} (\mbox{$v_{1D}=111^{+54}_{-31}$~km~s$^{-1}$}) in NGC~5813 and \mbox{$M=0.52^{+0.22}_{-0.15}$} (\mbox{$v_{1D}=121^{+51}_{-35}$~km~s$^{-1}$}) in NGC~4636. We obtain well bounded constraints,  and  upper or lower limits for 11 other systems:   NGC~1316, NGC~1404, NGC~3411, NGC~4261, NGC~4325, NGC~4374, NGC~4552, NGC~4649, NGC~4761, NGC~5044, and NGC~5846. For 3 galaxies, the pdfs show a preference for lower Mach numbers, but are too flat to derive meaningful upper limits;  for 6 galaxies the distributions are completely flat, thus we cannot constrain turbulent velocities in these objects with resonant scattering. For several of these galaxies it would be straightforward to obtain better resonant scattering measurements provided more data (Sect.~\ref{subsec:improv}).

\begin{figure*}
\begin{minipage}{180mm}
\includegraphics[width=0.5\columnwidth,trim={0mm 3mm 11mm 7mm},clip]{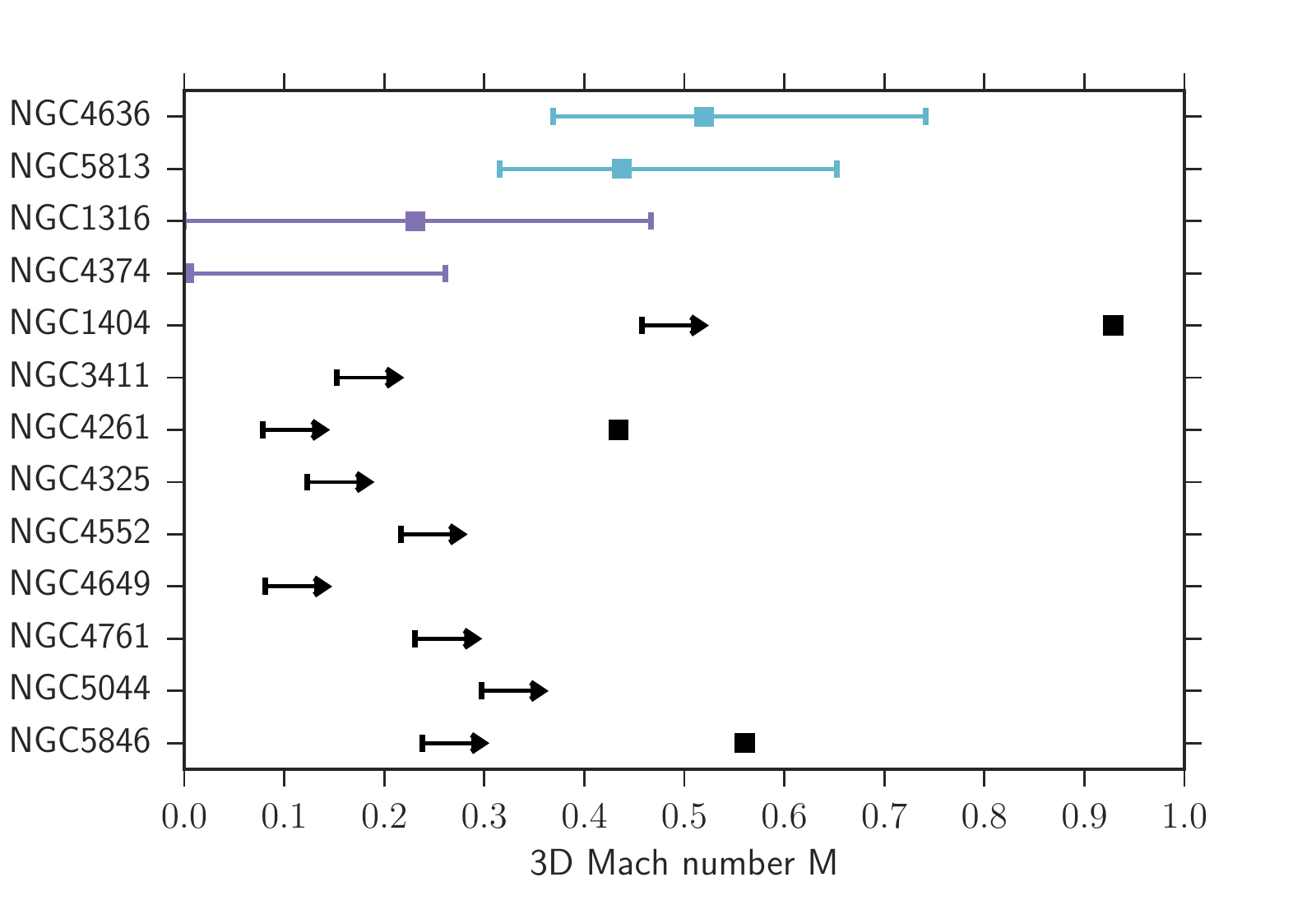}
\includegraphics[width=0.5\columnwidth,trim={0mm 3mm 11mm 7mm},clip]{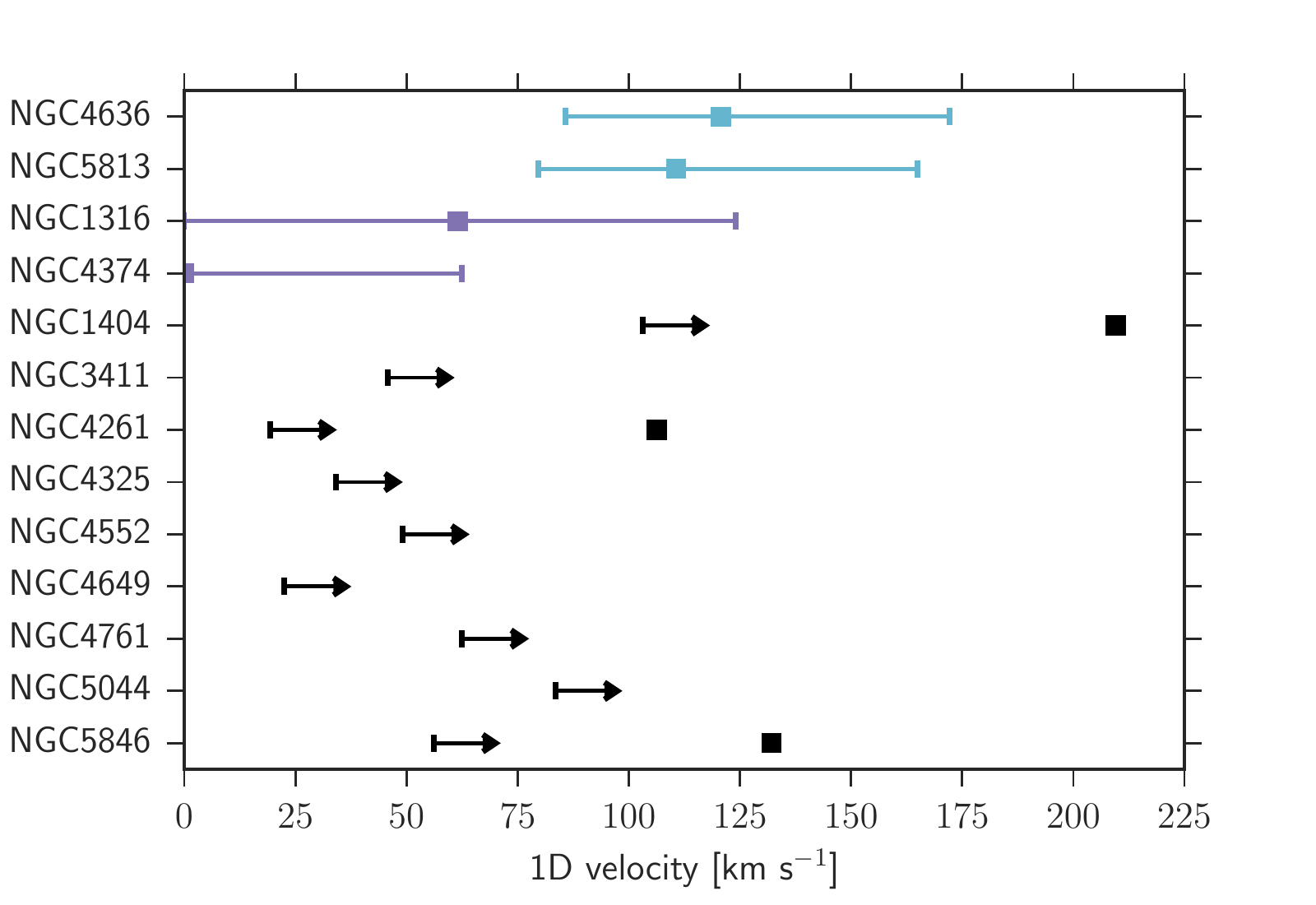}
\caption{Resonant scattering constraints on 3D Mach number (left panel) and  1D turbulent velocity amplitude (right panel) for galaxies in our sample. Blue and purple points show 1\sig limits for galaxies in groups I and II, while black show 2\sig upper limits of group III objects (full pdfs are shown in Figs.~\ref{fig:pdf} and \ref{fig:pdf1}; see Sect.~\ref{subsec:pdf} for details). Squares mark the most probable value (where available).}
\label{fig:sig_res}
\end{minipage}
\end{figure*}

Below we summarize the derived constraints, listed in Table~\ref{tab:results}, by dividing galaxies into 5 groups. Summary plot for groups I---III is shown in Fig.~\ref{fig:sig_res} (left), with group I shown in blue, group II in purple, and group III in black. The sensitivity of the results to the assumed maximum Mach number $M_{max}=1$ (the prior) is  discussed in Sect.~\ref{subsec:prior}. 

\begin{table*}
\begin{minipage}{180mm}
\caption{Constraints on turbulence in our sample of elliptical galaxies, divided into subgroups according to the availability of the  resonant scattering limit derived (Sect.~\ref{subsec:pdf}), and method used. RS: resonant scattering-only probability distributions (Sect.~\ref{subsec:pdf}); RS+LB: resonant scattering-plus-line broadening probability distributions (Sect.~\ref{subsec:combined}). Results are shown for: $M$ -- 3D Mach number, $v_{1D}$ -- line-of-sight turbulent velocity amplitude, and $\epsilon_{turb}/\epsilon_{th}$ -- turbulent to thermal pressure fraction. The \textit{RGS} measured flux ratio with 1\sig statistical uncertainty, along with the physical size of the \textit{RGS} extraction region (the slit width; cf. Fig.~\ref{fig:RGS_slit}) are also provided. Note that these measurements are performed using our fiducial model. We discuss in detail the effects of systematic uncertainties and model assumptions in Section~\ref{sec:unc}.}
\label{tab:results}
\centering
\begin{tabular}{@{ }cccccccc}
\hline
\hline
\multicolumn{8}{c}{I. Bounded constraints} \\
\hline
Source&  \textit{RGS} slit & Measured  & \multicolumn{2}{c}{RS-only 1$\sigma$/68\% limits} & \multicolumn{3}{c}{RS+LB 1$\sigma$/68\% limits} \\
name& width  &  flux ratio &  $M$ & $v_{1D}$  & $M$ & $v_{1D}$ &  $\epsilon_{turb}/\epsilon_{th}$  \\
        & [kpc]     &       &     & [km~ s$^{-1}$] &   &  [km~ s$^{-1}$]  &  [\%]\\
\hline
\hline
NGC 4636 &4.0 &  1.95  $\pm$  0.09   & $0.52^{+0.22}_{-0.22} $ & $ 121^{+51}_{-35} $   & $0.52^{+0.22}_{-0.15} $ &  $121^{+51}_{-36} $ & $12^{+15}_{-7} $ \\ \\ 
NGC 5813  &7.4 &  1.8 $\pm$ 0.1  & $0.44^{+0.21}_{-0.12} $ & $ 111^{+54}_{-31} $ & $0.37^{+0.12}_{-0.09} $ &  $94^{+29}_{-24} $ & $6.3^{+5.9}_{-2.7} $\\
\hline
\hline
\multicolumn{8}{c}{II. Upper limits} \\
\hline
Source &   \textit{RGS} slit& Measured&  \multicolumn{2}{c}{RS-only 1$\sigma$/68\% limits} & \multicolumn{3}{c}{RS+LB 1$\sigma$/68\% limits}\\
name & width &  flux ratio&    $M$ & $v_{1D}$  & $M$ & $v_{1D}$ & $\epsilon_{turb}/\epsilon_{th}$  \\
        &     [kpc]  &      &     & [km~ s$^{-1}$] &  &  [km~ s$^{-1}$]  &  [\%]\\
\hline
\hline
NGC 1316&4.4  &  1.8  $\pm$  0.1 & $0.23^{+0.24}_{-0.23} $& $61^{+63}_{-61}$  & $<0.28$ &  $<73 $ & $<5.1 $ \\ \\
NGC 4374 &3.2 &  2.1  $\pm$  0.2 &  $0_{-0}^{+0.26}$ & $0_{-0}^{+63} $ & $0_{-0}^{+0.19} $ &  $0_{-0}^{+47}$ & $< 3.2 $   \\ 
\hline \hline

\multicolumn{8}{c}{III. Lower limits} \\
\hline
Source &   \textit{RGS} slit& Measured & \multicolumn{2}{c}{RS-only 2$\sigma$/95\% limits} & \multicolumn{3}{c}{RS+LB 1$\sigma$/68\% limits}\\
name& &  flux ratio&   $M$ & $v_{1D}$  & $M$ & $v_{1D}$  & $\epsilon_{turb}/\epsilon_{th}$  \\
        &  [kpc]    &    &     & [km~ s$^{-1}$] & &  [km~ s$^{-1}$]  &  [\%]\\
\hline
\hline
NGC 1404 & 4.6 &   2.0  $\pm$  0.2&$>$ 0.46 &$>$ 103 & $1.02^{+0.58}_{-0.39} $ &  $230^{+130}_{-90} $ & $35^{+38}_{-18} $ \\ \\
NGC 3411 & 15  &  1.1 $\pm$  0.3& $>$ 0.15 &$>$ 46 & $0.58^{+0.66}_{-0.38} $ &  $170^{+200}_{-110} $ & $ < 43 $ 
\\ \\
NGC 4261 &1.1  &  1.9  $\pm$  0.5 &  $>$ 0.079 &$>$  19 & $0.29^{+0.43}_{-0.23} $ &  $71^{+105}_{-57} $ & $ < 24$   \\ \\
NGC 4325 &25 &  1.4  $\pm$  0.6 & $>$  0.12 &$>$  34 & $0.86^{+0.83}_{-0.57} $ &  $240^{+230}_{-160} $ & $ < 52 $  \\ \\
NGC 4552  &3.6&  1.6  $\pm$  0.3 &  $>$ 0.22 & $>$ 49 & $0.79^{+1.31}_{-0.49} $ &  $180^{+290}_{-110} $ & $8^{+46}_{-8} $  \\ \\
NGC 4649&4.0  &  1.3  $\pm$  0.2  &$>$ 0.081 & $>$  23& $0.26^{+0.32}_{-0.19} $ &  $71^{+91}_{-51} $ & $ < 16 $\\ \\
NGC 4761 & 14 &  1.3  $\pm$  0.1& $>$ 0.23 & $>$ 62 & $0.61^{+0.41}_{-0.33} $ &  $166^{+114}_{-90} $ & $ <39 $ \\ \\
NGC 5044&9.6  &  1.3  $\pm$  0.1 &$>$ 0.30 & $>$ 83 & $0.61^{+0.39}_{-0.28} $ &  $172^{+108}_{-79} $ & $8^{+35}_{-6} $ \\ \\
NGC 5846 & 7.4 &  1.7  $\pm$  0.2 & $>$ 0.24 &$>$ 56 & $0.32^{+0.25}_{-0.14} $ &  $77^{+58}_{-34} $ & $3^{+12}_{-3} $\\
\hline
\hline
\multicolumn{8}{c}{IV \& V. No resonant scattering limit} \\
\hline 
Source &  \textit{RGS} slit& Measured & \multicolumn{5}{c}{Reason behind }\\
name &   width [kpc] & flux ratio &   \multicolumn{5}{c}{the lack of limit}  \\
\hline
\hline
NGC 1332 & 4.4 &  2.7  $\pm$  0.8  & \multicolumn{5}{c}{qualitative description} \\ 
NGC 4406 &3.8 &  2.1  $\pm$  0.3  & \multicolumn{5}{c}{(see Sect.~\ref{subsec:pdf}.IV);} \\
NGC 4472 & 1.5 &  2.2  $\pm$  0.3  &  \multicolumn{5}{c}{additional data necessary} \\
\hline
NGC 507 & 15  &  1.0  $\pm$  0.4  & \multicolumn{5}{c}{ low concentration  }   \\
NGC 708  &16 &  1.4  $\pm$  0.2  & \multicolumn{5}{c}{of \Fe ions } \\
\hline
IC 1459  & 5.0 &  1.7  $\pm$  0.5  &   \multicolumn{5}{c}{very large scatter  } \\ 
NGC 533 & 32 &  2.1  $\pm$  0.7  &  \multicolumn{5}{c}{on measured line ratios;}  \\
NGC 1399  &4.4 &  1.3  $\pm$  0.2  & \multicolumn{5}{c}{for further improvements }  \\
NGC 4696   &10&  1.5  $\pm$  0.1  &  \multicolumn{5}{c}{additional data necessary }  \\
\hline
\hline
\end{tabular}
\end{minipage}
\end{table*}

\textit{I. Peaked distributions: bounded constraints.} Two galaxies,  NGC~4636 and NGC~5813, show clearly peaked distributions that allow us to  report the most probable value with 68~per~cent confidence (lower and upper limit; Fig.~\ref{fig:pdf}). Changing the Mach number prior does not affect the results significantly. 

\textit{II. Peaked distributions: low turbulence.} Two more objects have clearly peaked distributions: NGC~4374 and NGC~1316 (Fig.~\ref{fig:pdf1}).  Both galaxies have low turbulence, and the confidence intervals do not provide meaningful lower limits. However, since the only difference between the former and latter groups is that the distributions are shifted towards smaller Mach numbers, we treat them the same way as galaxies in group \textit{I} and report the most probable value with a 68~per~cent upper limit and a lower limit of 0. 

\textit{III. Lower limits.}  Resonant scattering technique is better suited for providing lower rather than upper limits on gas motions velocities: the stronger the turbulence, the smaller the line suppression (Sect.~\ref{subsec:rs}). Therefore the integrated flux ratio in the case of  large Mach numbers approaches the constant optically thin plasma case (cf. the blue and yellow lines in Fig.~\ref{fig:int_ratio}). Since pdfs  flatten at large $M$ (cf. Fig.~\ref{fig:pdf1}), our velocity constraints depend on maximum  Mach number $M_{max}$ assumed. We note  that lower limits are least sensitive to the choice of Mach number prior (Sect.~\ref{subsec:prior}).

For the majority of the objects in our sample (9; see Table~\ref{tab:results}) we are able to provide lower limits only (at 95~per~cent confidence; for pdfs see Fig.~\ref{fig:pdf1}).  Among these objects, three galaxies also show the most probable value (NGC~1404, NGC~4261, and NGC~5846). The mean lower limit is 0.178 (47~km~s$^{-1}$), excluding NGC~1404, which is an obvious outlier (lower limits:  $M>0.46$,  $v_{1D}>103$~km~s$^{-1}$).

\textit{IV. Low turbulence - qualitative description.} Three galaxies, NGC~1332, NGC~4406, and NGC~4472,  have high observed flux ratios, indicating that the optically thick line is suppressed due to resonant scattering. This points towards low turbulence, which is reflected in the pdfs of these objects (Fig.~\ref{fig:pdf2}). The effect itself is not strong, and the resulting probability distributions are too flat for deriving limits. However, we can see that the pdf at $M=0$ is approximately $\sim$5 times more than at $M=0.5$ or $M=1$   in NGC~4406,   $\sim 3$  times more likely in NGC~4472 and $\sim$2 times more likely in NGC~1332. 

\textit{V. No limits.} For 6 galaxies we are not able to derive any turbulence constraints. NGC~708 and NGC~507 have the largest gas temperatures in our sample and the resonant \Fe line is optically thin. Other objects (e.g. NGC~1399, NGC~4696) are more distant, thus their innermost regions, where most of the scattering occurs, have angular sizes significantly smaller than the \textit{RGS} aperture used. Current \textit{RGS} data does not allow us to measure line ratios reliably in such regions (Sect.~\ref{subsec:improv}).

\vspace{3ex}

\begin{figure}
\includegraphics[width=\columnwidth,trim={1.5mm 0mm 13mm 0},clip]{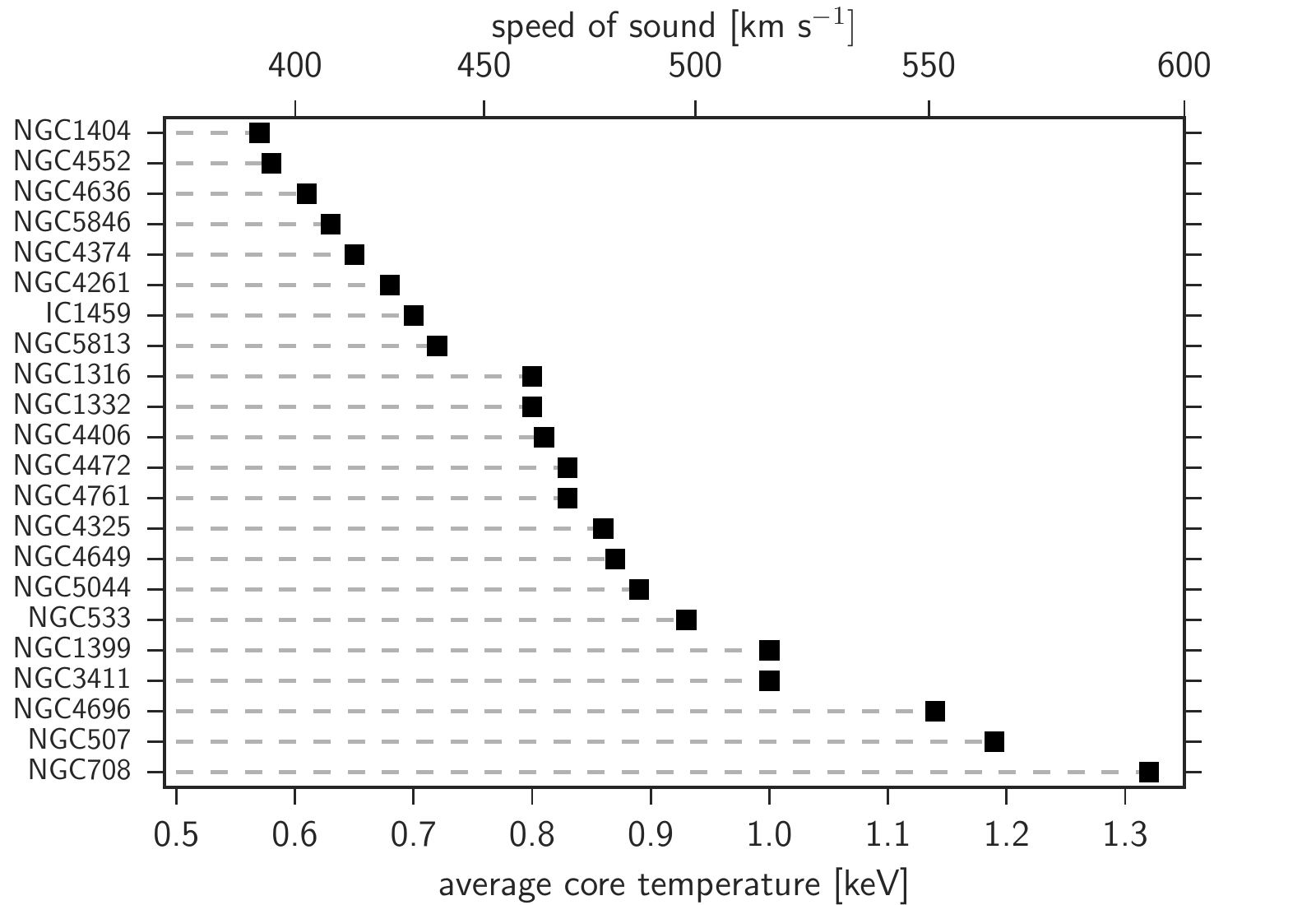}
 \caption{Average temperature within the \textit{RGS} extraction region (bottom axis), and respective sound speed (top axis; this axis has a square root scaling) calculated for all galaxies in our sample.} 
\label{fig:temp}
\end{figure}

The characteristic line-of-sight one-component turbulent velocity amplitude $v_{1D}$ can be derived directly from the 3D Mach number. If turbulence is isotropic, then:
\begin{equation}
v_{1D} = \frac{1}{\sqrt{3}} M c_s = \frac{1}{\sqrt{3}} M \sqrt{\frac{\gamma k T_e}{\mu m_p}}.
\end{equation}
The temperature $T_e$ that we use here is the mass weighted average electron temperature calculated within the spectral extraction region from the deprojected temperature profile. The temperatures of galaxies,  along with respective sound speeds, are shown  in Fig.~\ref{fig:temp} and Table~\ref{tab:sample}. The derived turbulent velocity constraints are listed in Table~\ref{tab:results}, and shown in the right panel of Fig.~\ref{fig:sig_res}. Typical lower limits are 20--60~km~s$^{-1}$. For a few objects the measured velocities are between 43--121~km~s$^{-1}$. 

\subsection{Resonant scattering--plus--line broadening}
\label{subsec:combined}

\begin{figure}
\includegraphics[width=\columnwidth,trim={0mm 2mm 12mm 0},clip]{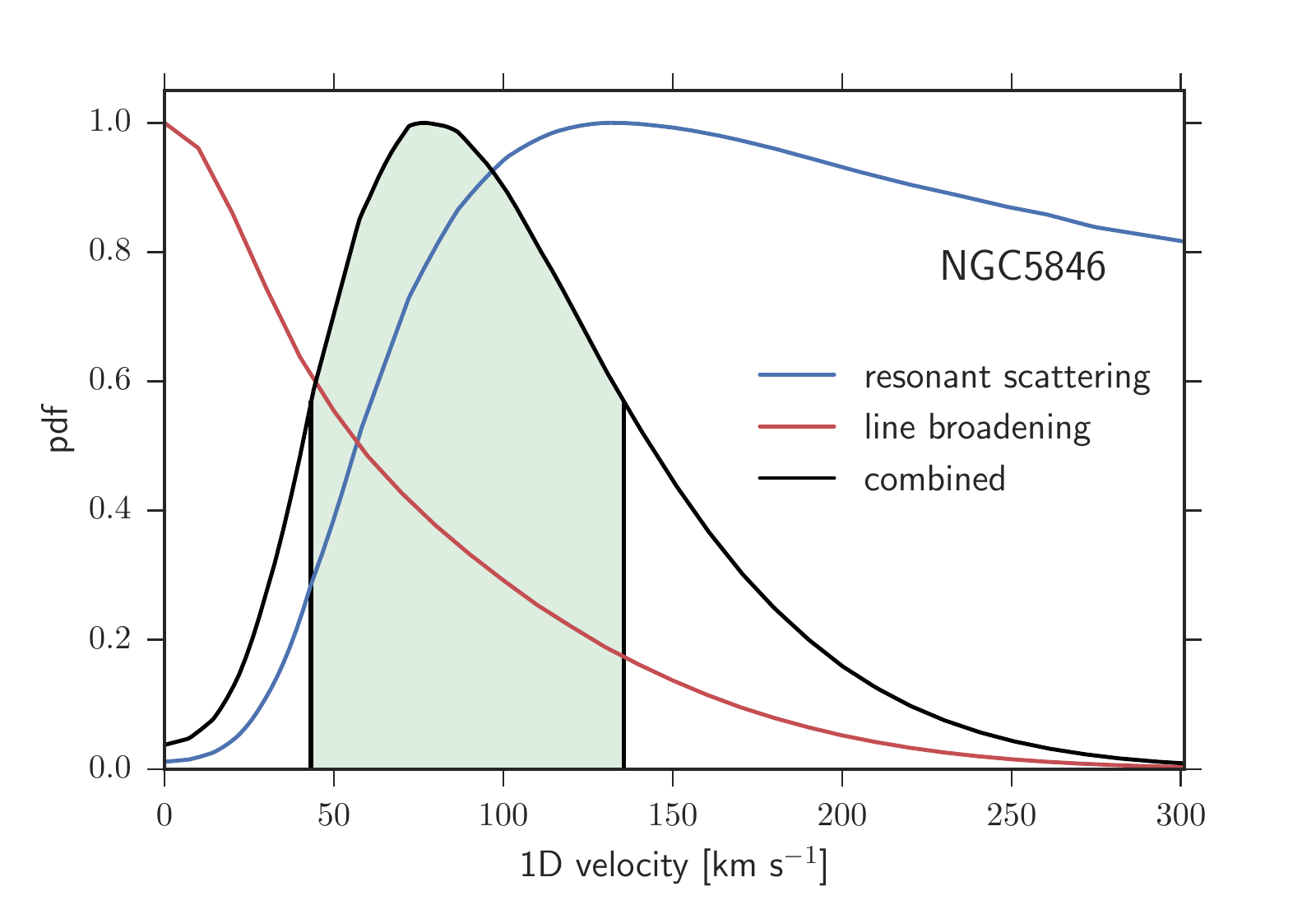}
\caption{Combined resonant scattering and line broadening measurement in NGC~5846. By using respective probability distributions (blue and red lines), a combined distribution is obtained (black line), which allows us to derive the most probable velocity with 1\sig lower and upper limits (green region). For details, see Sect.~\ref{subsec:combined}.}
\label{fig:LBRS}
\end{figure}

The velocity dispersion of the X-ray emitting gas can be measured directly through line broadening. Since the \textit{RGS} is a slitless spectrometer, the spatial extent of the galaxy core broadens the lines in addition to natural, thermal and turbulent broadening. Therefore, direct velocity measurements provide mostly upper limits \citep[e.g.][and Sect.~\ref{subsec:rgs}]{Pinto2015}. We obtained the line broadening velocity probability distributions by performing spectral fitting, varying the velocity and evaluating the C-statistics each time (Sect.~\ref{sec:velocity_broadening}). These pdfs are then combined by multiplying with the pdfs from the resonant scattering analysis (Sect.~\ref{subsec:pdf}). 

\begin{figure*}
\begin{minipage}{180mm}
\centering 
\includegraphics[width=0.49\columnwidth,trim={4mm 3mm 11mm 3mm},clip]{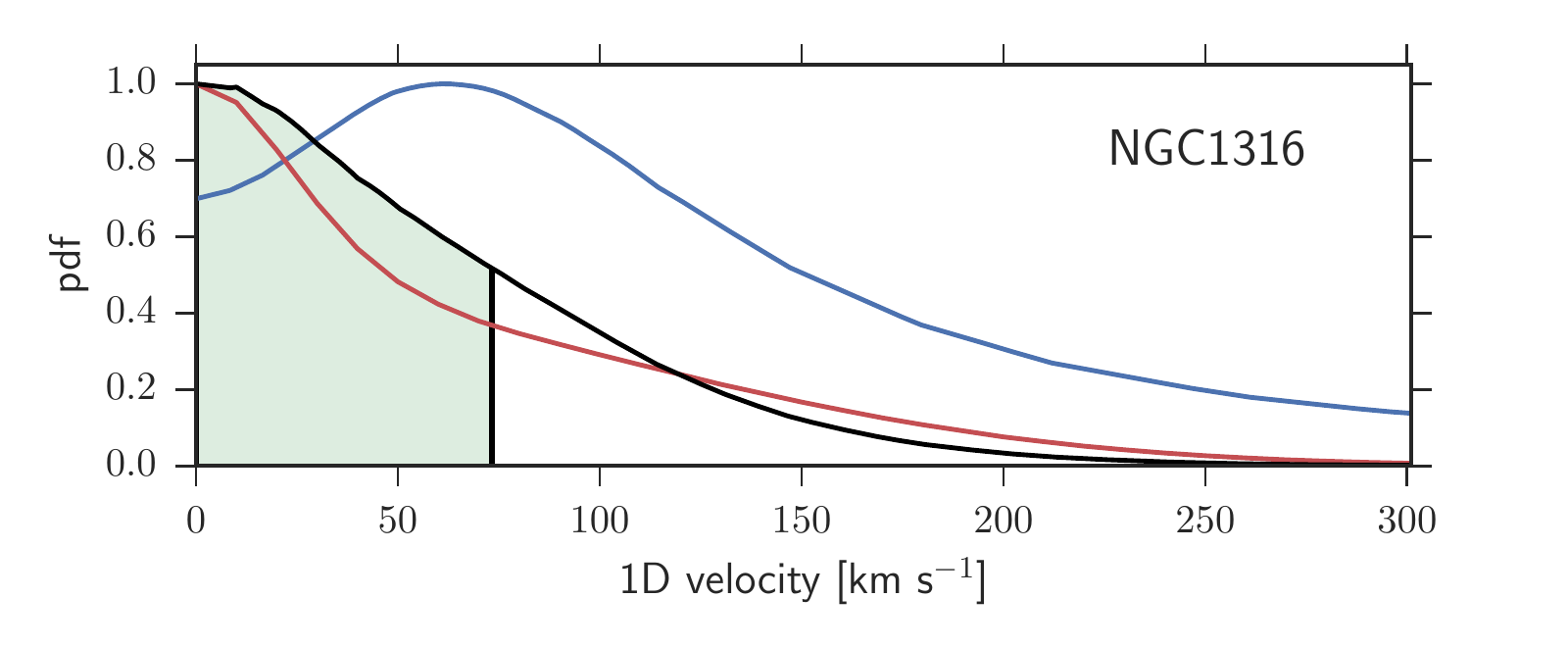}
\includegraphics[width=0.49\columnwidth,trim={4mm 3mm 11mm 3mm},clip]{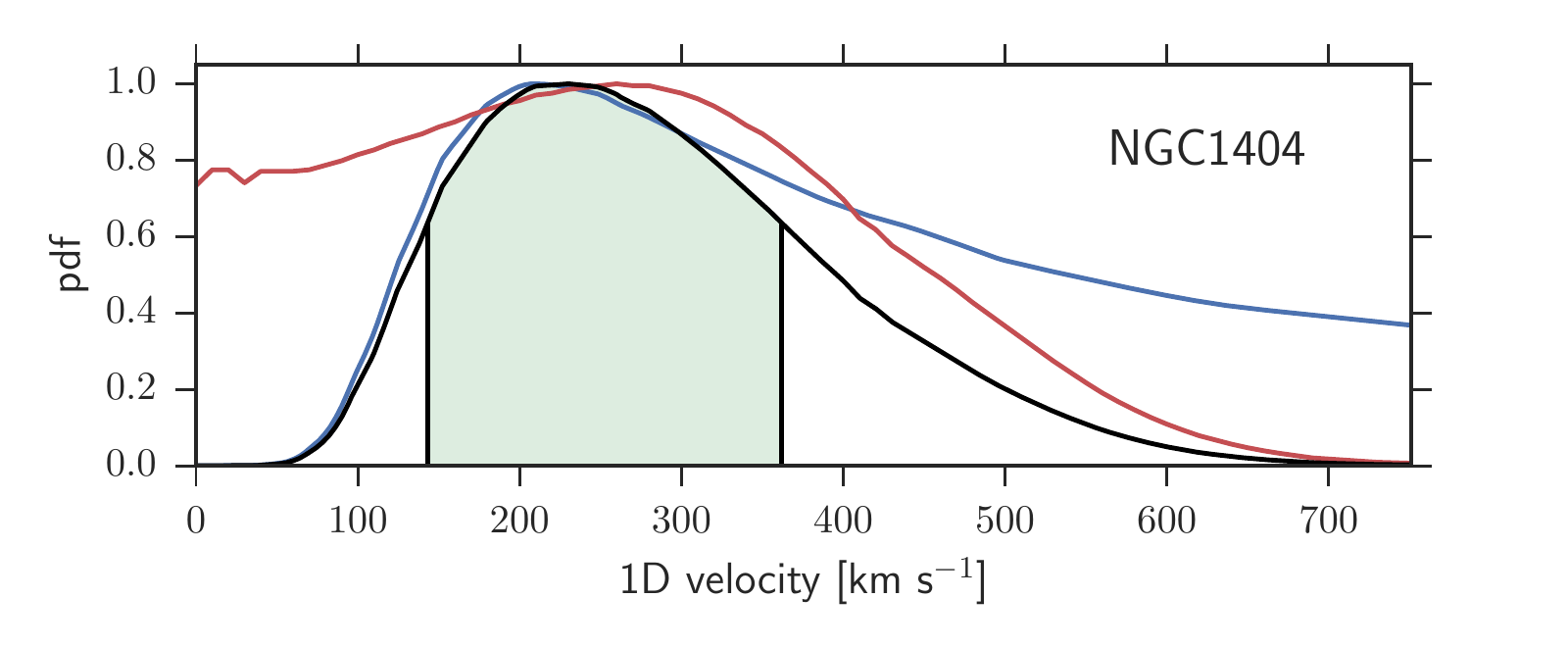}
\includegraphics[width=0.49\columnwidth,trim={4mm 3mm 11mm 3mm},clip]{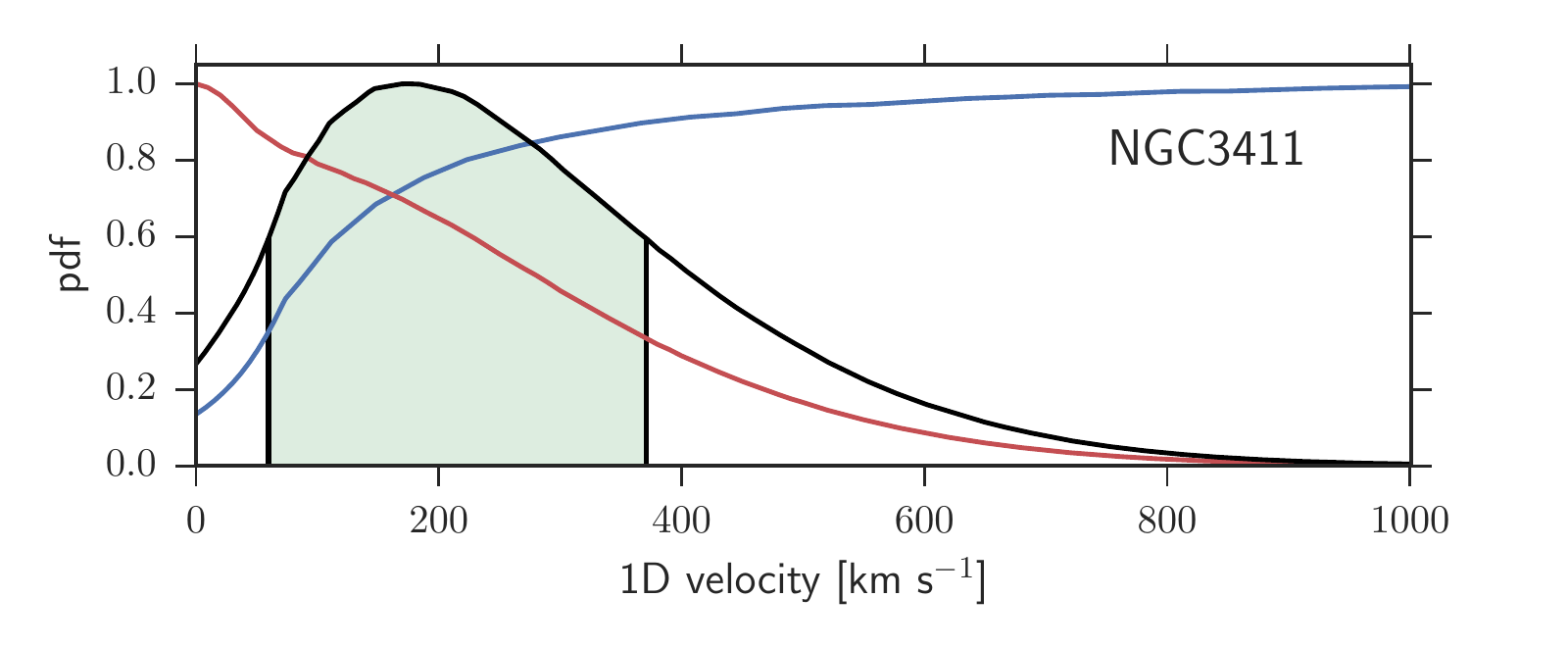}
\includegraphics[width=0.49\columnwidth,trim={4mm 3mm 11mm 3mm},clip]{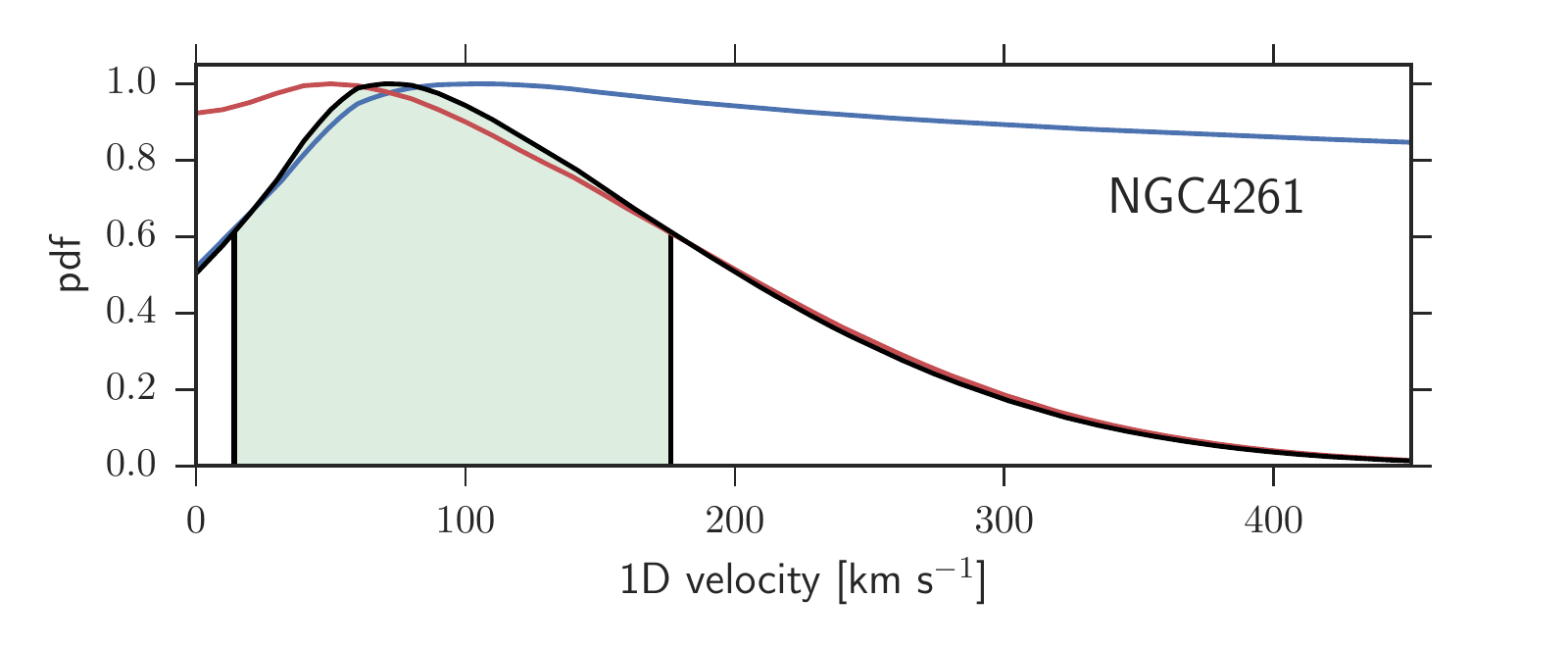}
\includegraphics[width=0.49\columnwidth,trim={4mm 3mm 11mm 3mm},clip]{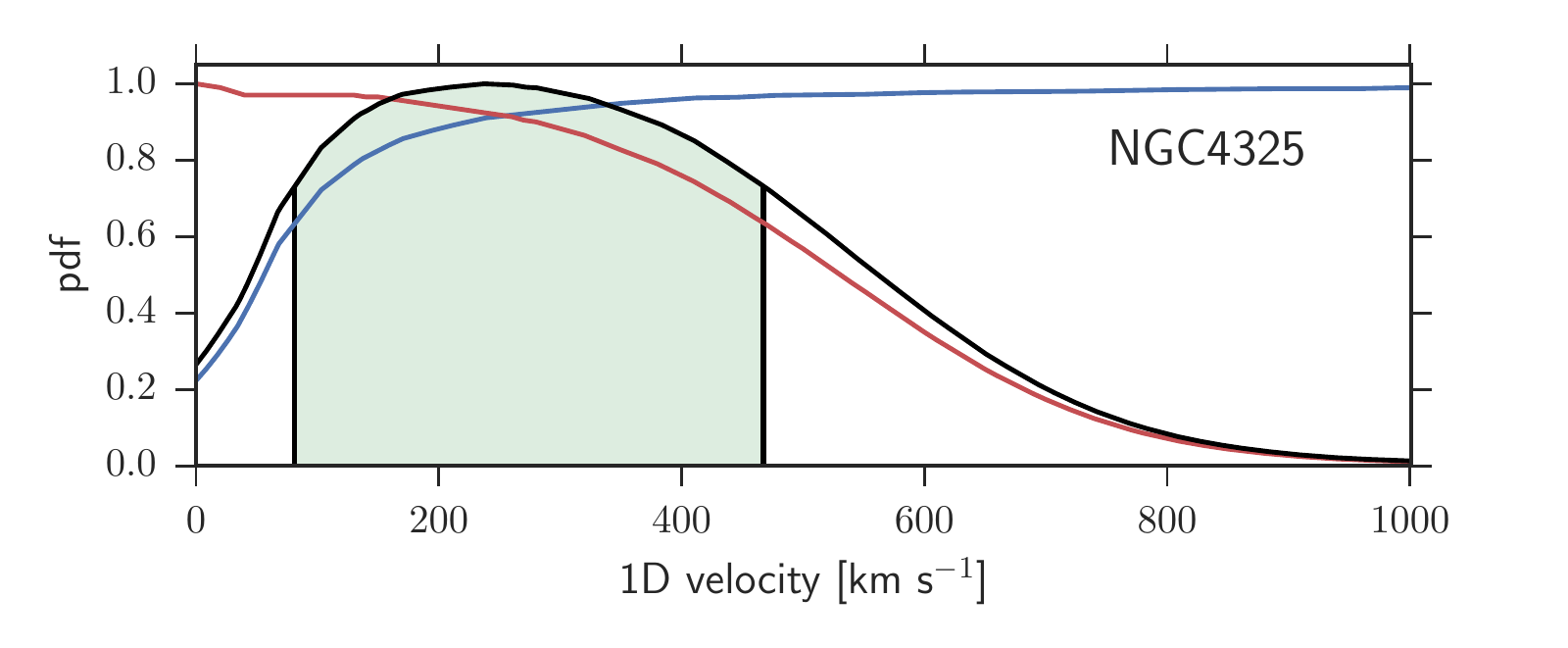}
\includegraphics[width=0.49\columnwidth,trim={4mm 3mm 11mm 3mm},clip]{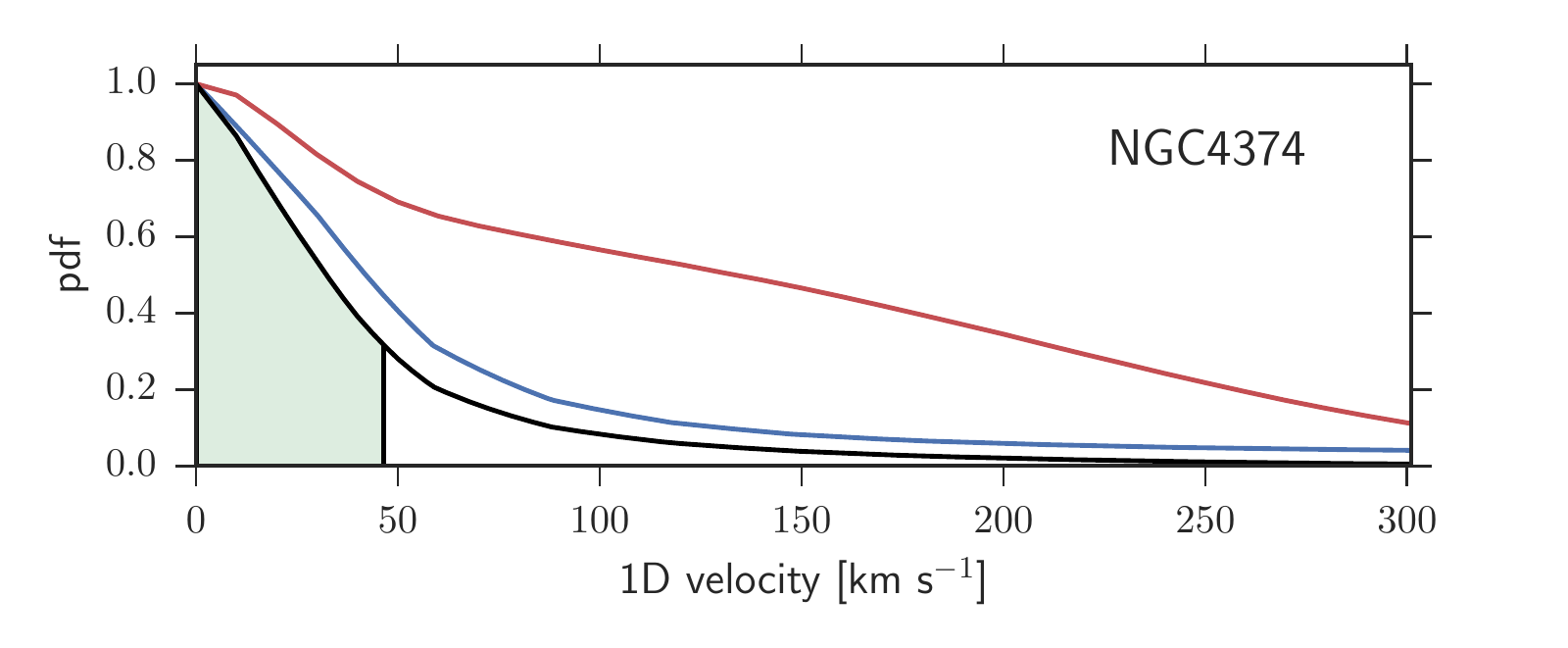}
\includegraphics[width=0.49\columnwidth,trim={4mm 3mm 11mm 3mm},clip]{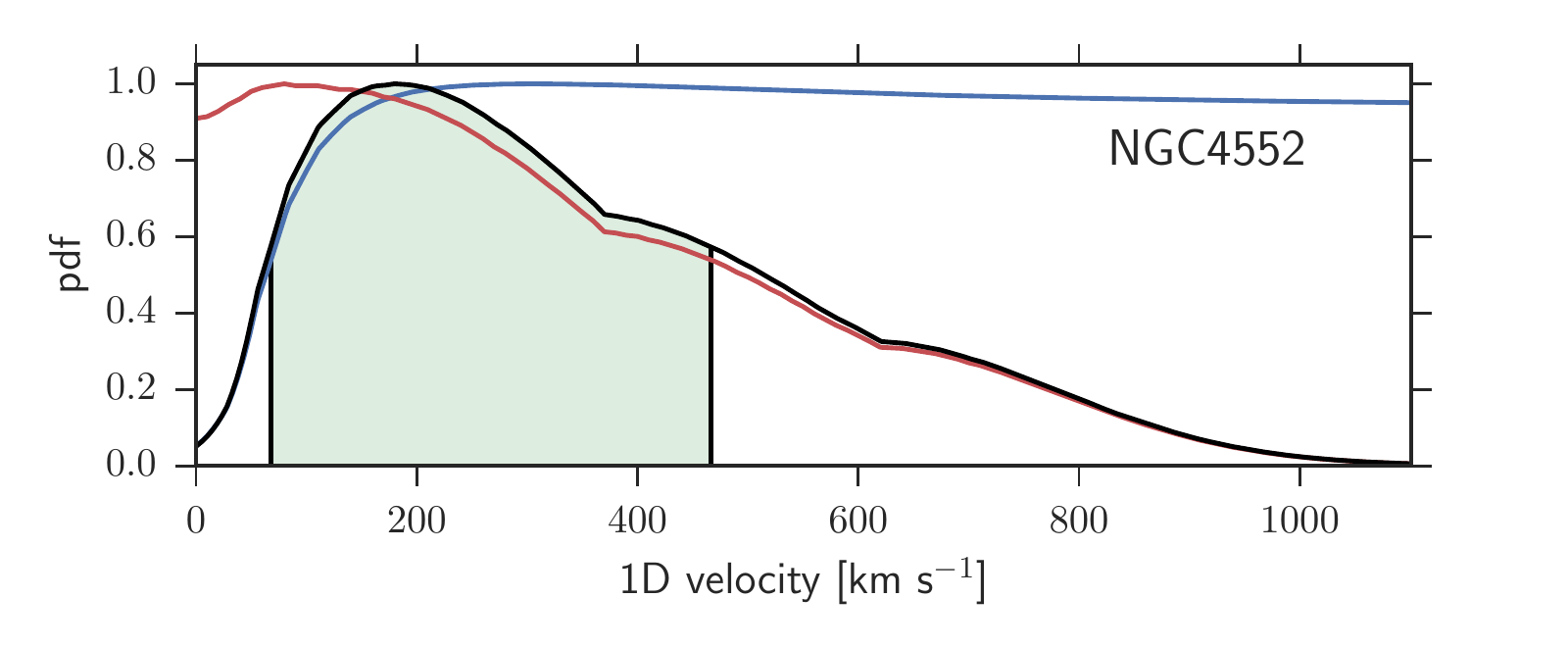}
\includegraphics[width=0.49\columnwidth,trim={4mm 3mm 11mm 3mm},clip]{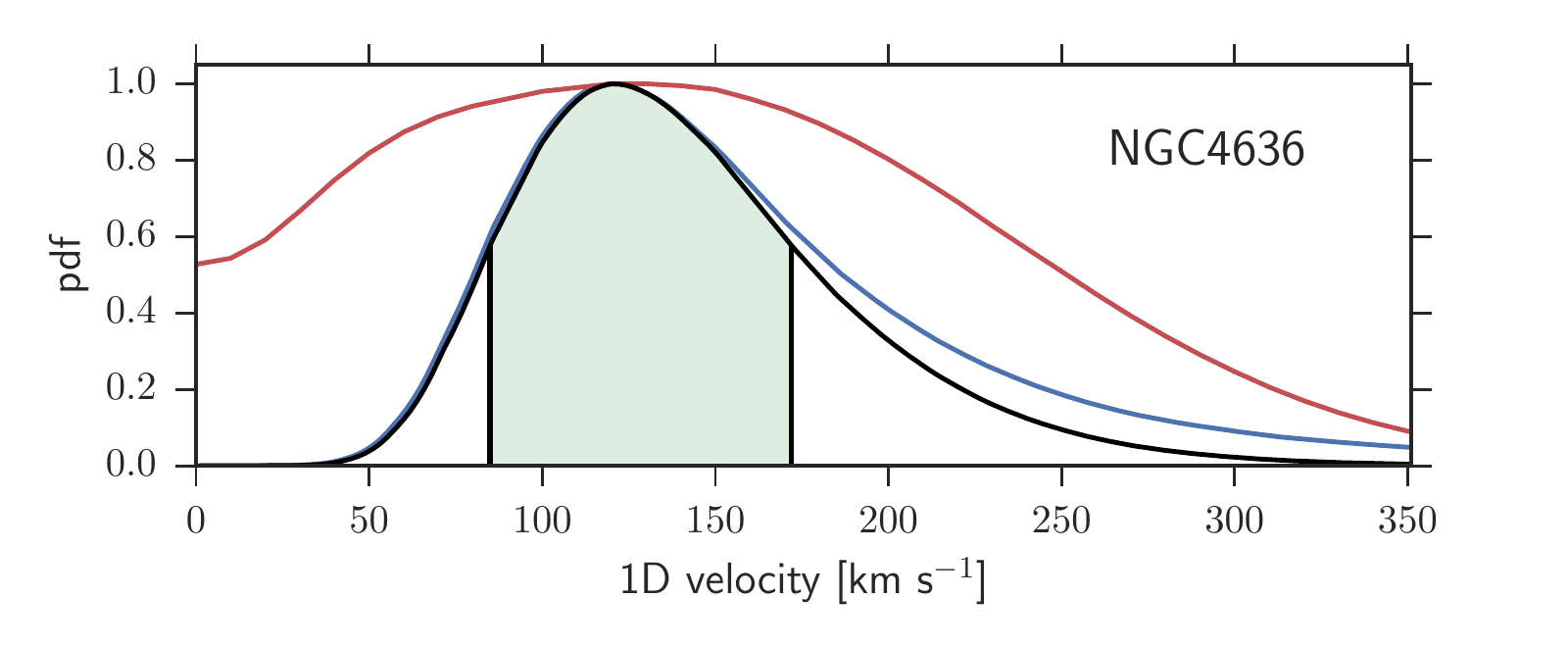}
\includegraphics[width=0.49\columnwidth,trim={4mm 3mm 11mm 3mm},clip]{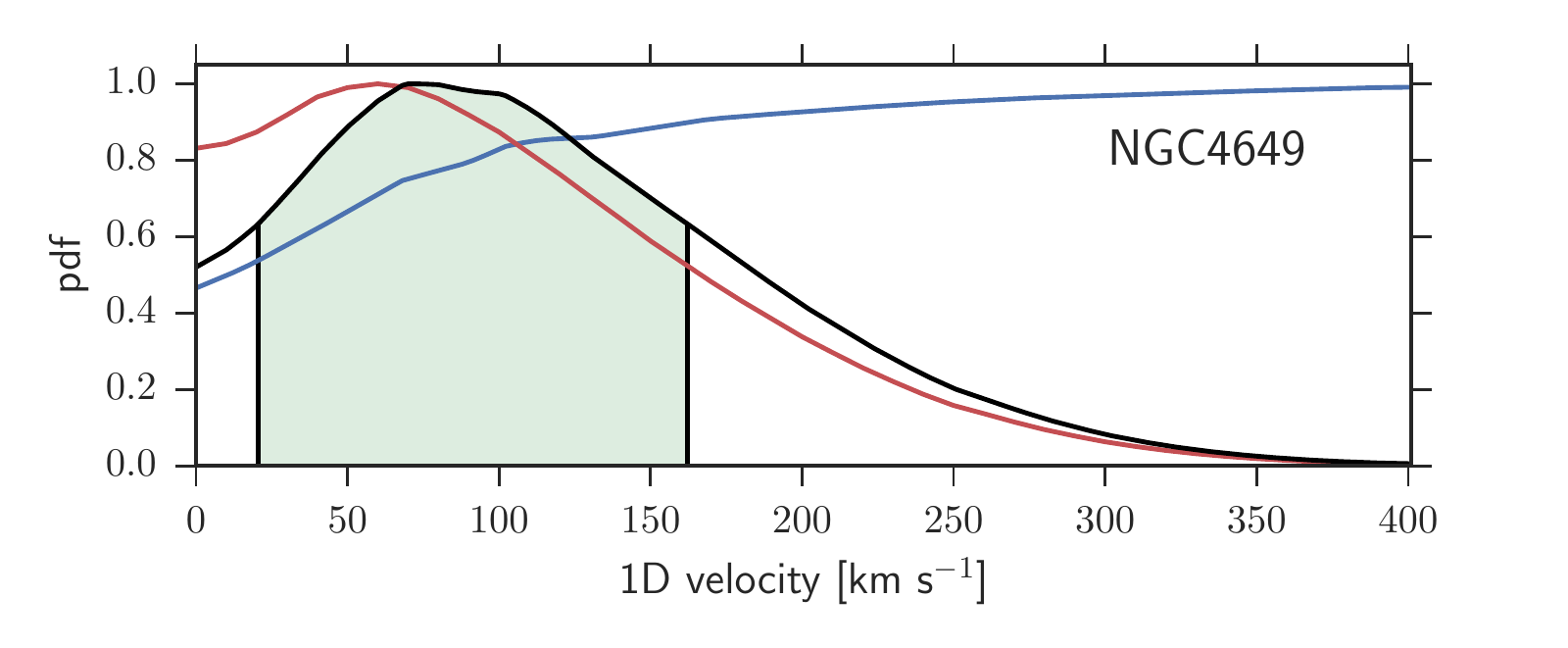}
\includegraphics[width=0.49\columnwidth,trim={4mm 3mm 11mm 3mm},clip]{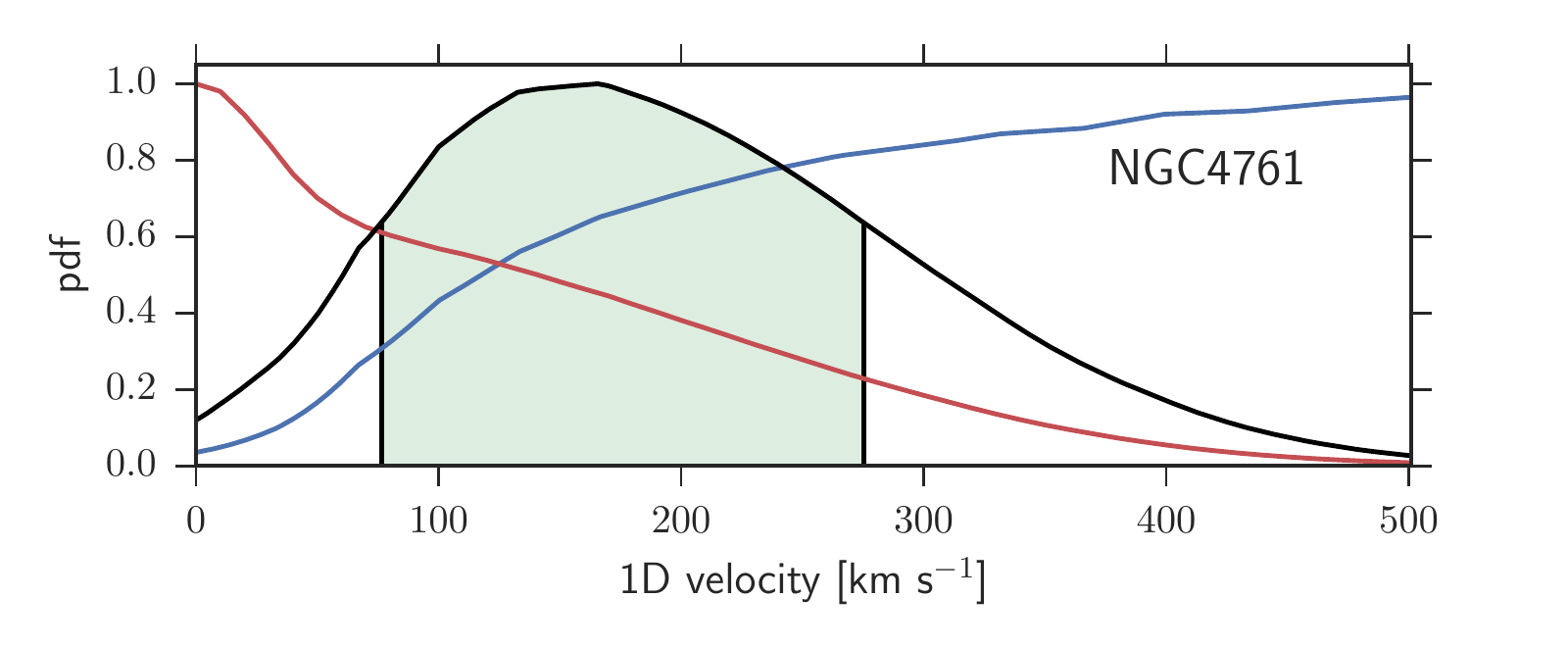}
\includegraphics[width=0.49\columnwidth,trim={4mm 3mm 11mm 3mm},clip]{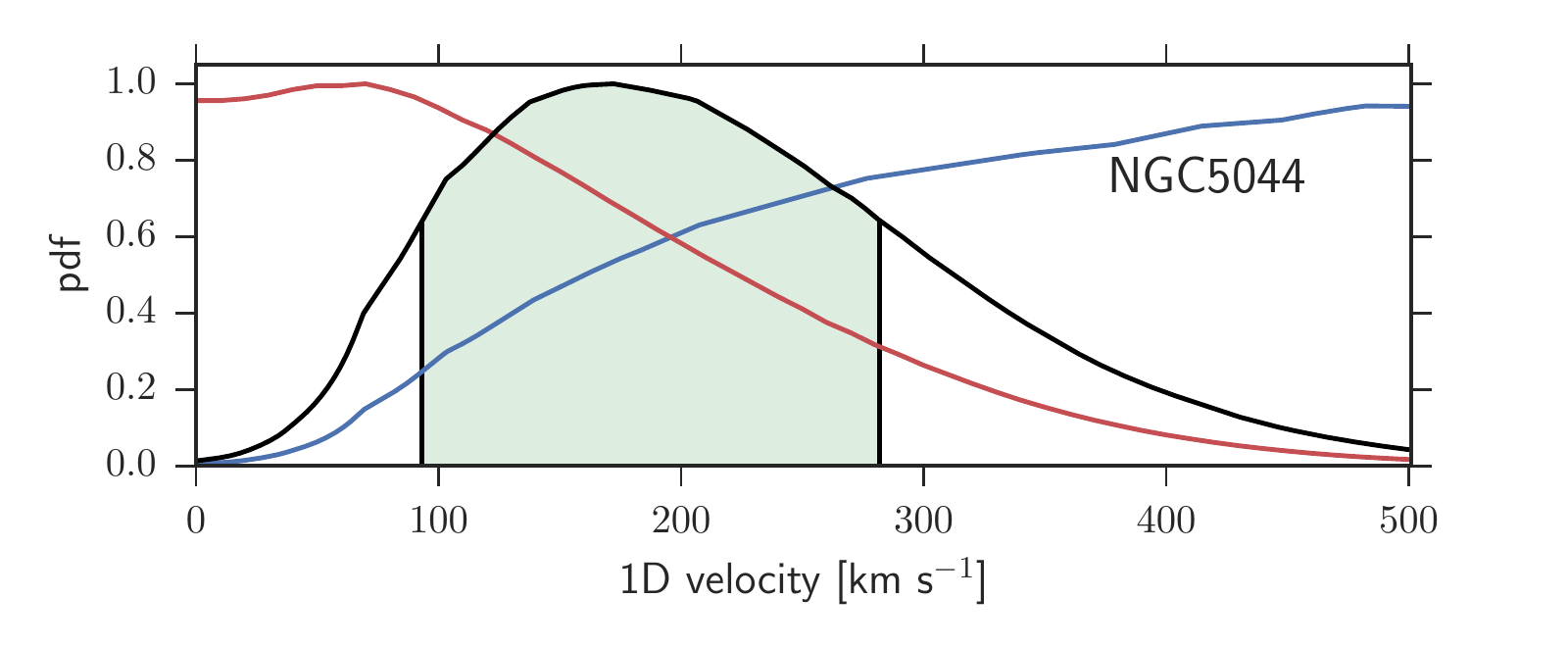}
\includegraphics[width=0.49\columnwidth,trim={4mm 3mm 11mm 3mm},clip]{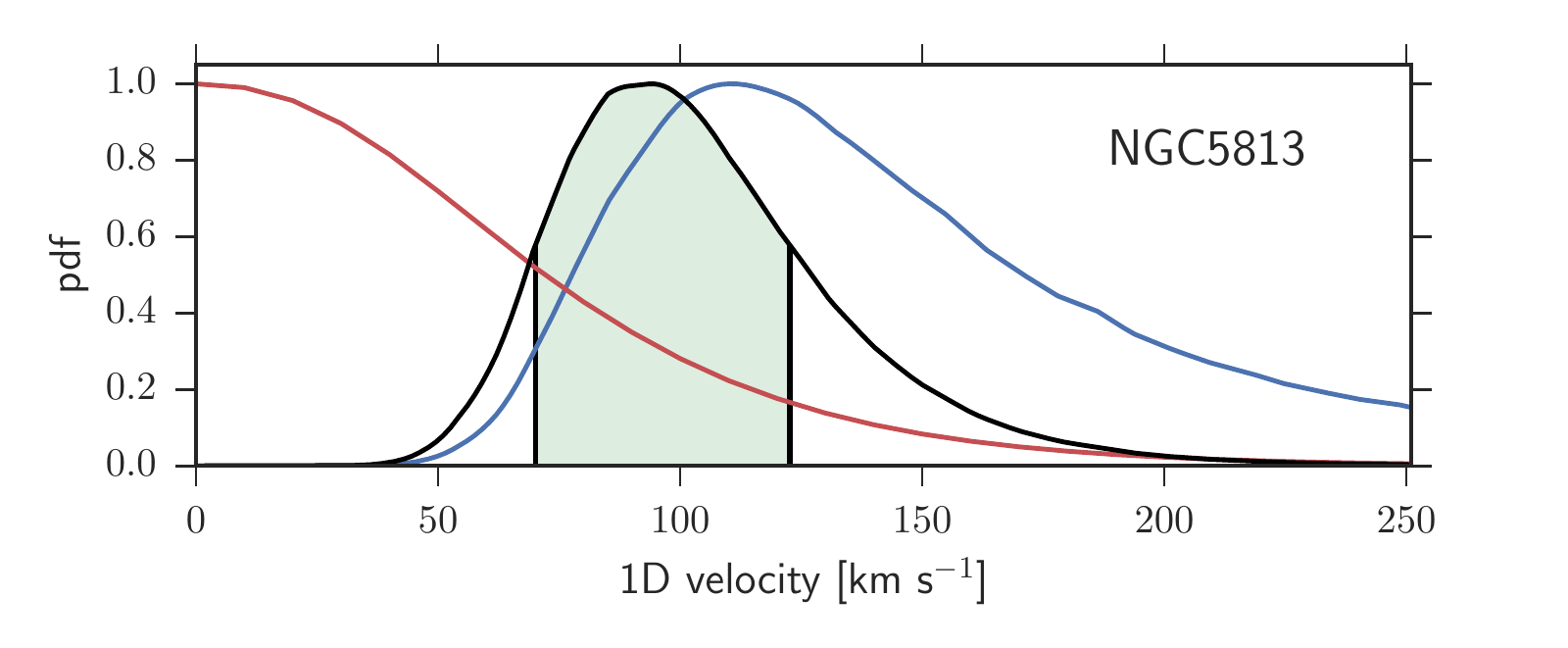}
\caption{Combined resonant scattering (blue) and line broadening (red) probability distributions as a function of one-component velocity  (black), with 1\sig confidence regions (green), shown  for the remainder of the sample. Derived constraints are listed in Table~\ref{tab:results}. See Sect.~\ref{subsec:combined} for details.}
\label{fig:lb_all} 
\end{minipage}
\end{figure*}

Fig.~\ref{fig:LBRS} illustrates the result of this procedure for NGC~5846. The blue line shows the resonant scattering pdf, which provides only a lower limit on the velocity, while the red line shows the line broadening pdf, which gives only an upper limit. The combined distribution, shown in black, is peaked and narrow, and provides a velocity measurement with 68 per cent uncertainties (green region). Analogous plots for the remaining galaxies can be found in Fig.~\ref{fig:lb_all}.

\subsection{Exploring the sample properties}
\begin{figure*}
\begin{minipage}{180mm}
\includegraphics[width=0.5\columnwidth,trim={0mm 2mm 11mm 7mm},clip]{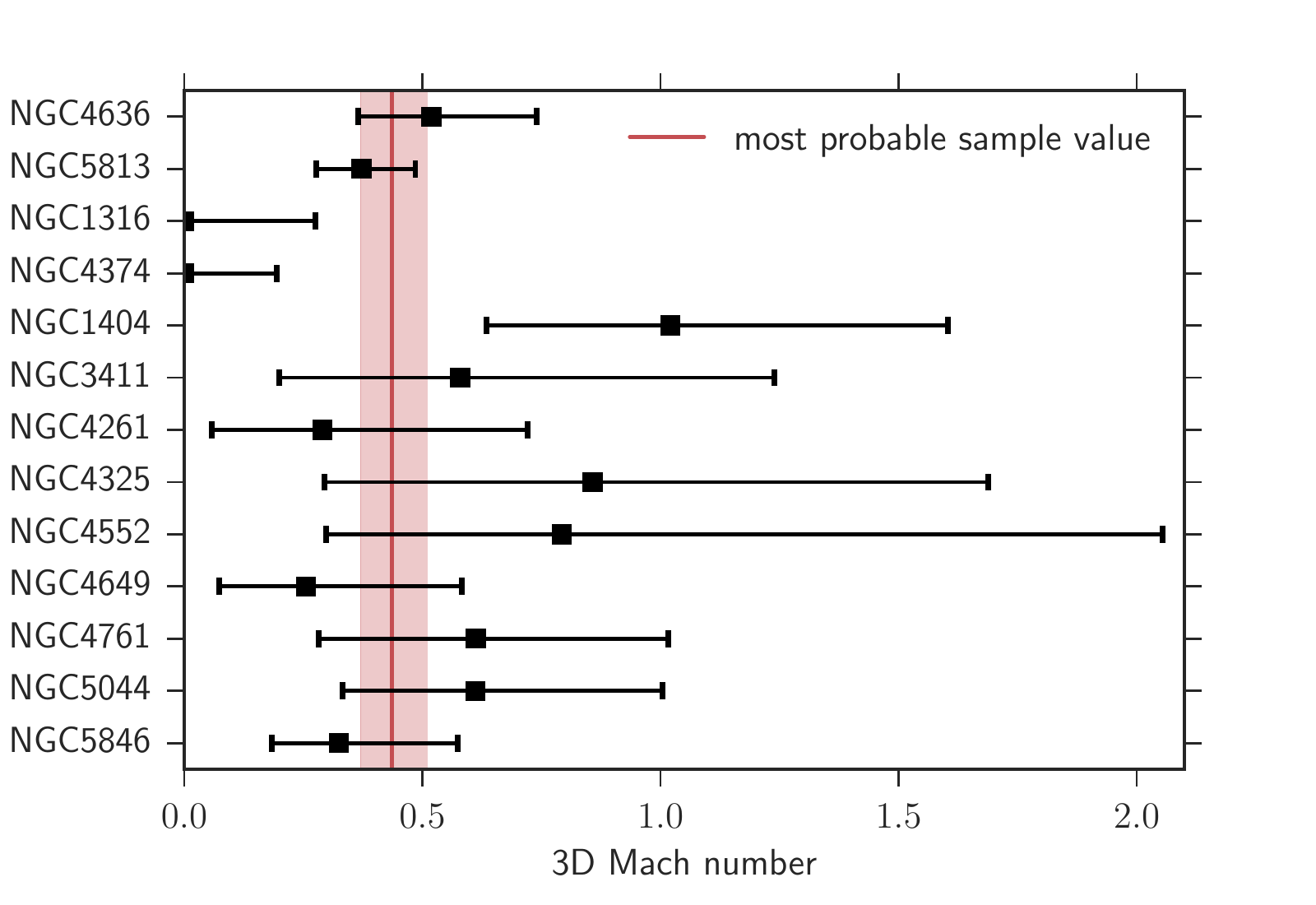}
\includegraphics[width=0.5\columnwidth,trim={0mm 2mm 11mm 7mm},clip]{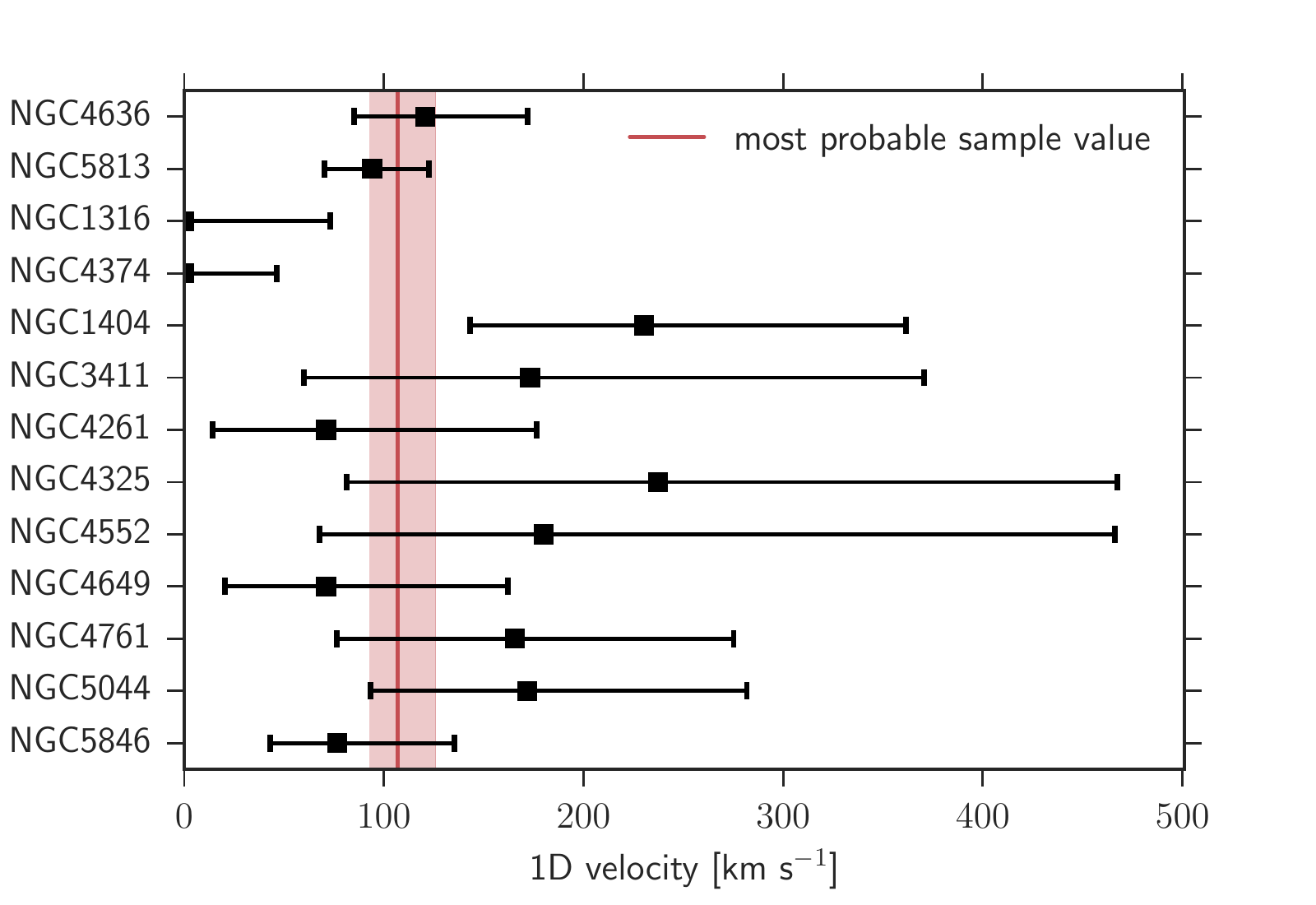}
\caption{Combined resonant scattering and line broadening constraints on 3D Mach number (left panel) and  1D turbulent velocity amplitude (right panel) for galaxies in our sample, with 1\sig uncertainties. Red lines and regions show best fit most probable sample values fitted to the measurements, $M_{mp}=0.44^{+0.07}_{-0.06}$ for Mach number and $v_{1D, mp}=107^{+19}_{-15}$ km~s$^{-1}$ for velocity. In both cases the data shows scatter around most likely values to be consistent with zero within current uncertainties.}
\label{fig:lb_res}
\end{minipage}
\end{figure*}

Using the combined distributions we are able to \textit{measure} velocities of gas motions in 13 galaxies in our sample. We have examined the distribution of measured velocity values, which, in principle, can provide key information on e.g. processes that are responsible for the motions.  A summary plot is shown in  Fig.~\ref{fig:lb_res} (see also Table~\ref{tab:results}). 

We fit the measured velocities and Mach numbers with single mean values, allowing for scatter, using Monte Carlo Markov Chain method. The resulting most probable values are $v_{1D, mp}=107^{+19}_{-15}$ km~s$^{-1}$ and $M_{mp}=0.44^{+0.07}_{-0.06}$ (shown with red lines and regions in Fig.~\ref{fig:lb_res}). In the case of both the velocity and Mach number, the scatter is consistent with zero with 1\sig upper limits of 38\kms~and 0.16, respectively. 

Note that by utilizing the line broadening pdfs, which provide upper limits, we derive these measurements with no prior maximum Mach number assumption, as opposed to the pure resonant scattering case (cf. Sect.~\ref{subsec:pdf}).  

\subsection{Turbulent pressure}
\label{subsec:press}

Knowing the characteristic Mach number, we can place constraints on the non--thermal pressure contributed by turbulence. The energy density of turbulence $\epsilon_{turb}$ is given by:
\begin{equation}
\epsilon_{turb} = \frac{3}{2} \varrho v_{1D}^2 = \frac{1}{2} \varrho c_s^2 M^2,
\end{equation}
where $\varrho$ is the gas mass density. On the other hand, the thermal energy density is:
\begin{equation}
\epsilon_{th} = \frac{3}{2} \frac{\varrho kT}{\mu m_p}.
\end{equation}
The fraction of the turbulent non--thermal to thermal pressure is a ratio of the two energy densities:
\begin{equation}
\frac{\epsilon_{turb}}{\epsilon_{th}} = \frac{\gamma}{3} M^2.
\end{equation}

\begin{figure}
\includegraphics[width=\columnwidth,trim={0mm 3mm 14mm 0},clip]{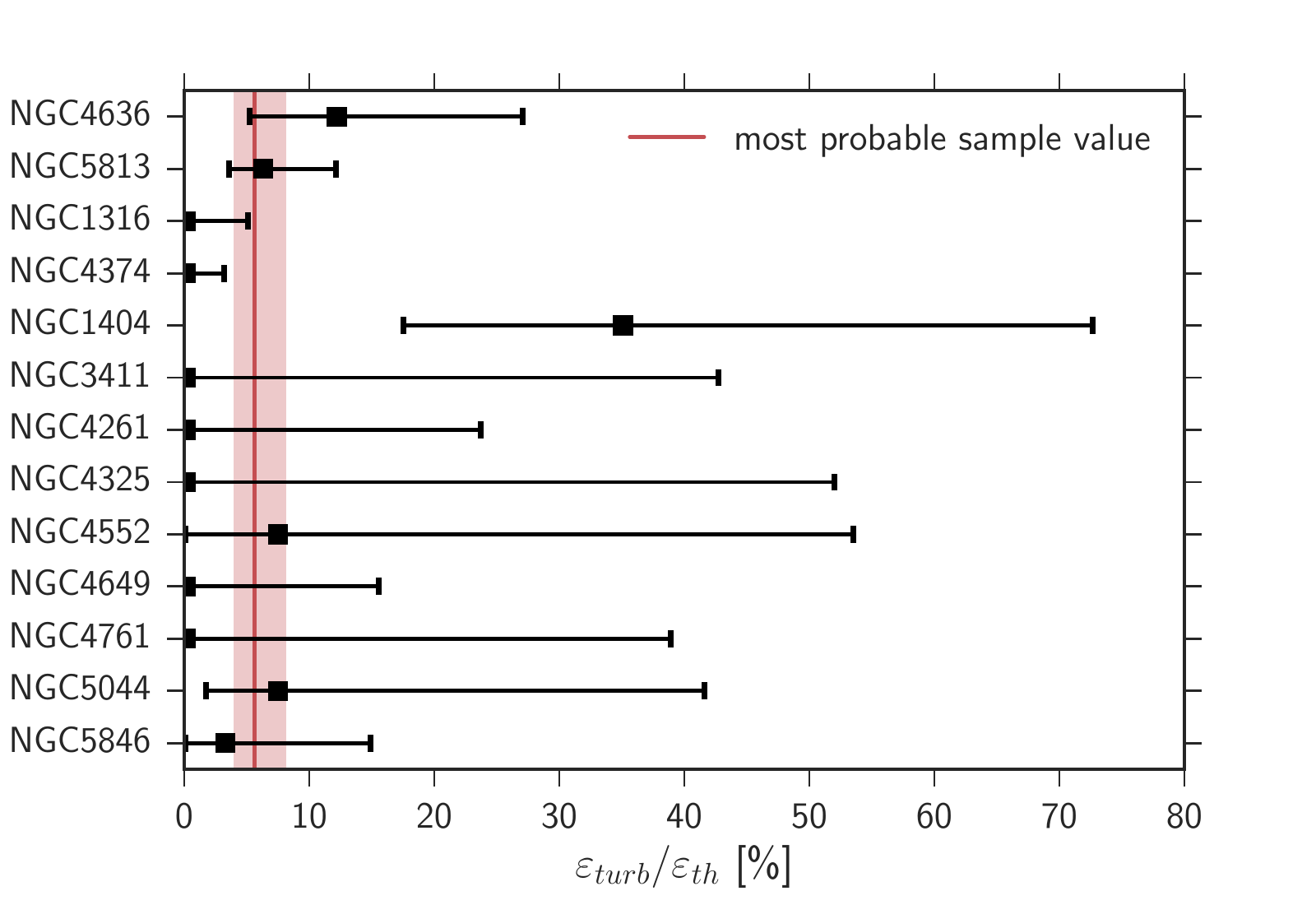}
\caption{Turbulent pressure support in sample galaxies (Sect.~\ref{subsec:press}), with 1\sig uncertainties based on combined resonant scattering and line broadening pdfs (Sect.~\ref{subsec:combined}). Red line and region show the best fit constant: 5.6$_{-1.5}^{+2.6}$~per~cent. Within current uncertainties, the data does not show any scatter around this single value.}
\label{fig:press}
\end{figure}

In order to measure the turbulent pressure support we used the combined resonant scattering and line broadening probability distributions of Mach numbers (Sect.~\ref{subsec:combined}), and calculated respective pressure distributions. In Fig.~\ref{fig:press} we show the resulting 68~per~cent constraints on pressure support (listed in Table~\ref{tab:results}). The data are consistent with no scatter around a single constant value of 5.6$_{-1.5}^{+2.6}$~per~cent, shown with a  red line  in Fig.~\ref{fig:press}. The 1\sig upper limit on scatter is 2.6~per~cent.

\section{Systematic uncertainties}
\label{sec:unc}
Our resonant scattering analysis makes a number of simplifying assumptions. Firstly, we assumed spherically symmetric models of galaxies, which is a crude approximation given that the morphology of some objects is quite disturbed  (e.g. filamentary structure, bubbles etc.; see Sect.~\ref{subsec:improv} for possible future improvements). Secondly, we fitted the deprojected spectra at a given radii with a single temperature model.  This might introduce a bias, since the  \Fe line emissivity and ionic fraction are strong functions of temperature (Fig.~\ref{fig:em_ionbal}). For some galaxies there is no evidence  that a multi-phase gas exists in a volume-filling way in their cores \citep[e.g. in NGC~5044 and NGC~5813,][]{dePlaa2012}, while for others it is likely the case \citep[e.g. in NGC~4696,][]{Sanders2016}. Finally, we have only considered isotropic turbulent motions, characterized by a single Mach number. If instead motions are anisotropic, the amplitude of line suppression due to scattering will change. Namely, if motions are predominantly in the radial direction, then for a given velocity amplitude the line suppression will be smaller than in the case of isotropic motions \citep{irina2011}. Measuring anisotropy effects will require spatial mapping with future microcalorimeter-like high spectral resolution data (Sect.~\ref{subsec:future}). 

Other key major systematic uncertainties are associated with the line spread function of \textit{RGS}, the abundance of heavy elements, current atomic data, charge exchange, and the choice of the maximal Mach number. We explore each of these in turn.

\subsection{\textit{RGS} instrumental uncertainty}
\label{subsec:RGSunc}

The Line Spread Function of the \textit{RGS} is known with accuracy of 5--10~per~cent, making the instrument  less sensitive to  low velocity line broadening. This uncertainty has  a  small effect on the measured line flux ratios. Therefore it does not affect the majority of our results, since most  pdfs at low velocities are dominated by the resonant scattering measurement, which relies on  the observed line flux ratios. 

There are two exceptions: NGC~1316 and NGC~4374, where the line broadening pdf  is important at low velocities. However, in these two galaxies the resonant scattering and line broadening independently provide similar constraints. If line broadening was not included, the upper limits on velocities would rise by $\sim$50~\kms~in NGC~1316 and $\sim$15~\kms~in NGC~4374 (cf. Table~\ref{tab:results}). Larger uncertainties in these galaxies would reduce the limit on the scatter around the  common sample velocity, leaving the best fit sample velocity unchanged. Therefore this effect does not affect our conclusions.

\subsection{Abundance of heavy elements}
\label{subsec:abun}

\begin{figure*}
\begin{minipage}{180mm}
\includegraphics[width=0.5\columnwidth,trim={4mm 3mm 11mm 7mm},clip]{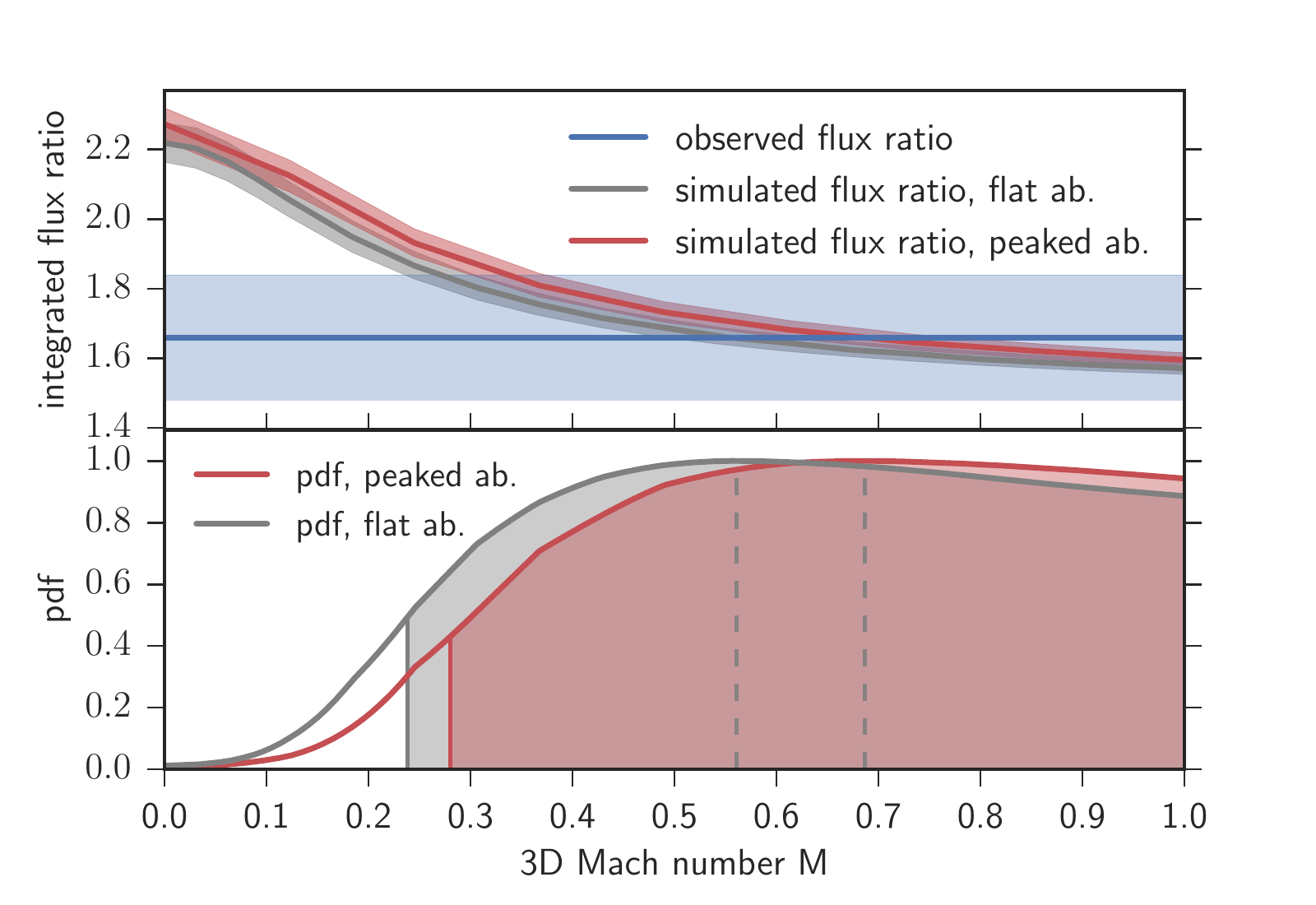}
\includegraphics[width=0.5\columnwidth,trim={4mm 3mm 11mm 7mm},clip]{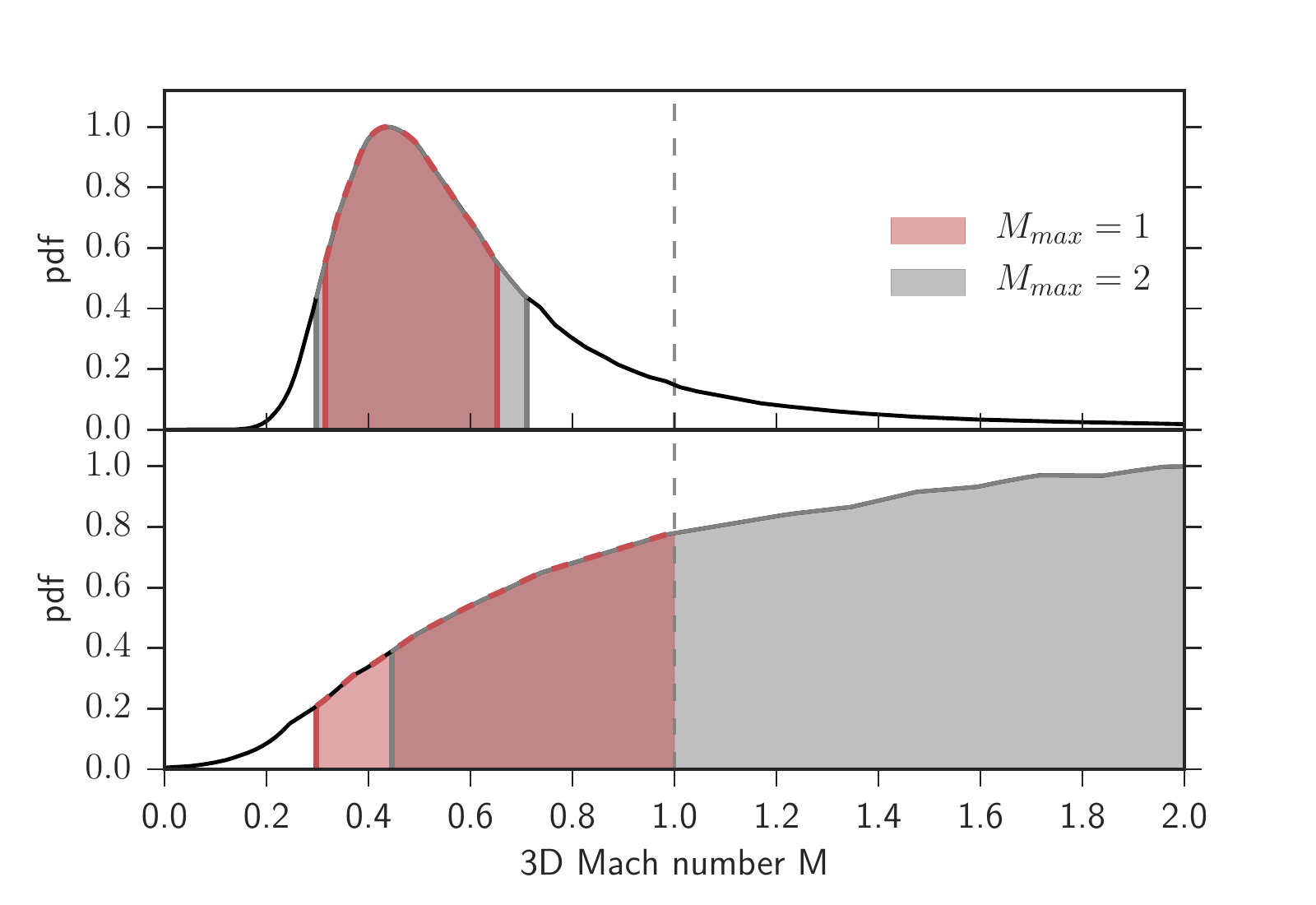}
\caption{\textit{Left}: Results for NGC~5846, assuming flat (grey; default) and peaked abundance models (red; Sect.~\ref{subsec:input}). Upper panel is the same as Fig.~\ref{fig:int_ratio}, and  lower panel is the same as Fig~\ref{fig:pdf}.   Lower limit derived with peaked abundance model is larger by 18~per cent (10 \kms),  and the most probable value is larger by 23~per~cent (29 \kms) than values inferred with flat abundance model (Sect.~\ref{subsec:abun}). \textit{Right}: Dependence of derived constraints on the maximum Mach number: the default case $M_{max}$=1 (grey dashed line), and $M_{max}$=2  case correspond to  the red and grey  confidence intervals respectively. Upper panel shows  1\sig limits in a peaked probability distribution in NGC~5813, where the lower limit changes from 0.32 to 0.3 (by 4~\kms), and the upper changes from 0.65 to 0.71 (by 12~\kms). Lower panel shows  2\sig limits in a non-peaked distribution in  NGC~5044, where the lower limit changes from 0.3 to 0.45 (by 34~\kms). See Sect.~\ref{subsec:prior} for discussion.}
\label{fig:prior}
\label{fig:fapa}
\end{minipage}
\end{figure*}

Obtaining robust abundance measurements for elliptical galaxies from current X-ray spectra  is  challenging. Thus, we conservatively choose to use a flat abundance, fixed at 0.5 Z$_{\sun}$, as our default model. This value is consistent with recent average iron abundance profile for group central galaxies in the CHEERS sample (Mernier et al. 2017, in prep.).

In order to test sensitivity of our measurements to this assumption, we repeated the calculations assuming a peaked abundance model (Eq.~\ref{eq:peaked}, Sect.~\ref{subsec:input}, and Fig.~\ref{fig:profiles}). For most galaxies the results only weakly depend on the adopted profile, with the difference $<20$~per~cent. We illustrate this in NGC~5846 (Fig.~\ref{fig:fapa}, left panel), in which the lower limit on Mach number is increased by 18~per~cent (10 \kms), and the most probable value by 23~per~cent (29 \kms) if the peaked abundance model is assumed instead of the constant value of 0.5 Z$_{\sun}$. This is smaller than  a nominal 68~per~cent  uncertainty almost by a factor of 2.

A few  objects show substantial differences, $\sim$20--50~per~cent (40--50 \kms;  NGC~1404, NGC~4261, NGC~4552, and NGC~4636). In these galaxies the physical size of the spectral extraction region is small compared to the half-light radius used to scale the peaked abundance profile, meaning that effectively the profile is flat within the extraction region with a value double the one adopted in the default flat model. This doubles the amount of \Fe ions present in the gas, leading to larger optical depths in considered lines (e.g. $\tau$ changing from  $\sim 6$ to $\sim$ 12 in NGC~4636), which  results in more efficient resonant scattering and slightly different constraints than in the default case. 

We note that abundance profiles with a central drop are sometimes observed in elliptical galaxies \citep[e.g.][]{abdrop}.  However, this drop might be a result of fitting a simplified model to a multiphase gas in the galactic core \citep[e.g.][]{Buote2003}. Therefore, we do not consider a model with a central drop in this work. 

\subsection{Atomic data}
\label{subsec:atdat}

The \Fe ion used in our analysis provides critical spectral diagnostics. However, due to the complicated inner structure of the ion, complete agreement between experimental results and theoretical predictions of the atomic properties has not yet been reached  \citep[but is expected to be achieved when high order effects are included in calculations, e.g.][]{Brown2006}.

The two major current astrophysical plasma codes, APEC and SPEX, predict slightly different optically thin ratios  of the lines considered in this work, e.g.  for a 0.6~keV plasma, the ratio $\sim$1.5 is predicted by APEC and $\sim$1.3  by SPEX. Here we use APEC ver. 2.0.2 for our fiducial models\footnote{Using the latest AtomDB ver. 3.0.7 changes results by $\sim$1~per~cent.}, because the respective atomic properties of \Fe are closer to the experimental measurements from EBIT and LCLS \citep{Brown2001,Bernitt2012}. 

Future experiments providing smaller uncertainties on measured atomic properties, accompanied by precise calculations, are needed to match the precision of X-ray microcalorimeter-type missions, in order to fully enable the range of spectral diagnostics of the \Fe ion. 

\subsection{Charge exchange}

Charge exchange between ions and neutral atoms or molecules could potentially mimic the effect of resonant scattering, since it enhances the forbidden line with respect to the resonant line. However, the magnitude of the effect is expected to be small in comparison with current uncertainties. \cite{Pinto2016} applied a simple charge exchange model to the elliptical galaxy NGC~4636, concluding that it did not affect the \Fe spectral lines, but only lines of ions tied to colder gas, e.g.  O~{\scriptsize VII}. Therefore we do not include this process in our current models.

\subsection{Maximal Mach number}
\label{subsec:prior}

Deriving uncertainties on measured Mach numbers using resonant scattering probability distributions requires us to adopt a prior on the 3D Mach number range (Sect.~\ref{subsec:pdf}). Motivated by numerical simulations and measured upper limits, we choose to consider only subsonic turbulence, $M_{max}=1$. In some cases, gas motions can be sufficiently strong to suppress the resonant scattering effect, i.e. the line ratio becomes consistent with the optically thin case. In such situations, the pdfs will be flat at large Mach numbers, and  our constraints will therefore depend on the assumed maximal Mach number. 

This is illustrated in the right side of Fig.~\ref{fig:prior}, where the upper panel shows 68~per~cent constraints in a peaked pdf in NGC~5813, and the lower panel shows 95~per~cent lower limits in a non-peaked pdf in NGC~5044, for two different maximal Mach numbers: $M_{max}=1$ (our default) and $M_{max}=2$ (red and grey regions respectively). The lower limit in NGC~5813 is least affected by this assumption: the difference between the default case and larger $M_{max}$ is  only  5~per~cent (4~\kms),   while the upper limit increases by 9~per~cent (12~\kms).  Uncertainties in  other galaxies with peaked distributions behave similarly. In NGC~5044 the 2\sig lower limit increases  by  50~per~cent (34~\kms), which is the case for most galaxies with flatter pdfs. Objects least affected are the ones with very peaked distributions, such as NGC~1404 (22~per~cent/18~\kms~change) and NGC~5846 (27~per~cent/12~\kms~change). 

Note that the combined resonant scattering and line broadening pdfs (results in Figs.~ \ref{fig:lb_res} and~\ref{fig:press})  are not affected by this assumption (Sect.~\ref{subsec:combined}).


\section{Discussion}
\label{sec:disc}
\subsection{Comparison with previous measurements}

Resonant scattering in three galaxies (NGC~4636, NGC~5044, and NGC~5813) has been previously examined  using a similar approach, and the same data and codes, by \cite{WERNER2009} and \cite{dePlaa2012}. The main difference with respect to these works is in comparison of simulations and observations, and in combining of resonant scattering and line broadening constraints. Despite these differences, the results on resonant scattering are consistent with the previous works. 

Our measurements also agree with estimations of shock-driven turbulent velocities  in NGC~5813 \citep{Randall2015}, as well as with previous upper limits measured via line broadening with \textit{RGS} data \citep{sanders2011,Sanders2013,Pinto2015}. 

Interestingly, the cold interstellar medium (ISM) velocity dispersions mapped with C~{\scriptsize II} in the cores of NGC~4636, NGC~5044, NGC~5813, and NGC~5846, match the hot gas velocity dispersions measured in this work, supporting the argument that the cold ISM cools out of the hot gas \citep{Werner2014}.

A recent \textit{Hitomi}  measurement in the inner 30~kpc region of the Perseus Cluster core (NGC~1275) showed velocity broadening of 187$\pm$13~km~s$^{-1}$ and a non-thermal pressure contribution of $\sim$4 per cent \citep[8 per cent accounting for sheer motions;][]{Hitomi_Perseus}. This velocity is  close to the best fit sample values of 107$_{-15}^{+19}$~\kms, and 5.6$_{-1.5}^{+2.6}$~per~cent non-thermal pressure support measured in this work. The sample--averaged velocity is slightly lower than the one in Perseus, but also the regions we probe in this work are smaller (few kpc vs. few 10s kpc). 

Our inferred  non-thermal pressure contribution is consistent with observations of the cores of other giant ellipticals \citep[NGC~1399 and NGC~4486,][]{Churazov2008}. Its low value encourages the use of hydrostatic equilibrium approximation for galaxy mass measurements with X-rays. 

\subsection{Comparison with numerical simulations}

Detailed hydrodynamical simulations of AGN feedback by \cite{Gaspari2012} reproduce a level of subsonic turbulence consistent with our measurements. Within the inner few kpc, the time-averaged velocity dispersion varies from $\sim$60--200~\kms~in simulations, while our typical averaged results provide velocity $\sim$93--126~\kms. Similarly, \cite{Valentini2015} find hot gas velocity dispersions of 100--200~\kms~in the cores of simulated massive elliptical galaxies. A somewhat different conclusion was recently reached by \cite{Yang2016}, where the kinetic energy in AGN feedback simulations is only $\sim$1~per~cent of the gas thermal energy (we find this fraction to be 5.6$_{-1.5}^{+2.6}$~per~cent). 

The missing element in these  simulations of AGN feedback is the effect of cosmological merging, which  triggers additional gas motions  \citep[as shown in galaxy cluster cosmological simulations by e.g.][]{Lau2009,Vazza2011}.

\subsection{Radiative cooling and turbulent heating balance}

The inferred gas velocity values measured in all objects in our sample are comparable, and there is no detectable scatter around the best fit mean value (albeit with large statistical uncertainty). This may hint at a common origin of the velocity field, which could be sourced by a quasi--continuous feedback mechanism, with the  most obvious candidate energy source being AGN, although additional contributions from e.g. merging activity may also be present \citep[e.g.][]{Ascasibar2006, ZuHone2010}.

If indeed turbulence is produced by  a feedback activity, it is interesting to estimate whether the turbulent energy dissipation  is sufficient to balance the radiative cooling. By adopting an average value  of the normalized cooling function $\Lambda_n \approx 1.5\cdot 10^{-23}$, which is representative for the temperatures and metallicities of considered galaxies, we can estimate the  Mach number of turbulence $M_l$, associated with spatial scales of motions $l$, which provides the balance \citep{Irina2014}:

\begin{equation}
M_{l} = 0.13  \left( \frac{n_e}{10^{-2}\text{cm}^{-3}}\right)^{1/3} \left(\frac{c_s}{1000 \text{\kms}}\right)^{-1} \left(\frac{l}{10  \text{ kpc}}\right)^{1/3}.
\end{equation}

A typical galaxy core electron number density in our sample is 0.05~cm$^{-3}$, and sound speed is 450~\kms. Both the typical size of the \textit{RGS} extraction region and the effective length along the line of sight \citep[distance that contributes 50~per~cent of the flux at a given projected radius, see][]{Zhuravleva2012}, are about $\sim$ 5~kpc. If our measured velocities are dominated by gas motions on such scale, then  $M_l\sim$0.4 should provide the cooling--heating balance. If, instead, scale associated with the measured velocity is smaller (larger), e.g. 1~kpc (10~kpc), then $M_l\sim$0.2 (0.5).  Our measurements give an averaged $M\sim$0.44, meaning that the  turbulence should typically be enough to offset the radiative cooling in the cores of massive elliptical galaxies. 

Interestingly, we measured (or have strong hints of) weak turbulence  in the most relaxed giant elliptical galaxies, NGC~4261, NGC~4472, and NGC~4649 \citep[as identified by e.g.][]{Werner2012}, while galaxy with the largest measured velocity, NGC~1404, is very dynamically active, moving within and interacting with the ICM of the Fornax cluster \citep[e.g.][]{Machacek2005, Su2017}. We also found a few potential exceptions in our sample, where we see the indications of an insufficient level of turbulence to balance cooling in the inner few kpc region. However, any firm conclusions require additional \textit{RGS} data.

\subsection{Further improvements}
\label{subsec:improv}
\label{subsec:turnover}

The statistical uncertainty on our measurements comes from the convolved errors from the \textit{RGS} line flux measurements and the deprojected thermodynamic profiles from \textit{Chandra} data. In most galaxies the \textit{RGS} error is the dominant one and could be reduced by acquiring additional XMM-\textit{Newton} data. The exceptions are   NGC~1316 and NGC~4636, which require deeper \textit{Chandra} observations.

The resonant scattering signal is stronger if considered in even smaller regions, where the optical depth is the largest. However, using smaller regions leads to lower photon counts and larger statistical uncertainties. Therefore, this approach  also requires additional data for most objects, except for   NGC~4261, NGC~4374, NGC~4472, and NGC~5044, where current data are sufficient to maximize the scattering signal. 

\cite{Ahoranta2016} studied spatial the variation of the optically thin to thick line ratio in NGC~4636, suggesting high turbulence in the southern, and low in the northern region of the galaxy. In order to perform quantitative analysis of this asymmetry, it is necessary to perform Monte Carlo simulations similar to these discussed in this paper. However, such analysis requires more complex models of the galaxy geometry,  relaxing the spherical symmetry assumption, which are beyond the scope of this paper.

\subsection{Future microcalorimeter-type missions}
\label{subsec:future}
\begin{figure*}
\begin{minipage}{180mm}
\includegraphics[width=0.5\columnwidth,trim={4mm 2mm 12mm 7mm},clip]{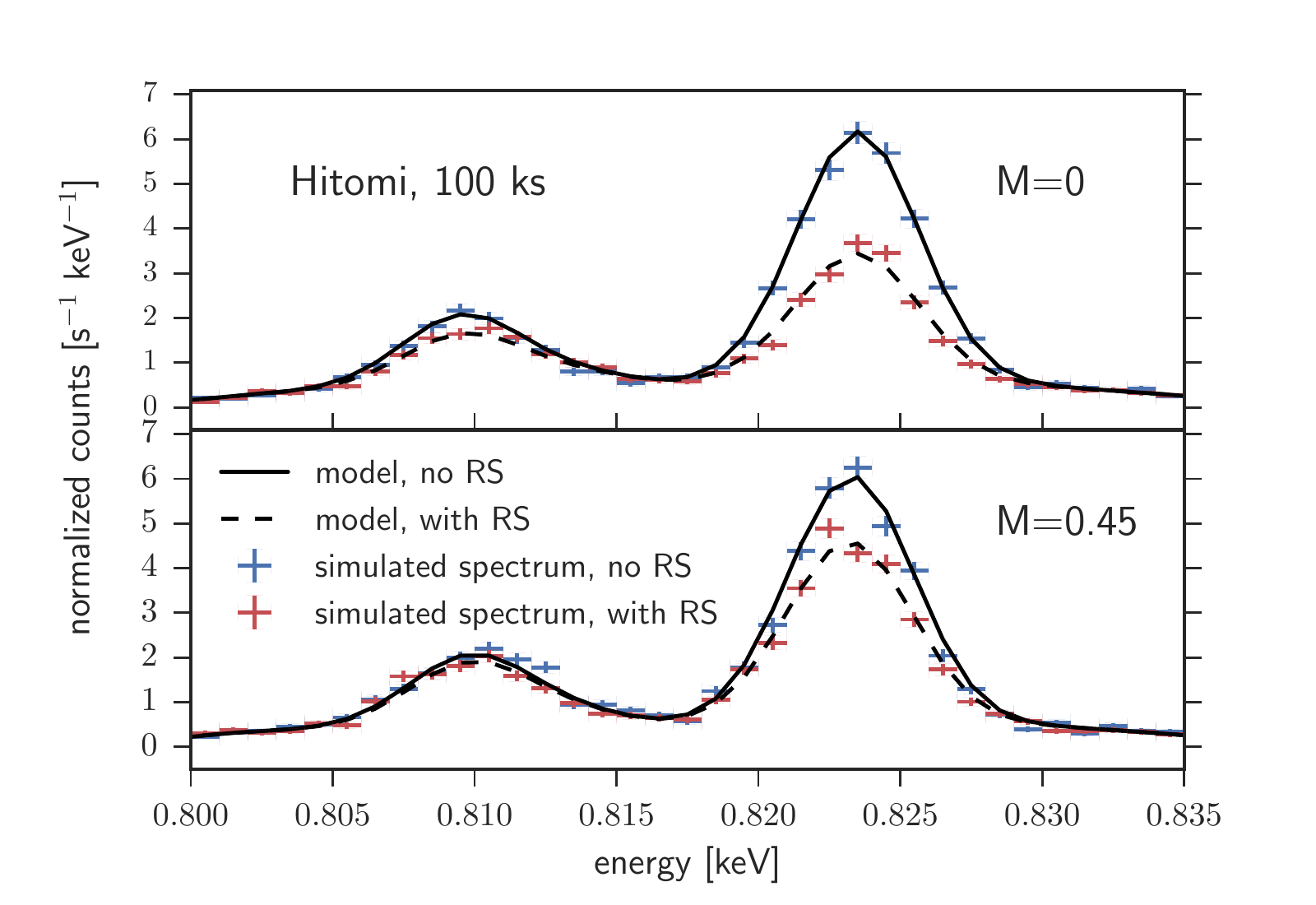}
\includegraphics[width=0.5\columnwidth,trim={4mm 2mm 12mm 7mm},clip]{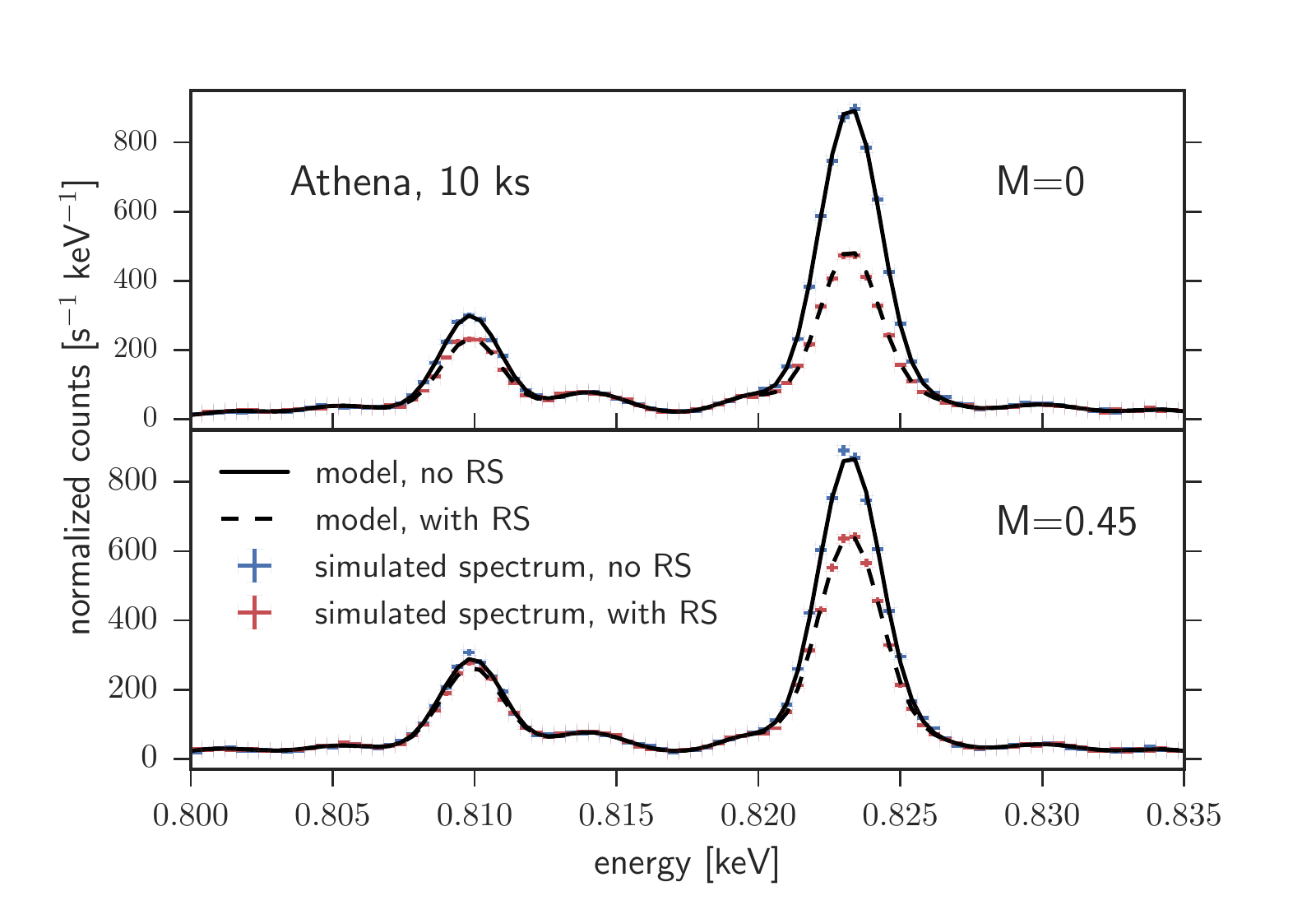}
\caption{Simulated spectrum around the \Fe 15.01 \AA\ (0.826~keV) line in the center of NGC~4636, as would be seen by a single \textit{Hitomi}  pixel  ($\sim$ 5~eV resolution; left panel) and by \textit{Athena} in the same region ($\sim$ 2.5~eV resolution; right panel). Upper panels show the case of no turbulence, and lower panels show the case of 3D Mach number of 0.45. Dashed and solid black lines show  model spectra  with and without resonant scattering (RS) respectively. Red and blue points show simulated normalized photon counts for a 100 ks and a 10~ks observation in the case of \textit{Hitomi} and \textit{Athena} respectively, with and without resonant scattering. Note that for both of the instruments the spectral  line broadening may be challenging to measure robustly.}
\label{fig:hitomi}
\end{minipage}
\end{figure*}

Future X-ray missions with microcalorimeter-type detectors, such as e.g. \textit{Athena}\footnote{\url{http://www.the-athena-x-ray-observatory.eu/}}, \textit{Lynx}\footnote{\url{https://wwwastro.msfc.nasa.gov/lynx/}}, and \textit{Hitomi Recovery Mission},  should deliver large volumes of spatially resolved data with superb spectral resolution. As shown by the \textit{Hitomi} observation of the Perseus Cluster \citep{Hitomi_Perseus}, this will allow for direct measurements of gas bulk and turbulent velocities in the centers of clusters, directly from line widths and Doppler shifts, and indirectly through resonant scattering. For cooler objects, such as galaxies, with average temperature $\sim$1~keV, it might still be challenging to measure velocities through line broadening with $\sim$2.5-5~eV resolution microcalorimeters. Typical velocity found in this work (107~\kms) causes the \Fe lines broadening of $\sim$0.5~eV, while even sonic motions would cause a sub-eV broadening in a cool, 0.75~keV gas. However, the microcalorimeter data will not suffer from the major systematic of current best instrument (\textit{RGS}; the spatial line broadening), allowing for better resonant scattering and line broadening studies. 

We show simulated \textit{Hitomi} (left) and \textit{Athena} (right) spectra of NGC~4636 in Fig.~\ref{fig:hitomi} for  no turbulence (upper panels), and substantial turbulence ($M=0.45$, as measured in this work; lower panels). Resonant scattering causes suppression in the line center of $\sim$~45 per cent for $M=0$ and $\sim$~25 per cent for $M=0.45$, which can be clearly seen with just a 100~ks \textit{Hitomi} or a 10~ks \textit{Athena} observation. 


\section{Conclusions}
\label{sec:concl}

In this work we have performed a detailed analysis of the resonant scattering effect in nearby, giant elliptical galaxies, thereby constraining  the turbulent velocities of the hot gas in their cores. By combining the velocity constraints from resonant scattering with upper limits from spectral line broadening, we  obtain significantly improved constraints on gas velocities, Mach numbers, and non-thermal pressure support. Our analysis combines high-resolution XMM-\textit{Newton} \textit{RGS} spectra, \textit{Chandra} imaging and numerical simulations of the resonant scattering effect in these galaxies.  Our main findings are as follows:
\begin{itemize}
\item Using the resonant scattering effect only, we measure 1D turbulent velocities  in the cores of 4 massive elliptical galaxies ranging between 0--120 \kms~(3D Mach numbers 0--0.5), and place lower limits of 20--100 \kms~(3D Mach numbers 0--0.5) in a further 9 systems. 
\item By combining line broadening and resonant scattering measurements, we  obtain interesting velocity constraints for 13  galaxies. The sample best fit  values are: 1D velocity dispersion   $\sim 110$~\kms, 3D Mach number $\sim 0.45$, and non-thermal to thermal pressure fraction $\sim 6$~per~cent. The scatter around these values is rather small, consistent with zero within uncertainties.
\item Consistency with a single velocity supports a common gas heating mechanism in massive elliptical galaxies. 
\item For the majority of objects the measured turbulence is enough to balance the radiative cooling, assuming conservative estimates of spatial scales associated with the measured motions.
\item The inferred non-thermal pressure contribution in the cores of giant galaxies is encouraging for hydrostatic mass measurements.
\item In the near term, our analysis could be improved by acquiring additional XMM-\textit{Newton} and \textit{Chandra} data, and by advancing the models. 
\item In the long term, this approach can be used to interpret future high-resolution observations with  \textit{Hitomi Recovery Mission},   \textit{Athena}, and \textit{Lynx}. We highlighted the major sources of systematic uncertainties, e.g. abundance measurements and atomic data, which have to be addressed to enable full exploitation of the microcalorimeter data.
\end{itemize}


\section*{Acknowledgements}
This work is based on observations obtained with XMM-\textit{Newton}, an ESA science mission funded by ESA Member States and USA (NASA). CP acknowledges support from ERC Advanced Grant Feedback 340442 and new data from the awarded XMM-\textit{Newton} proposal ID 0760870101. AO thanks J. Sanders for providing deprojected data for NGC~4696.

\bibliographystyle{mnras}
\bibliography{bib} 
%
%
\appendix
\section{Resonant scattering results for the remainder of the sample}
\label{app:all}

In Fig.~\ref{fig:pdf1} we present resonant scattering--only results for the remaining galaxies, which have not been shown in the main text. For details, see Sects.~\ref{subsec:pdf}.

\begin{figure*}
\begin{minipage}{180mm}
\includegraphics[width=0.5\columnwidth,trim={4mm 2mm 12mm 7mm},clip]{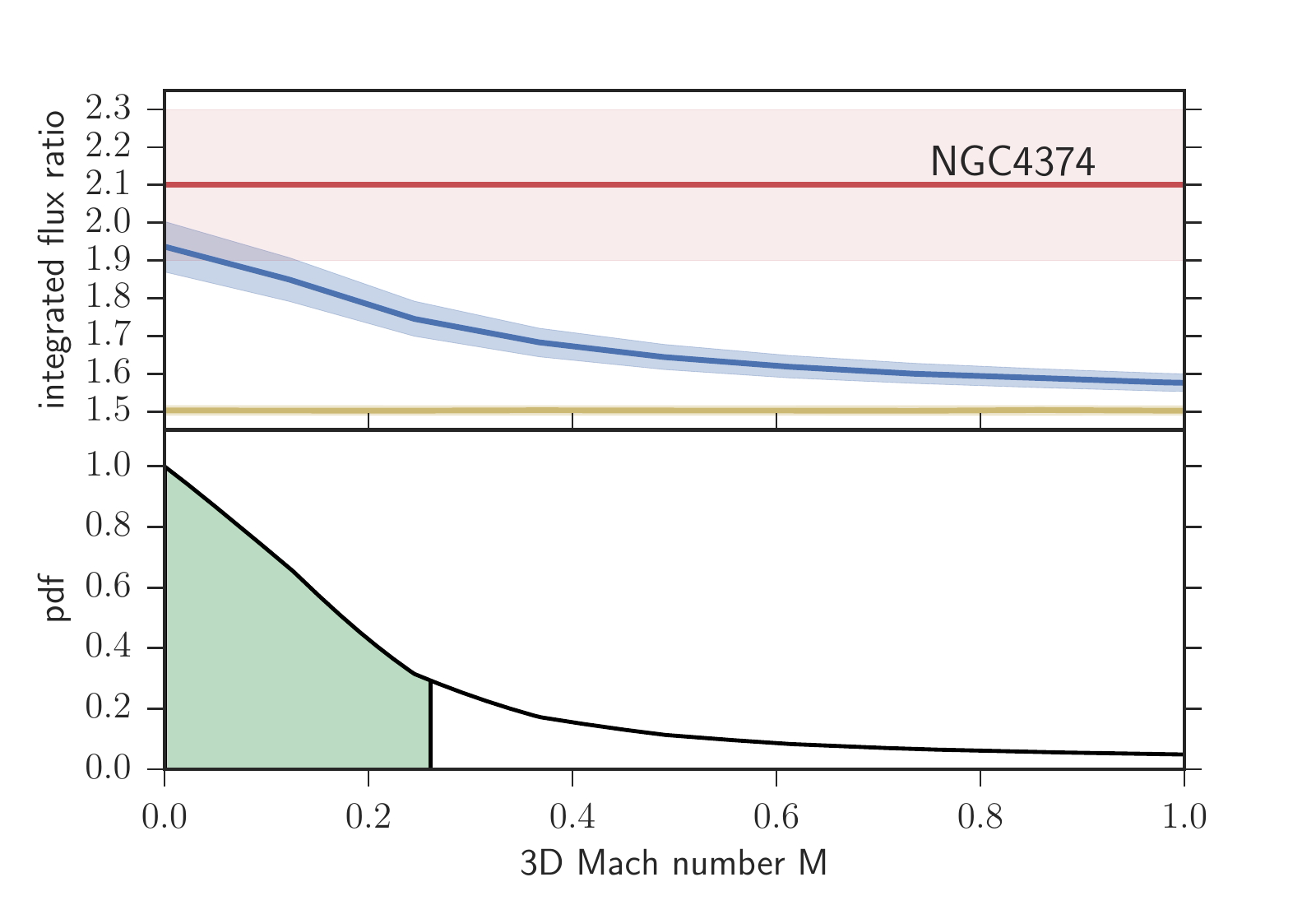}
\includegraphics[width=0.5\columnwidth,trim={4mm 2mm 12mm 7mm},clip]{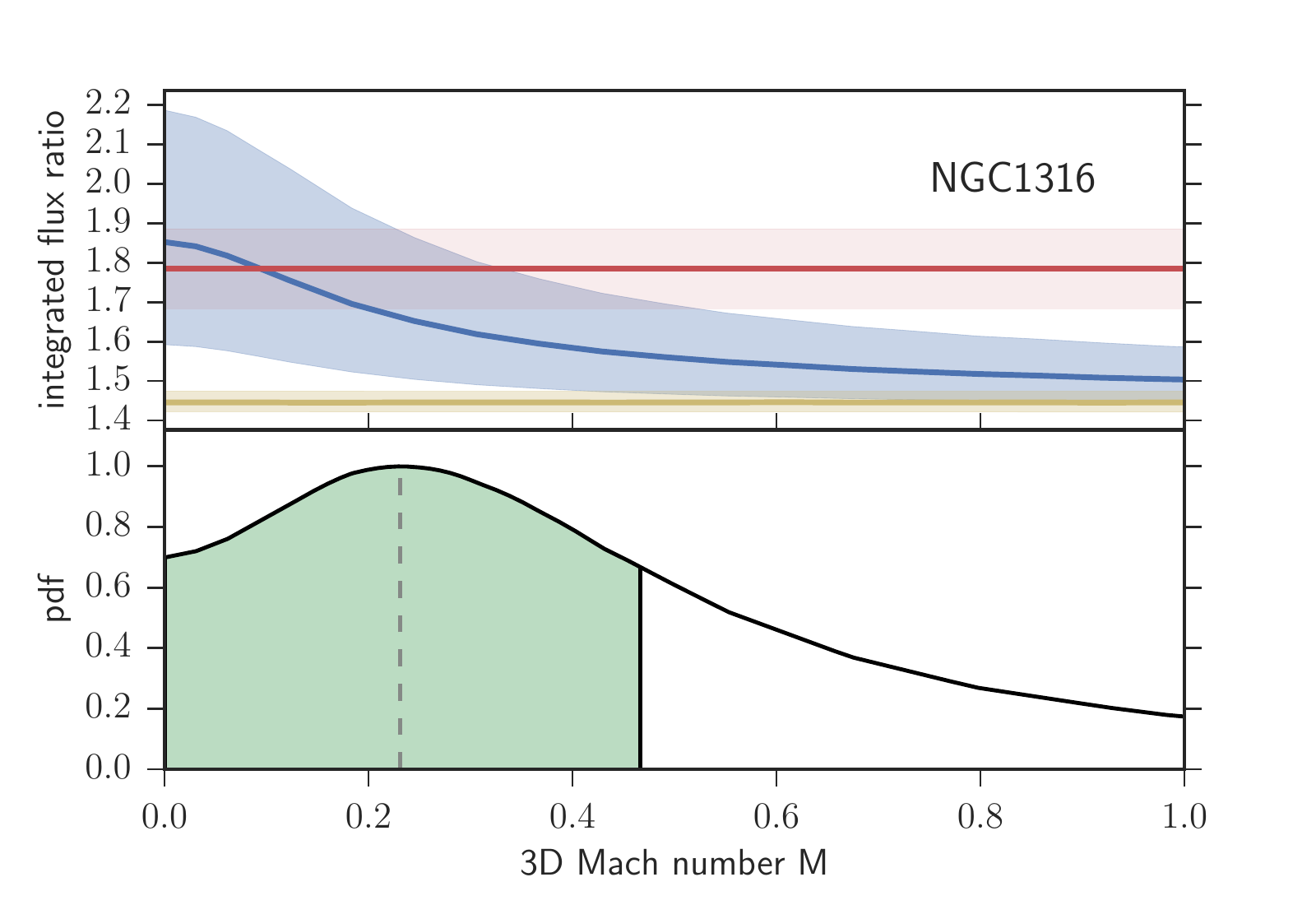}
\includegraphics[width=0.5\columnwidth,trim={4mm 2mm 12mm 7mm},clip]{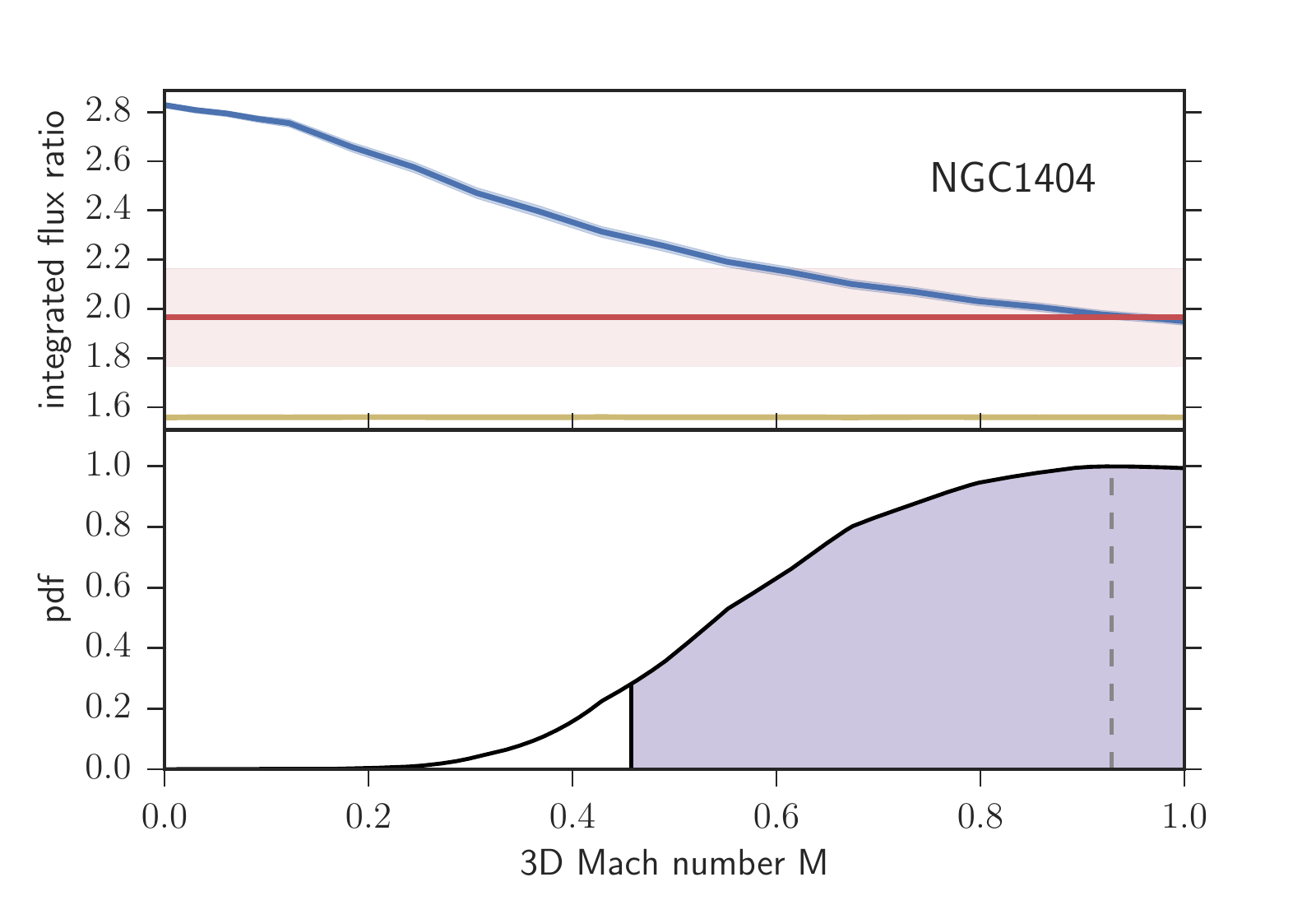}
\includegraphics[width=0.5\columnwidth,trim={4mm 2mm 12mm 7mm},clip]{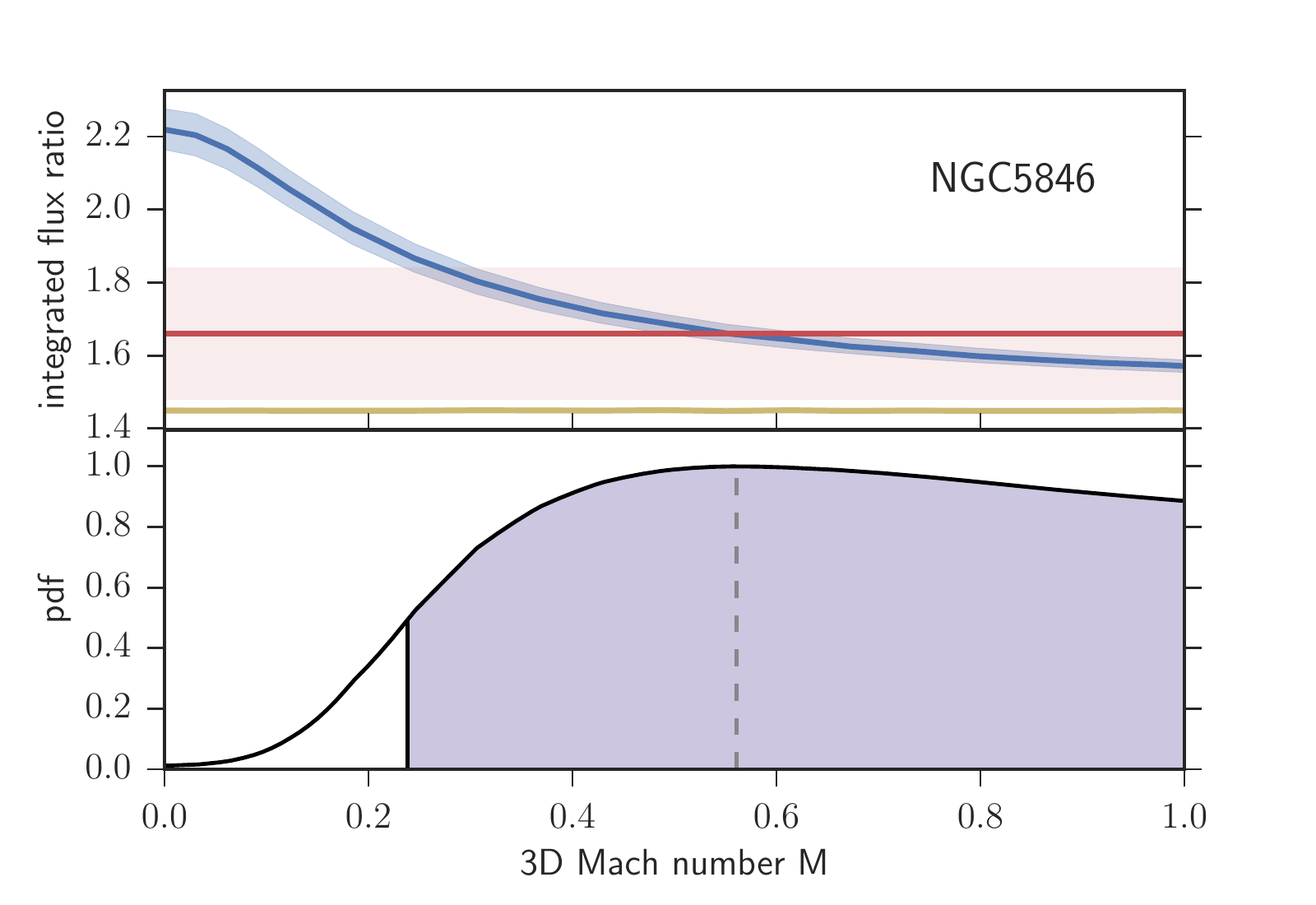}
\includegraphics[width=0.5\columnwidth,trim={4mm 2mm 12mm 7mm},clip]{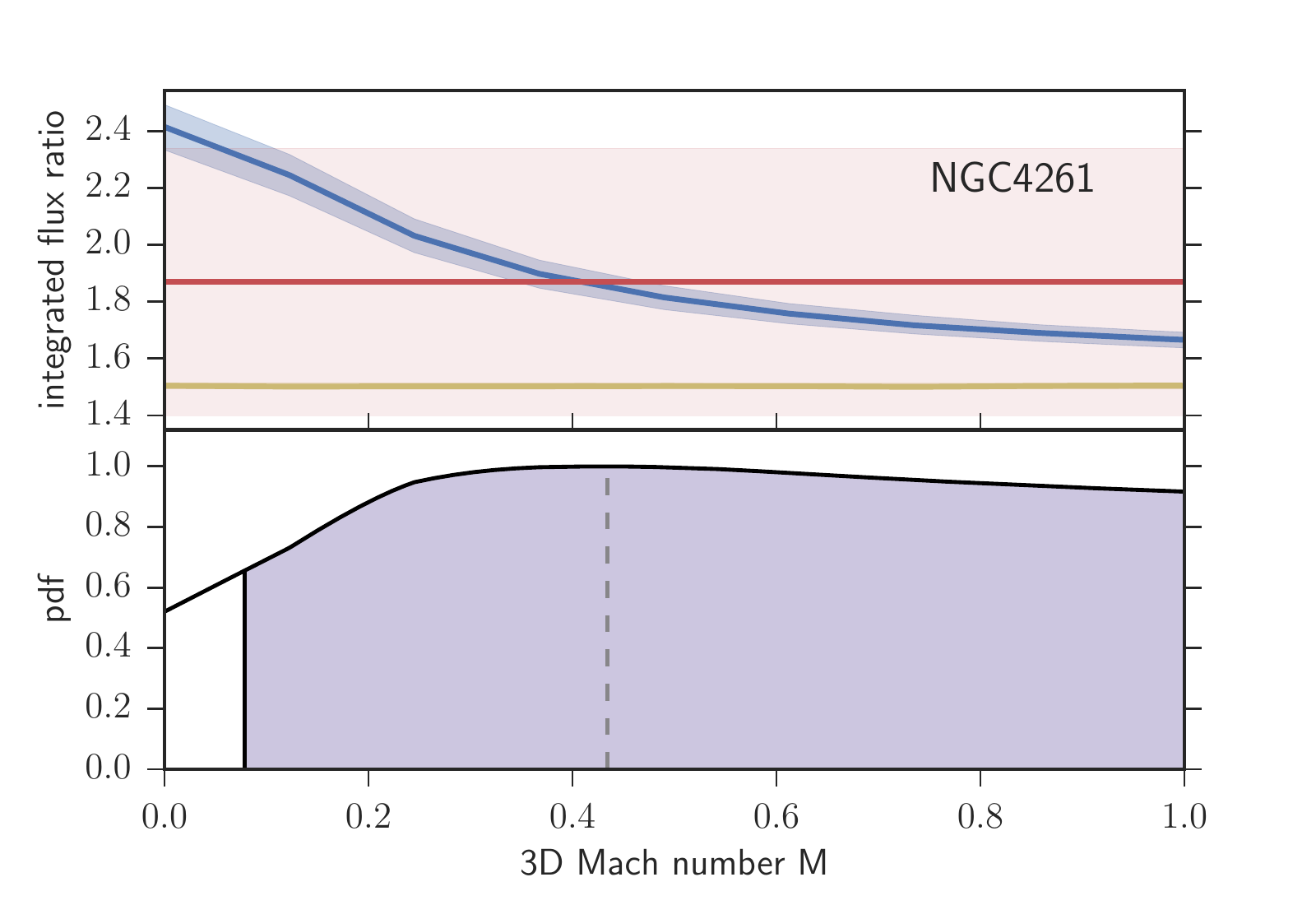}
\includegraphics[width=0.5\columnwidth,trim={4mm 2mm 12mm 7mm},clip]{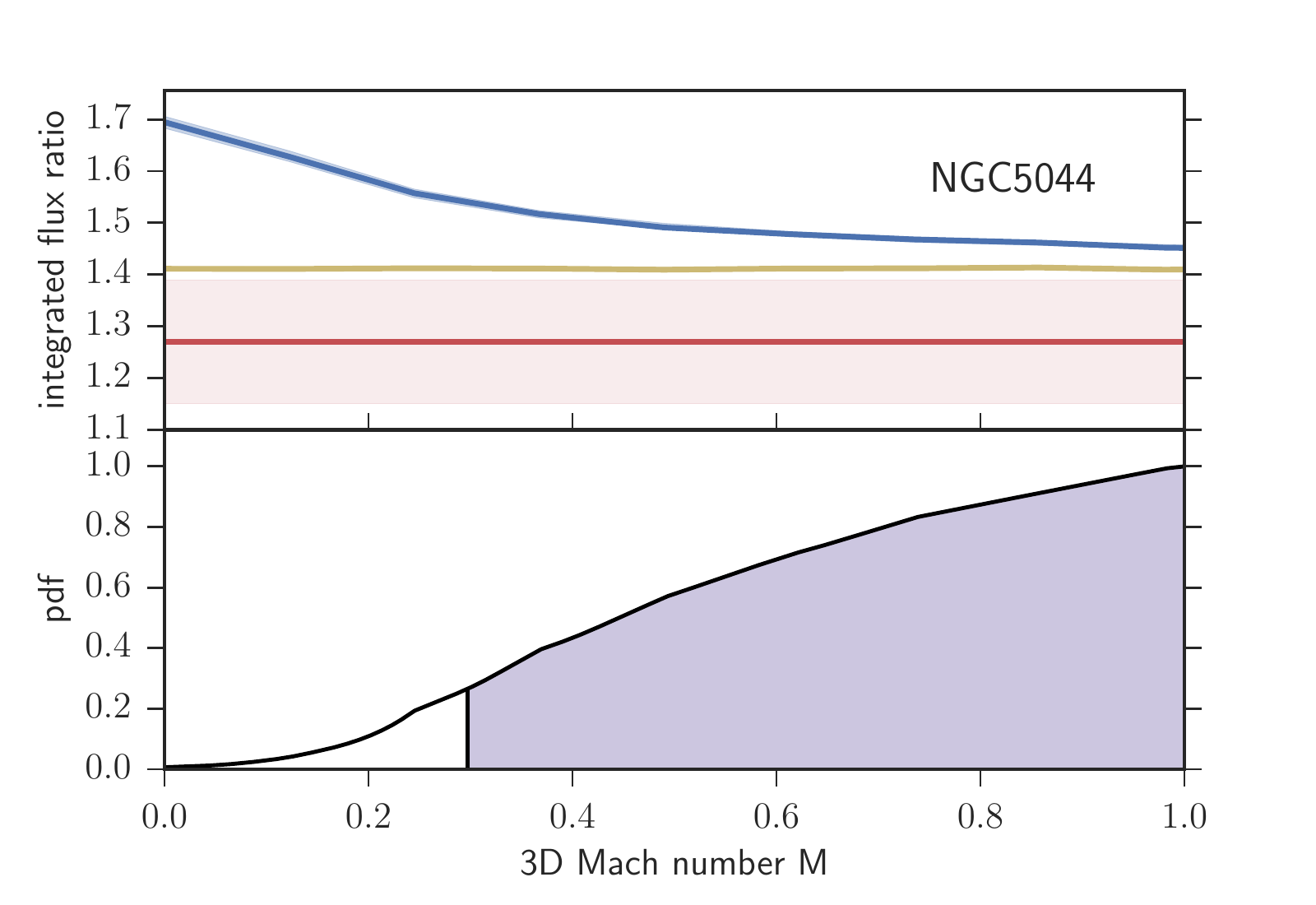}
\caption{Results for the remaining galaxies in our sample. \textit{Upper panels:} Comparison of the observed flux ratio (red line) with one simulated given different turbulent motions (blue line; see Sect.~\ref{subsec:ifr} for details). The limit of optically thin lines is shown in yellow. Respective colored regions show 1\sig uncertainties. \textit{Lower panels:} Mach number probability density function resulting from convolution of models and data shown in upper panel. Grey dashed line shows the most probable value, while green and purple regions mark the 1\sig and 2\sig constraints respectively. The figure continues onto the next page. See Sect.~\ref{subsec:sub} for details.}
\label{fig:pdf1}
\label{fig:pdf2} 
\end{minipage}
\end{figure*}
\begin{figure*}\ContinuedFloat
\begin{minipage}{180mm}
\centering
\includegraphics[width=0.48\columnwidth,trim={4mm 2mm 12mm 7mm},clip]{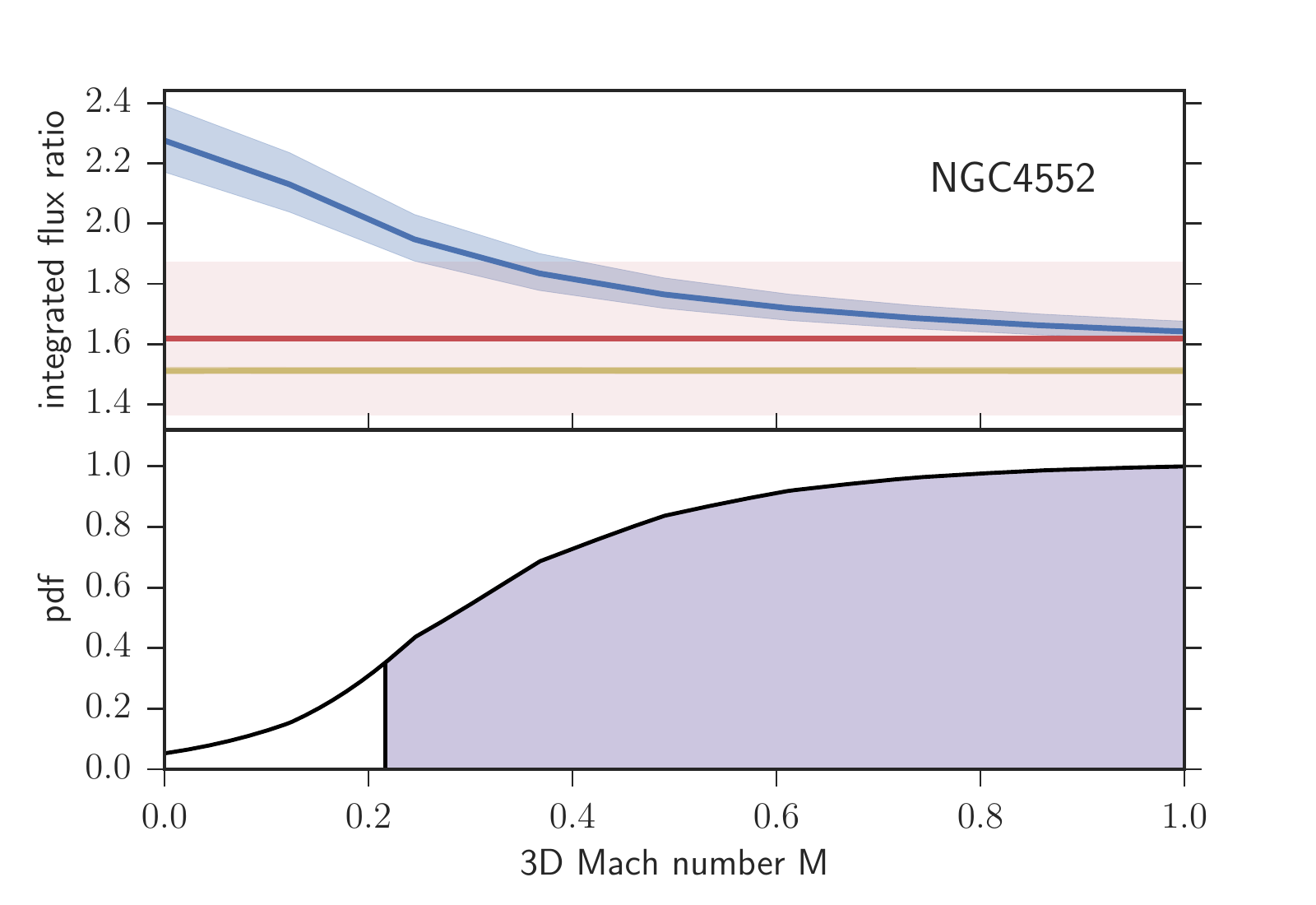}
\includegraphics[width=0.48\columnwidth,trim={4mm 2mm 12mm 7mm},clip]{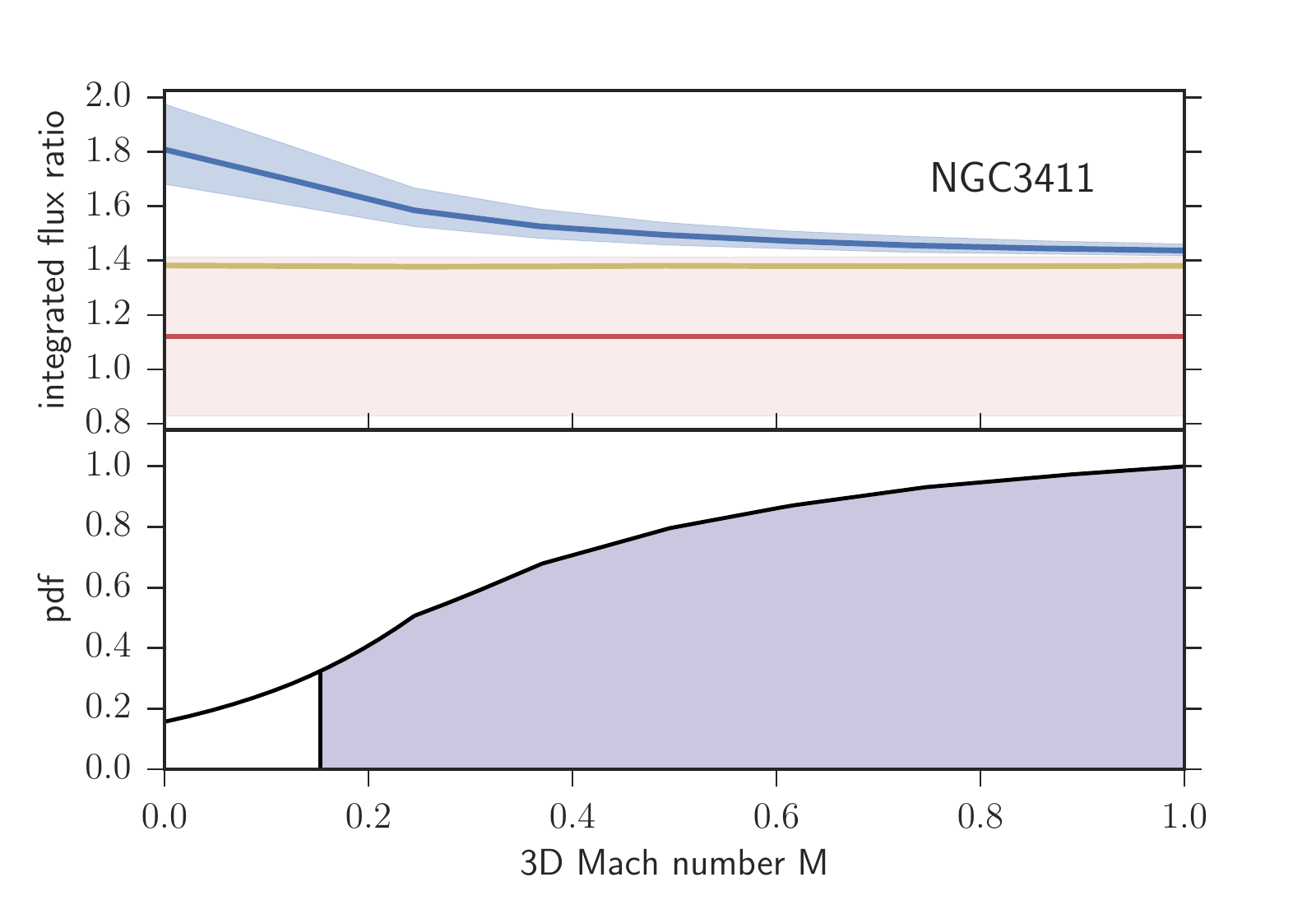}
\includegraphics[width=0.48\columnwidth,trim={4mm 2mm 12mm 7mm},clip]{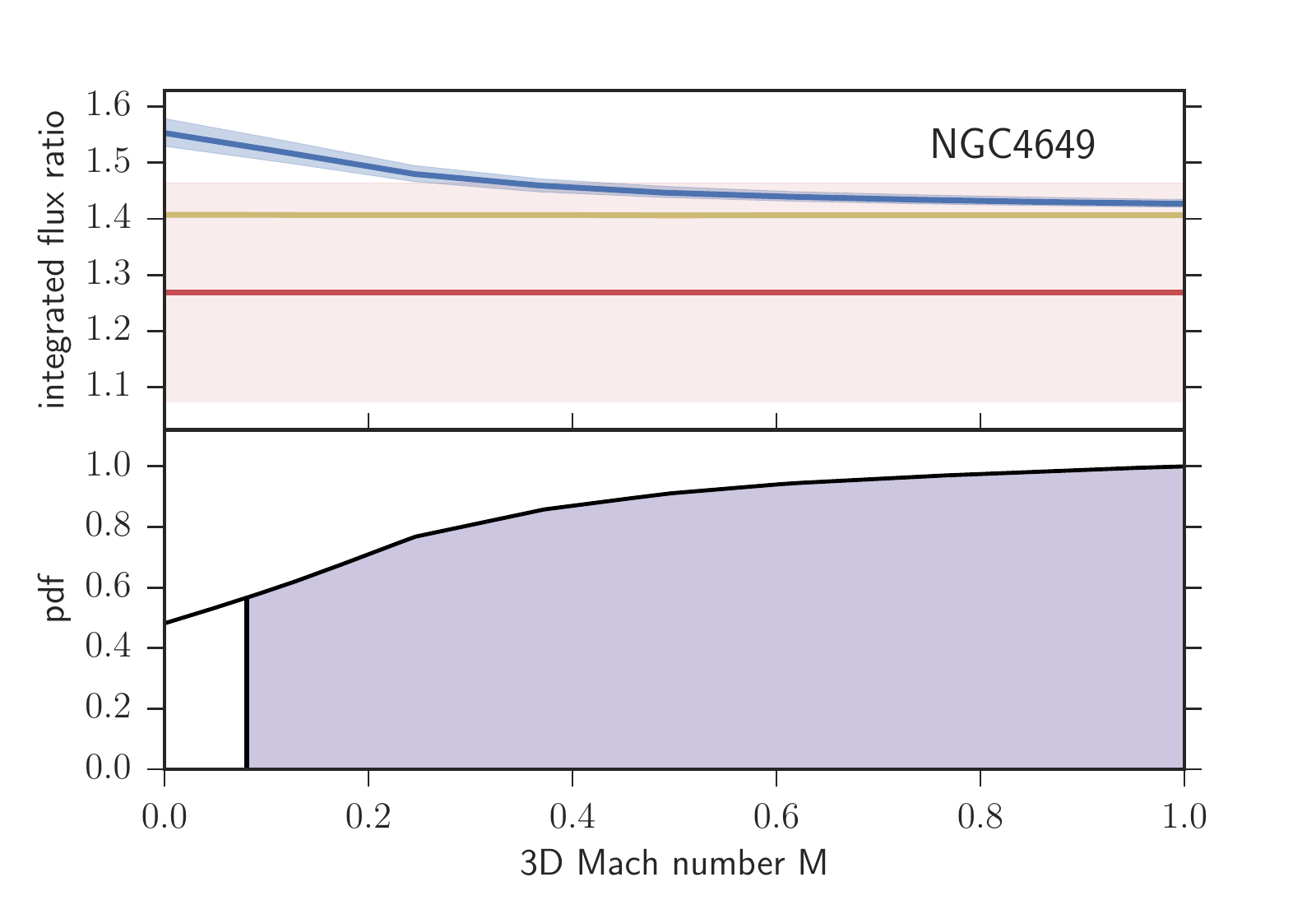}
\includegraphics[width=0.48\columnwidth,trim={4mm 2mm 12mm 7mm},clip]{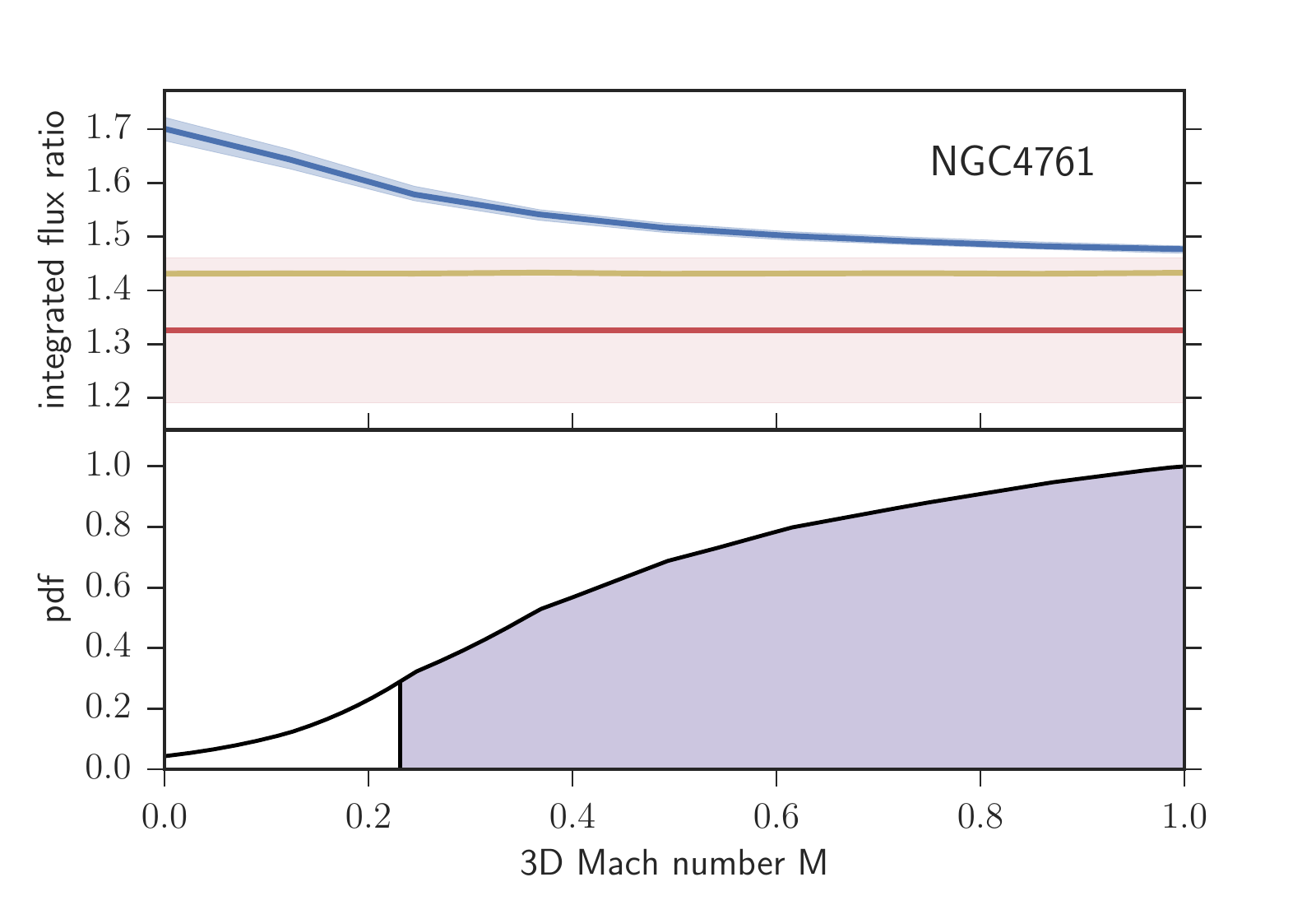}
\includegraphics[width=0.48\columnwidth,trim={4mm 2mm 12mm 7mm},clip]{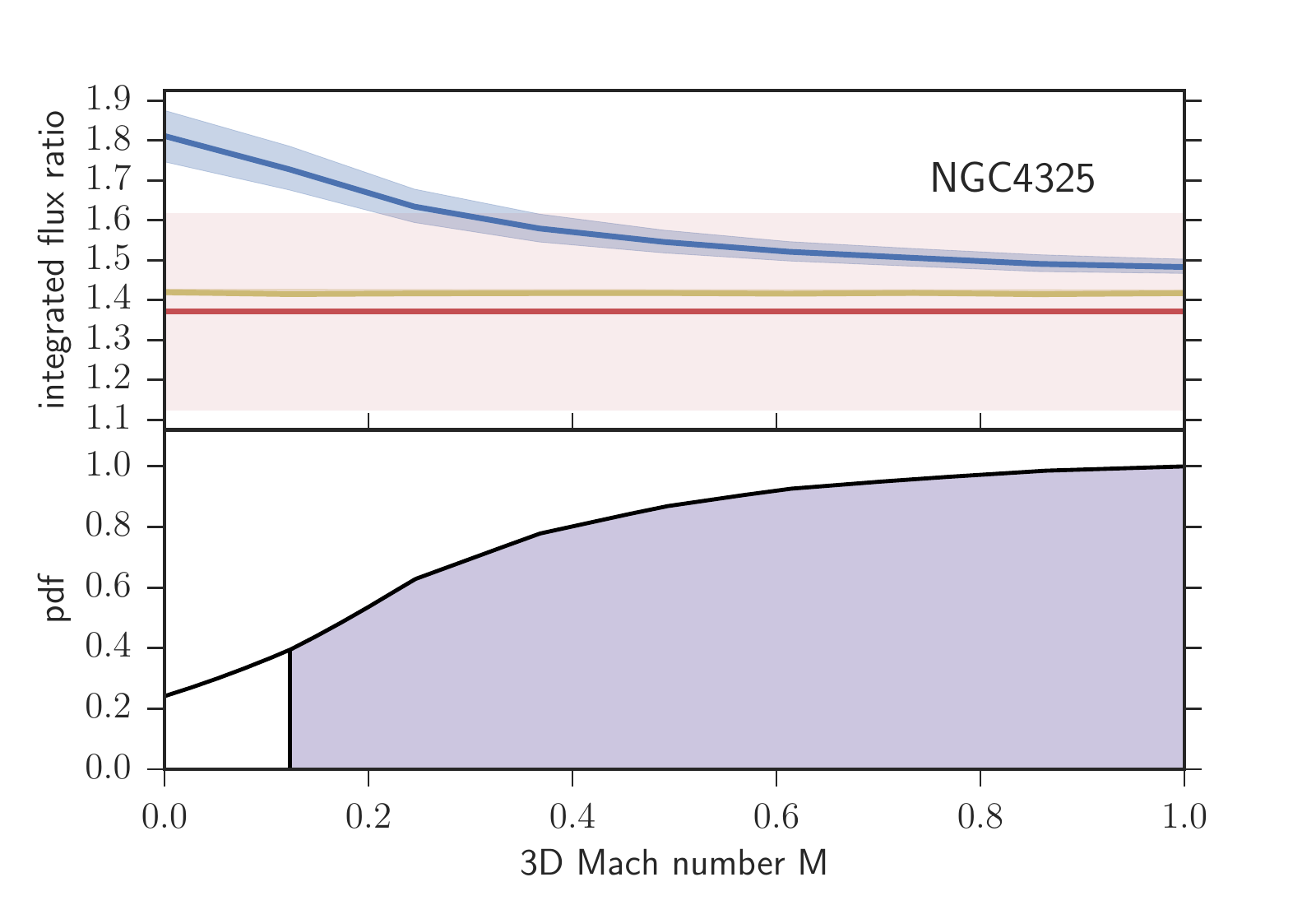}
\includegraphics[width=0.48\columnwidth,trim={4mm 2mm 12mm 7mm},clip]{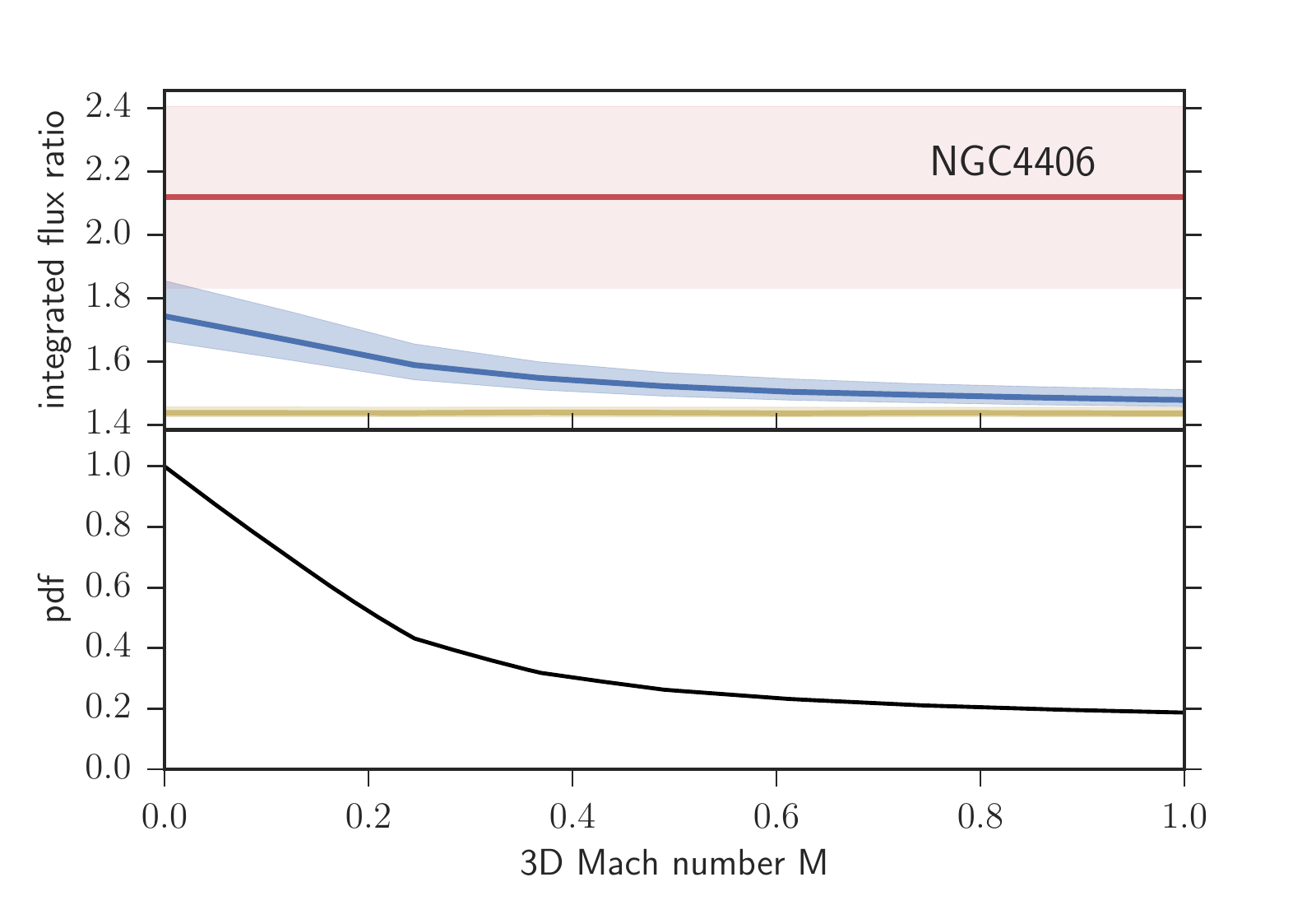}
\includegraphics[width=0.48\columnwidth,trim={4mm 2mm 12mm 7mm},clip]{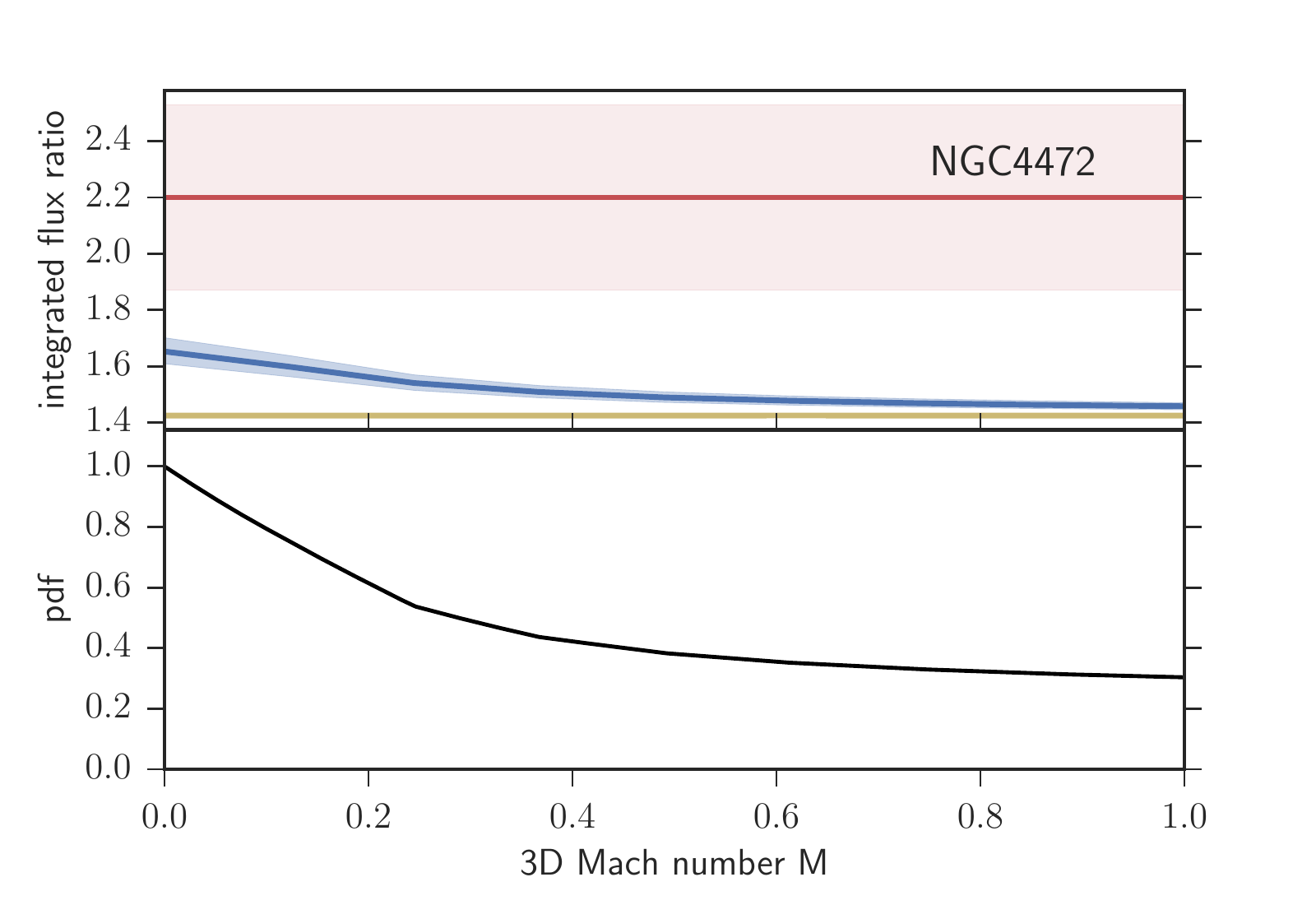}
\includegraphics[width=0.48\columnwidth,trim={4mm 2mm 12mm 7mm},clip]{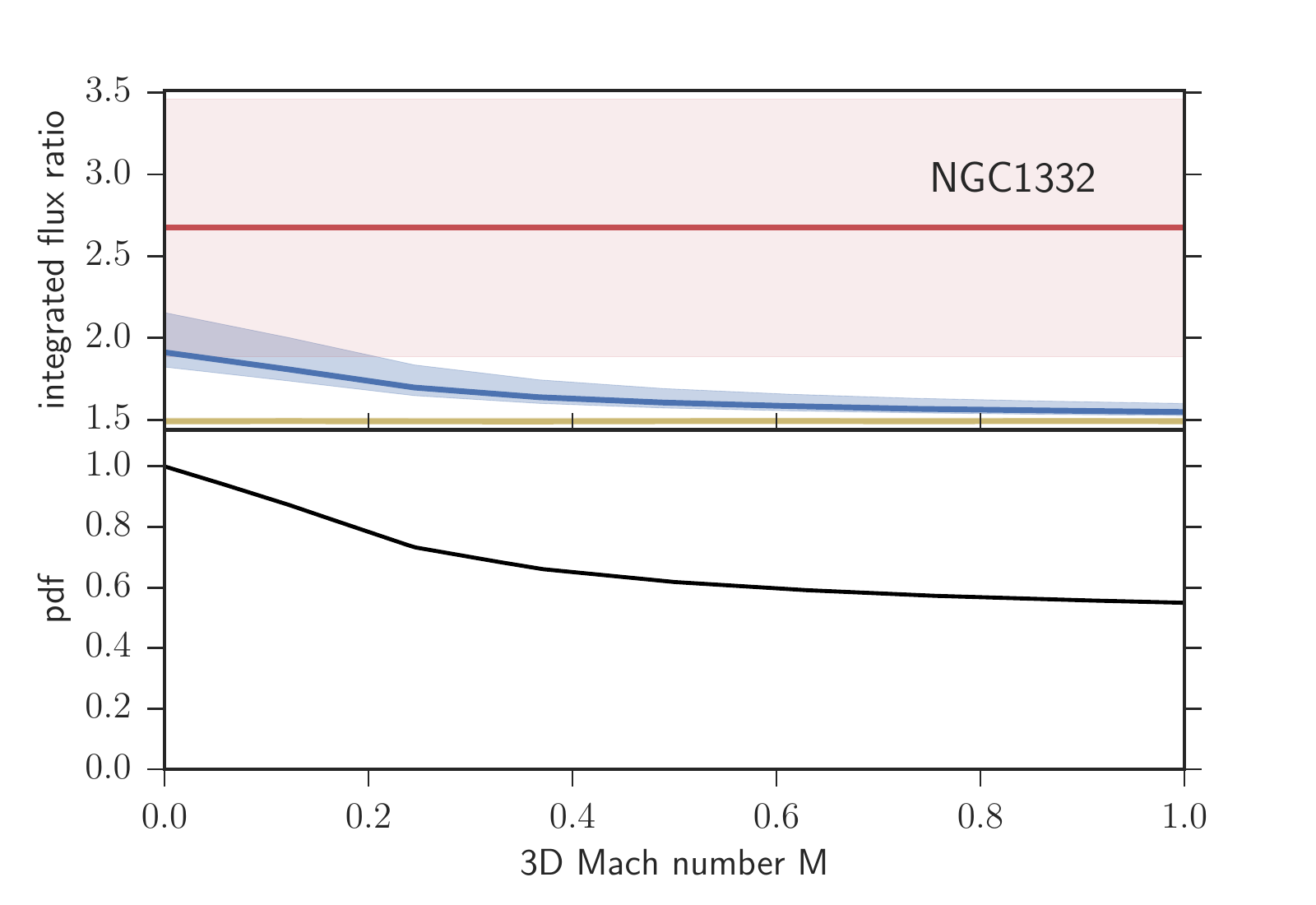}
\caption{ continued.}
\end{minipage}
\end{figure*}

\bsp	
\label{lastpage}
\end{document}